\documentclass[trackchanges,preprint2]{aastex63}

\newcommand{\rom}[1]{\MakeUppercase{\romannumeral#1}}
\usepackage{comment}
\usepackage{hyperref}
\revised{March 20, 2021}
\accepted{March 23, 2021}
\submitjournal{AJ}
\turnoffeditone

\shorttitle{{\rm Be}+{\rm sdO} Binaries}
\shortauthors{Wang et al.}
\begin{document}

\title{The Detection and Characterization of Be+sdO Binaries \\
  from HST/STIS FUV Spectroscopy}

% Author information
\correspondingauthor{Luqian Wang}
\email{lwang@chara.gsu.edu, gies@chara.gsu.edu, gpeters@usc.edu}
\email{ygoetberg@carnegiescience.edu, drewski@nmsu.edu}
\email{lester@chara.gsu.edu, steve.b.howell@nasa.gov}

\author[0000-0003-4511-6800]{Luqian Wang}
\affiliation{Center for High Angular Resolution Astronomy and Department of Physics and Astronomy\\
Georgia State University, P.O. Box 5060\\
Atlanta, GA 30302-5060, USA}
\affiliation{Yunnan Observatories, CAS, P.O. Box 110, Kunming 650011, Yunnan, China}

\author[0000-0001-8537-3583]{Douglas R. Gies}
\affiliation{Center for High Angular Resolution Astronomy and Department of Physics and Astronomy\\
Georgia State University, P.O. Box 5060\\
Atlanta, GA 30302-5060, USA}
%\email{gies@chara.gsu.edu}

\author[0000-0003-4202-269X]{Geraldine J. Peters}
\affiliation{Space Science Center, Department of Physics and Astronomy, University of Southern California, Los Angeles, CA 90089-1341, USA}
%\email{gpeters@usc.edu}

\author[0000-0002-6960-6911]{Ylva G\"{o}tberg}
\affiliation{Carnegie Observatories, 813 Santa Barbara Street, Pasadena, CA 91101, USA}
%\email{ygoetberg@carnegiescience.edu}

\author[0000-0001-9984-0891]{S.\ Drew Chojnowski}
\affiliation{Apache Point Observatory and New Mexico State University, P.O.\ Box 59, Sunspot, NM 88349-0059, USA}
%\email{drewski@nmsu.edu}

\author[0000-0002-9903-9911]{Kathryn V. Lester}
\affiliation{NASA Ames Research Center, Moffett Field, CA 94035, USA}
%\email{lester@chara.gsu.edu}
%\nocollaboration{1}

\author[0000-0002-2532-2853]{Steve B. Howell}
\affiliation{NASA Ames Research Center, Moffett Field, CA 94035, USA}
%\email{steve.b.howell@nasa.gov}
%\nocollaboration{1}

% Abstract
\begin{abstract}
The B-emission line stars are rapid rotators that were probably 
spun up by mass and angular momentum accretion through mass transfer 
in an interacting binary.  Mass transfer will strip the donor star 
of its envelope to create a small and hot subdwarf remnant. 
Here we report on {\it Hubble Space Telescope} / STIS 
far-ultraviolet spectroscopy of a sample of Be stars
that reveals the presence of the hot sdO companion through the 
calculation of cross-correlation functions of the observed and 
model spectra.  We clearly detect the spectral signature of the 
sdO star in 10 of the 13 stars in the sample, 
and the spectral signals indicate that the sdO stars
are hot, relatively faint, and slowly rotating as predicted by models. 
A comparison of their temperatures and radii with evolutionary 
tracks indicates that the sdO stars occupy the relatively 
long-lived, He-core burning stage.  Only one of the ten detections 
was a known binary prior to this investigation, which emphasizes 
the difficulty of finding such Be+sdO binaries through optical 
spectroscopy.  However, these results and others indicate that
many Be stars probably host hot subdwarf companions. 
\end{abstract}

%% Keywords should appear after the \end{abstract} command. 
%% See the online documentation for the full list of available subject
%% keywords and the rules for their use.

%%%%%%%%%%%%%%%%%%%
\section{Introduction} \label{sec:intro}

% Binarity & evolutionary scenarios
Close binaries are common among massive stars, and growing evidence suggests that 
many massive stars are evolutionary products of post-interacting binary systems that 
have experienced mass transfer \citep{demink2014}. There are two main routes of 
binary interaction.  In the case where the companion star has much lower mass, 
the more massive component will evolve and fill its Roche lobe, and may engulf 
the companion in a common envelope stage that leads to a merger.  Alternatively, 
for systems with more comparable masses, the more massive, donor star will transfer
its mass and angular momentum to the gainer star through Roche lobe overflow. 
The orbit shrinks and promotes further mass transfer until the two components reach 
comparable masses. Then continuing mass transfer will expand the orbit until the 
donor star is stripped of its outer envelope, leaving a hot core with a size much 
smaller than its Roche lobe.  The gainer star will become a rapid rotator. 

% Be stars & binary fraction 
Be stars are B-type main sequence stars, and their optical spectra display broad 
H$\alpha$ emission that originates in an outflowing, circumstellar decretion disk. 
They are rapidly rotating stars, and their projected rotational velocities $V\sin{i}$ 
can reach up to 70 to 80$\%$ or larger of their critical rotational velocities 
\citep{yudin2001, huang2010}. \citet{pols1991} and \citet{shao2014} argue that Be stars 
were spun up through close binary interactions.  If so, then many Be stars will have 
companions that are the remnants of the mass donor.  The final stripped donor star may 
explode to create a neutron star in a Be X-ray binary, or it will evolve into a faint, 
hot, low-mass subdwarf (sdO) star and create a Be+sdO binary system. \citet{pols1991} 
predict that majority of the evolved companions in Be binaries are He stars (sdO remnants). 
Recently \citet{bodensteiner2020a} showed that Be stars lack normal B-type companions, in stark
contrast to regular B-type stars, and they argue that this implies that many Be stars
are the products of binary interaction (either mergers or binaries with faint evolved 
companions).  

% General description of sdO stars
\edit1{The sdO spectral designation encompasses a wide range of hot objects with diverse evolutionary histories. Many sdO stars occupy the extreme horizontal branch in the H-R diagram, and they typically have effective temperatures in the range of 35 kK to 55 kK, gravities $\log{g}$ spanning the range of 5.1 and 6.4 (c.g.s\ units), and luminosities from 10 to 100 times solar luminosity \citep{heber2009,heber2016}. Most of the sdO stars are probably descendants of red giant stars, although some have hot temperatures and high luminosities similar to those of the central stars of planetary nebulae \citep{heber1988,rauch1991}. The sdO stars with temperatures below 40~kK are probably He-core burning objects, while the hotter ones have entered a subsequent He-shell burning stage. The sdO companions of Be stars must have formed by stripping through binary interactions, and they typically have larger masses and radii than the general population of low-mass sdO stars.} 

% Known detections of Be+sdO binaries
The sdO components of Be binary systems are faint and hard to detect, but they are hot and contribute more flux 
in the shorter wavelength part of the spectrum due to their high temperature.  Thus, 
searches for their spectral features are more favorable in the far-ultraviolet (FUV).  The first detection of 
the spectrum of an sdO star in a Be binary $\phi$ Per was reported by \citet{thaller1995} using the 
\emph{International Ultraviolet Explorer} (\emph{IUE}) FUV spectroscopy. They estimated that 
the sdO companion contributes $\sim$12$\%$ of the UV flux of the Be star.  This result was 
confirmed by \citet{gies1998} from FUV observations with \emph{Hubble Space Telescope} 
(\emph{HST}), and the sdO-to-Be flux ratio was estimated to be 16$\%$ at 1374 \AA. 
The binary orbit of this system was subsequently resolved through long baseline interferometry 
with the CHARA Array by \citet{mourard2015}.  Continuing searches for sdO companion stars using 
\emph{IUE} spectroscopy led to the discovery of another three Be+sdO systems: 
FY~CMa \citep{peters2008}, 59~Cyg \citep{peters2013}, and 60~Cyg \citep{wang2017}. 
The sdO companions in these systems contribute $\sim4\%$ of the FUV flux. 
One fainter sdO companion was found in HR~2142,
which has an estimated flux ratio of only $\sim 1\%$ \citep{peters2016}.  
\citet{wang2018} conducted an \emph{IUE} survey study to search for additional Be+sdO systems, 
and they found another 12 candidates in a sample of 264 stars that were not known binaries. 
\citet{chojnowski2018} reported a detection of a sdO companion through an investigation of a 
large set of optical spectra in Be binary system HD~55606, and several other systems are 
suspected of hosting a hot companion from clues in their optical spectra 
\citep{koubsky2012, koubsky2014, harmanec2020}. The detected sdO companions have temperatures in the range of 42 to 53 kK and masses between 0.7 and 1.7$\ M_\odot$. 

% Importance of these systems: impact on ISM, model construction, and stellar evolution
These Be+sdO binaries are important for studies of massive star evolution, because a 
large fraction of massive stars have nearby stellar companions \citep{sana2012} and 
binary interactions play a key role in their destinies \citep{demink2014}. 
Hot sdO stars may contribute significantly to the UV flux of stellar populations 
\citep{han2010, gotberg2019} and constitute an important contribution to 
spectral synthesis models \citep{eldridge2017}. Those sdO stars with a mass above the 
Chandrasekhar limit may explode as hydrogen deficient supernovae (SN Ib and SN Ic; 
\citealt{eldridge2013}). Binary systems at the high mass end may also be closely related 
to the progenitors of neutron star pairs that can merge to create bright gravitational 
wave sources \citep{tauris2017}. 

% Connections to LB-1 and HR 6819
\edit1{Be+sdO binaries may also be related to several systems that are claimed to be black hole binaries. \citet{liu2019} found that the LB-1 system consists of what appears to be a B-giant star with an invisible companion. More recently, \citet{rivinius2020} reported that HR~6819 also hosts a B-giant orbiting an unseen companion. In both cases, the mass of the faint component is much larger than that of the B-giant, so that the companions are possibly black holes but without the X-ray emission that characterizes the massive X-ray binaries. Also, in both cases Balmer line emission is present, the defining criterion of Be stars. Subsequent work indicates that both the emission lines \citep{liu2020,gies2020} and faint, broad absorption lines \citep{shenar2020,bodensteiner2020b,elbadry2021} show the small-amplitude, reflex orbital motion around the B-giant component. If the Be star component has normal mass, then the B-giant component must be an unusual, low mass object. Furthermore, the atmosphere of the B-giant component in LB-1 is 
enriched in He and N \citep{irrgang2020,shenar2020}, and a N enrichment is found for HR~6819 \citep{bodensteiner2020b,elbadry2021}. These enrichments are associated with He-core burning through the 
CNO-cycle, and their presence suggests that plasma from the core now occupies their atmospheres.  Thus, the B-giant star in both cases may be the stripped remnant of mass transfer caught during a short transition stage leading to a Be+sdO system (Section 7).} 

% Continue work of IUE survey study & outline of this paper
It is vitally important to obtain FUV spectroscopy of Be+sdO systems, because 
analysis of their rich spectra offers us the means to determine 
the physical parameters of the stripped-down remnants.  
Here we report on new FUV spectroscopy from the \emph{HST} / 
Space Telescope Imaging Spectrograph (STIS) of the
sample of candidate Be+sdO binaries from the work of \citet{wang2018} and 
\citet{chojnowski2018}.  We describe the FUV observations and associated reduction details 
in Section 2. In Section 3, we present measurements of the radial velocities of the Be stars.
These are used in Section 4 to remove the Be star contribution to cross-correlation functions
(CCFs) of the observed spectrum with a model spectrum for the hot component. 
We detect the CCF signal from a hot sdO component in 10 of the 13 systems.    
In Section 5, we describe our methodology to determine the physical properties 
of the sdO stars for the detections.  We use model fits of the observed spectral energy 
distributions in Section 6 to determine the angular diameters of the Be stars, 
and these are used with the estimated distances and radius ratios to estimate 
the radii of each stellar component. Section 7 presents a comparison of the derived 
temperatures and radii of the sdO components with evolutionary tracks.  We summarize 
our work in Section 8.  

%%%%%%%%%%%%%%%%%%%%%%%%%%%%%%%%%
\section{{\it HST}/STIS FUV Spectroscopy} \label{sec:observations}

The targets of this investigation were identified as candidate Be+sdO binaries 
by \citet{wang2018} through an examination of cross-correlation functions (CCFs)
of high dispersion, Short Wavelength Prime (SWP) camera spectra from {\it IUE} with 
a model spectrum with an assumed temperature of 45~kK.  These 12 targets all 
displayed narrow CCF peaks (unlike the broader Be star features), and the 
CCF peaks were generally velocity variable (in cases with multiple spectra), 
indicating Doppler shifts from orbital motion.  The final target HD~55606 was 
discovered to be a Be+sdO binary with a 93.76 day period by \citet{chojnowski2018}, 
who observed narrow components of \ion{He}{1} lines from the sdO component in 
high quality optical spectra. Table 1 summarizes the properties of the sample.
It lists the Henry Draper Memorial (HD) catalog number, star name, spectral classification, 
and physical properties of the Be stars, including the effective temperature ($T_{\rm eff}$), 
gravity ($\log g$), and projected rotational velocity ($V \sin i$), as well as source 
references for these estimates. We also include in last column an estimate of $V \sin i$ 
that we applied in creating models of the Be star spectrum to account for its contribution 
to the CCF analysis (see Section 4).

% Table 1: Observations
\begin{deluxetable*}{lclcccccccc}
\rotate
%\tabletypesize{\scriptsize}
\tablenum{1}
\tablecaption{List of targets\label{tab:observations}}
\tablewidth{0pt}
\tablehead{
\colhead{HD} & \colhead{Star} & \colhead{Spectral} & \colhead{Reference} & \colhead{$T_{\rm eff}$}  & \colhead{Reference} & \colhead{$\log{g}$} &  \colhead{Reference} & \colhead{$V\sin{i}$} & \colhead{Reference} & \colhead{$V\sin{i}$ (\emph{HST)}} \\
\colhead{Number} & \colhead{Name} & \colhead{Classification} & \colhead{} & \colhead{(K)}  & \colhead{} & \colhead{(cm s$^{-2}$)} & \colhead{} &  \colhead{(km s$^{-1}$)} & \colhead{} & \colhead{(km s$^{-1}$)}
}
\startdata
 29441  & V1150 Tau    & B2.5 \rom{5}ne & 1 & 20350 &  2 & 4.0 & 11 & 338 &  3 & 380 \\ 
 43544  & HR 2249      & B3 \rom{5}     & 1 & 21500 &  4 & 3.9 &  4 & 260 &  7 & 260 \\ 
 51354  & QY Gem       & B3 \rom{5}e    & 1 & 20000 & 11 & 4.0 & 11 & 330 &  5 & 330 \\ 
 55606  & MWC 522      & B2 \rom{5}nnpe & 6 & 27350 &  7 & 4.3 &  7 & 335 &  7 & 335 \\ 
 60855  & V378 Pup     & B3 \rom{4}     & 1 & 20000 & 11 & 4.0 & 11 & 244 &  8 & 300 \\ 
113120  & LS Mus       & B2 \rom{4}ne   & 4 & 22800 &  7 & 3.7 &  7 & 339 &  7 & 460 \\ 
137387  & $\kappa$ Aps & B2 \rom{5}npe  & 4 & 23950 &  7 & 4.0 &  7 & 250 &  7 & 380 \\ 
152478  & V846 Ara     & B3 \rom{5}npe  & 4 & 19800 &  7 & 3.7 &  7 & 295 &  7 & 300 \\ 
157042  & $\iota$ Ara  & B2.5 \rom{4}e  & 1 & 25860 &  7 & 4.2 &  7 & 340 &  7 & 320 \\ 
157832  & V750 Ara     & B1.5 \rom{5}e  & 9 & 25000 &  9 & 3.9 & 11 & 277 & 10 & 250 \\ 
191610  & 28 Cyg       & B3 \rom{4}e    & 1 & 20470 &  7 & 3.7 &  7 & 300 &  7 & 300 \\ 
194335  & V2119 Cyg    & B2 \rom{3}e    & 1 & 25600 &  7 & 4.3 &  7 & 360 &  7 & 360 \\ 
214168  & 8 Lac A      & B1 \rom{4}e    & 1 & 27380 &  7 & 4.1 &  7 & 300 &  7 & 360 \\ 
\enddata
%\tablenotetext{}{  }
\tablecomments{Indices of references: (1) \citet{slettebak1982} (2) \citet{hohle2010} (3) \citet{jaschek1982} (4) \citet{levenhagen2006} (5) \citet{halbedel1996} (6) \citet{chojnowski2018} (7) \citet{zorec2016} (8) \citet{huang2006} (9) \citet{lopes2011} (10) \citet{yudin2001} (11) Estimated values based upon spectral classification.}
\end{deluxetable*}

We obtained FUV spectra of the 13 Be+sdO candidate systems with the \emph{Hubble Space Telescope} 
and Space Telescope Imaging Spectrograph (STIS) \citep{kimble1998}.  
The spectra were made with the MAMA detector and the E140M echelle grating. 
The spectra cover a wavelength range of $1144 - 1710$ \AA\ with a resolving power of $R=45,800$
\citep{riley2019}. Most of the stars are bright and required the use of the {\tt 0.2X0.05ND} or 
other aperture with a neutral density filter in order to meet the bright flux limits
of the MAMA detector.  The sole exception was the fainter star, HD~55606, where 
the default {\tt 0.2X0.2} aperture was used.  The observations were made for each target in 
three, single orbit visits between 2019 August and 2020 February at intervals of days
to months.  We usually obtained a signal-to-noise ratio of 30 per pixel or better 
in the best exposed parts of the spectrum. 

The FUV spectra were reduced, and the wavelength and flux were calibrated, following the 
standard STIS pipeline \citep{sohn2019}.  We subsequently combined the echelle orders onto 
a single uniform wavelength grid on a $\log \lambda$ scale with an equivalent pixel size of 7.5 km~s$^{-1}$. 
A number of prominent interstellar medium lines, such as 
\ion{Si}{4} $\lambda\lambda$1394,1403, and \ion{C}{4} $\lambda\lambda$1548,1550, 
were removed from the spectra by linear interpolation across the features. 
We also removed the geocoronal Ly$\alpha$ $\lambda1216$ line by interpolation.  
The spectra were rectified in flux by applying a 
spline fit to relatively line-free regions.  The final working product was a rectified 
spectrum matrix as a function of wavelength and heliocentric Julian date (HJD).  
Figure~1 displays the observed spectra of a cooler Be star (HD 43544) and a hotter Be star (HD 55606). 
The observed spectrum is shown in black, and it is compared to model spectra for the 
Be star (offset by $+1.0$ and shown in blue) and the sdO star (offset by $+2.5$ and shown in green).
These kinds of models are used in the CCF analyses described below (see Sections 3 and 4).

\begin{figure*}[ht!]
\gridline{\fig{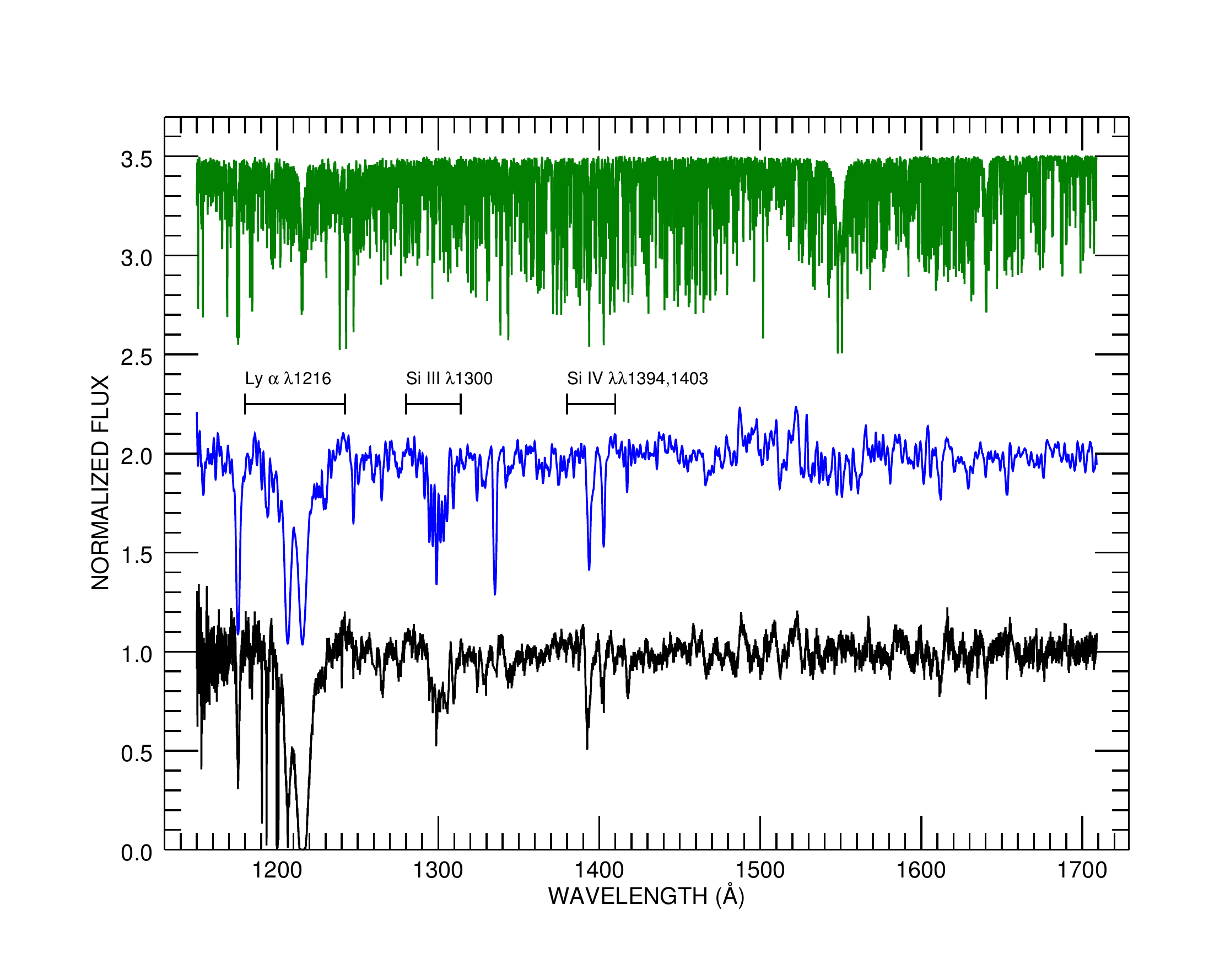}{0.5\textwidth}{}
          \fig{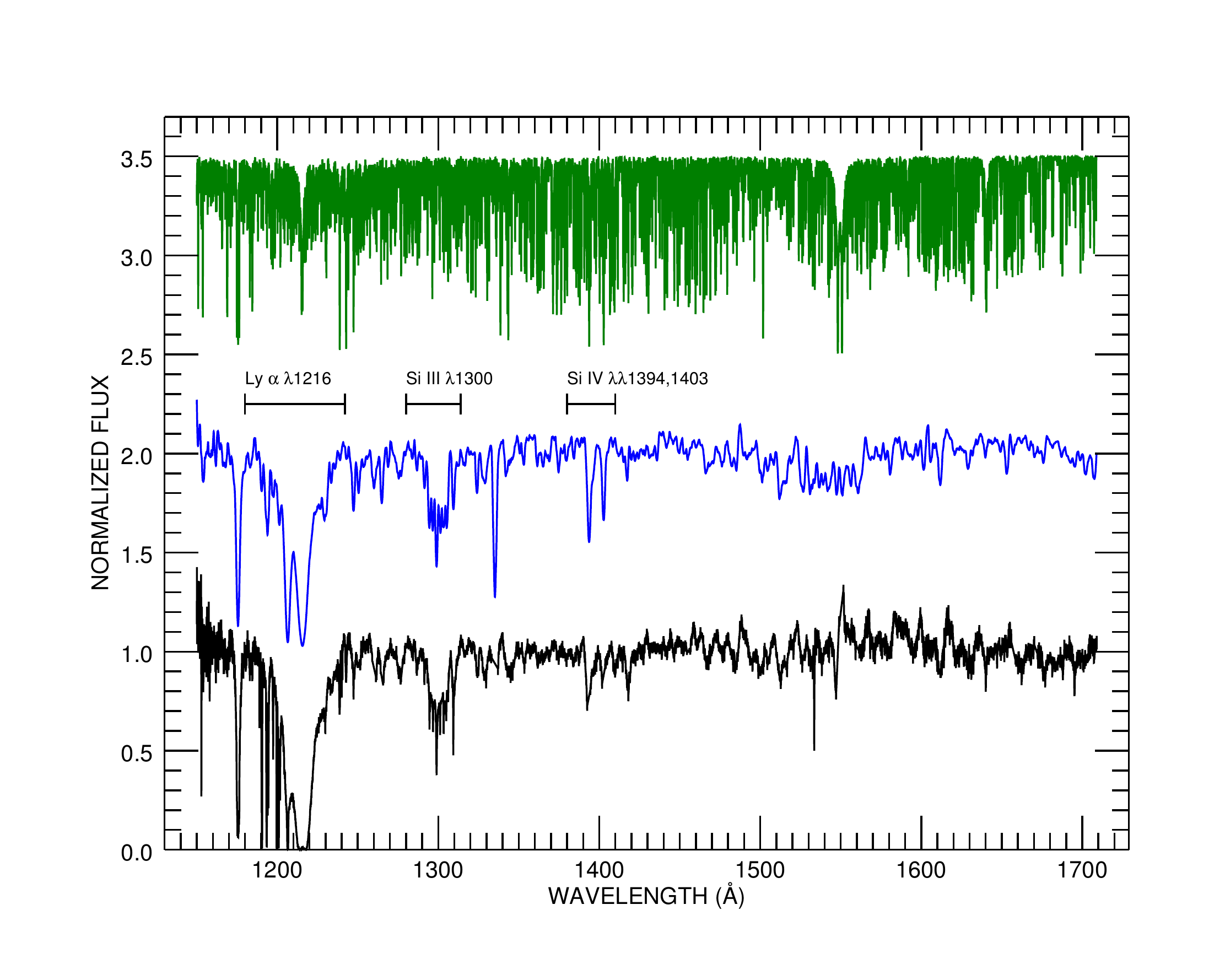}{0.5\textwidth}{}
          }
\caption{Left panel: HST/\emph{STIS} spectrum of HD 43544 observed on HJD 2458707.4 (black). The TLUSTYB Be model spectrum is offset by +1.0 (blue) for parameters $T_{\rm eff} = 21500$ K, $\log{g} = 3.9$, and $V\sin{i} = 260$ km s$^{-1}$ adopted from \citet{levenhagen2006}. The TLUSTY model spectrum of the sdO component is offset by +2.5 (green) for $T_{\rm eff} = 45000$ K. Right panel: HST/\emph{STIS} spectrum of HD 55606 observed on HJD 2458756.7. The model spectra are plotted in the same format as HD 43544, except adopting parameters of the Be component, $T_{\rm eff} = 27350$ K, $\log{g} = 4.3$, and $V\sin{i} = 335$ km s$^{-1}$, adopted from \citet{zorec2016}.}
\label{fig:FUV spectra}
\end{figure*}

% Figure 1: HST spectrum plot
\placefigure{fig:FUV spectra}
%\begin{figure*}[h!]
%\includegraphics[width=\textwidth]{spectra_HD43544.eps}
%\caption{HST/\emph{STIS} spectrum of HD 43544 observed on night of HJD 2458707.4 (black). The TLUSTYB Be model spectrum is offset by +1.0 (blue) for parameters $T_{\rm eff} = 21500$ K, $\log{g} = 3.9$, and $V\sin{i} = 260$ km s$^{-1}$ adopted from \citet{levenhagen2006}. The TLUSTY model spectrum of the sdO component is offset by +2.0 (green) for $T_{\rm eff} = 45000$ K. }
%\label{fig:CCFs of FUV spectra}
%\end{figure*}

\pagebreak

%%%%%%%%%%%%%%%%%%%%%%%%%%
\section{Radial Velocities of the \texorpdfstring{B\MakeLowercase{e}}{Be} Stars} \label{sec: Be CCFs}

Our main goal in this work is to find the spectral line patterns that correspond 
to the faint, sdO companion star in these targets.  It is very difficult to identify
individual spectral lines from the sdO star because they are weak in the combined 
spectrum and blended with the rotationally broadened lines of the Be star. 
Instead, we calculate the cross-correlation functions (CCFs) of the observed 
spectra with a model spectrum for an adopted hot temperature for the sdO star. 
The CCF effectively multiplexes the spectral signal from all the faint lines into
a single high S/N CCF for investigation. However, there is usually some correlation 
between the spectral features of the Be star and the hot model spectrum, so that 
the CCF appears as a composite of a broad component from the Be star and a 
generally narrow component from the sdO star.  We need to remove the Be star 
component from the CCF in order to isolate the part of CCF due to the sdO 
spectral features.  The first step in this process is to register the 
Be component in velocity space, and in this section we describe how the 
radial velocities of the Be stars were measured. 

The radial velocities of the Be stars were determined by forming the CCF
of the observed spectrum with a model for the Be star itself. 
The Be star model spectrum was constructed from the TLUSTY BSTAR2006 grid 
\citep{lanz2007} by interpolation in solar abundance model spectra 
for the effective temperature ($T_{\rm eff}$) and surface gravity ($\log{g}$)
given in Table 1.  The interpolated spectrum was convolved with a 
rotational broadening function for the projected rotational velocity 
($V \sin i$) from Table~1 and an interpolated, linear limb darkening 
coefficient from Kurucz LTE models derived by \citet{wade1985}. 
The resulting CCFs are very broad (Fig.~2) and their central 
peaks may be influenced by correlation between the Be model and the 
the central CCF peak from the sdO star.  Consequently, we measured 
the Be star velocity by finding the wing bisector position using the 
Gaussian sampling method described by \citet{shafter1986}.  
The radial velocity uncertainties were calculated using the maximum 
likelihood from \citet{zucker2003}, which is associated with the 
second derivative of the CCF peak, number of wavelength points within the CCF peak, 
and the peak height.  Tests show that these errors are comparable to 
those obtained by selecting bisectors above and below the default value
of $50\%$ peak height.  We report our radial velocity measurements and 
their uncertainties in columns 3 and 4, respectively, of Table 2. 

\begin{figure*}
\includegraphics[width=\textwidth]{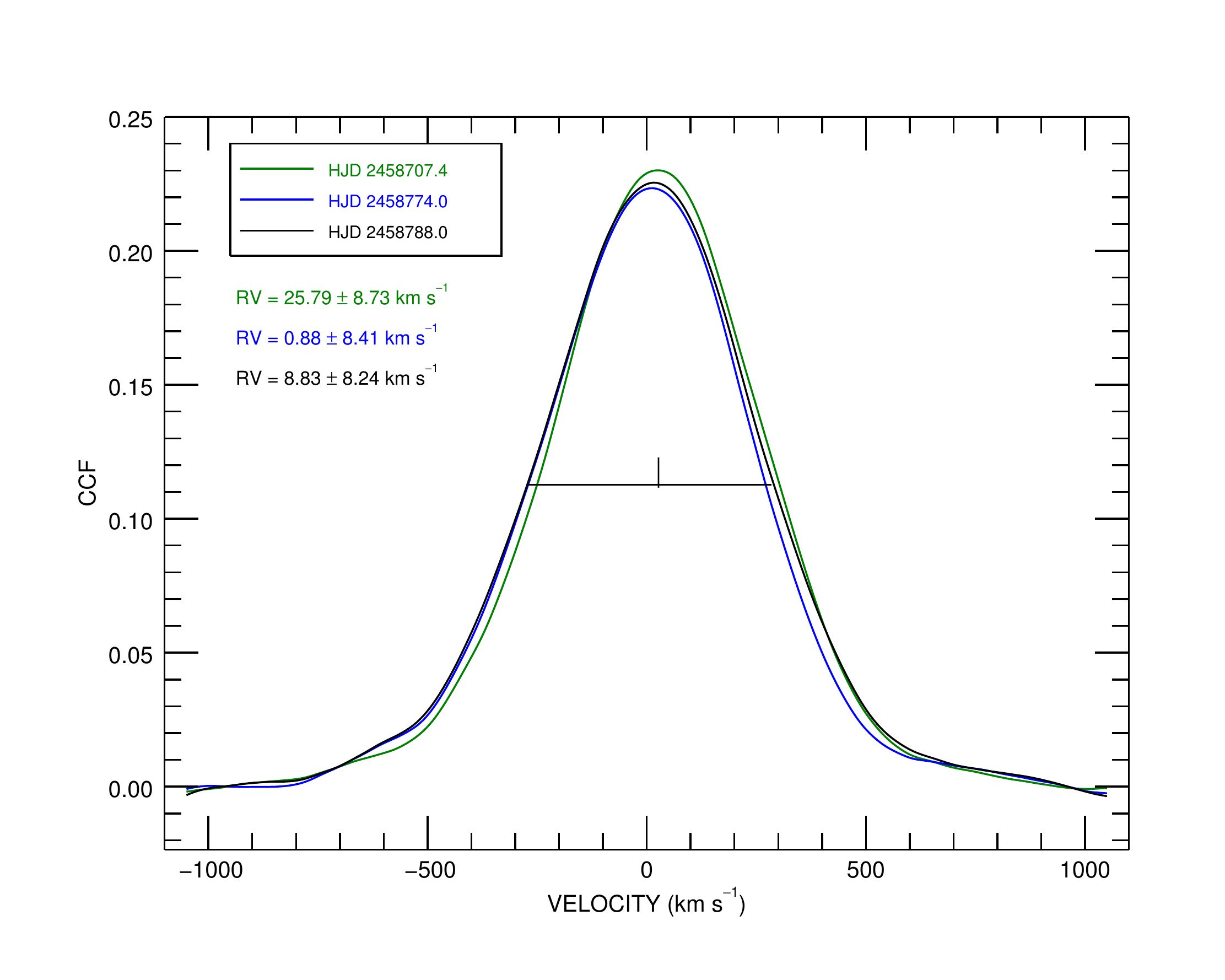}
\caption{CCFs of observed FUV spectra of HD 43544 cross correlated with a TLUSTY BSTAR2006 Be model spectrum. Radial velocities were measured from the wings at 50\% peak height of the CCFs by adopting the bisector technique as described in \citet{shafter1986}. The horizontal black line indicates the 50\% peak height of the CCFs for the spectrum obtained on HJD 2458788.0, and the associated RV is shown as the tick mark.}
\label{fig:CCFs of FUV spectra}
\end{figure*}

% Figure 2: CCFs of observation w/ Be component
\placefigure{fig:CCFs of FUV spectra}

% Table 2: RV measurements 
\begin{deluxetable*}{lccccc}
\tablenum{2}
\tablecaption{Radial Velocity Measurements of Be+sdO systems \label{tab: RVs}}
\tablewidth{0pt}
\tablehead{
\colhead{HD} & \colhead{Date} & \colhead{$V_{Be}$} & \colhead{$\sigma_{Be}$} & \colhead{$V_{sdO}$} & \colhead{$\sigma_{sdO}$} \\
\colhead{Number} & \colhead{(HJD-2,400,000)} & \colhead{(km s$^{-1}$)} & \colhead{(km s$^{-1}$)} & \colhead{(km s$^{-1}$)} & \colhead{(km s$^{-1}$)} 
}
\startdata
29441 & 58769.2148 &    \phn$-$2.1 &      11.1 &      \phs17.1 &       3.3 \\ 
29441 & 58717.5000 &   \phn\phs5.3 &      11.1 &      \phs31.9 &       2.9 \\ 
29441 & 58741.0742 &   \phn\phs1.1 &      11.0 &      \phs38.9 &       3.1 \\ 
43544 & 58707.3750 &      \phs25.8 &   \phn8.7 &       $-$37.9 &       1.5 \\ 
43544 & 58774.0469 &   \phn\phs0.9 &   \phn8.4 &      \phs77.9 &       1.5 \\ 
43544 & 58788.0156 &   \phn\phs8.8 &   \phn8.2 &      \phs54.0 &       1.7 \\ 
 51354 & 58748.7578 &      \phs48.2 &      10.6 &       $-$58.8 &       6.4 \\ 
 51354 & 58762.1953 &      \phs36.5 &      10.6 &       $-$30.7 &       8.8 \\ 
 51354 & 58787.9570 &      \phs56.5 &      11.3 &      \phs42.6 &       6.4 \\ 
 55606 & 58756.6992 &      \phs42.7 &      13.0 &   \phn\phs0.3 &       2.2 \\ 
 55606 & 58888.5820 &      \phs32.5 &      12.7 &      \phs20.4 &       2.3 \\ 
 55606 & 58899.5078 &      \phs34.4 &      12.6 &      \phs65.9 &       2.3 \\ 
 60855 & 58804.5664 &      \phs59.1 &   \phn6.9 &      \phs32.9 &       2.5 \\ 
 60855 & 58810.5977 &      \phs55.6 &   \phn7.0 &      \phs37.2 &       2.4 \\ 
 60855 & 58855.4219 &      \phs53.6 &   \phn7.0 &      \phs52.9 &       2.3 \\ 
113120 & 58768.0898 &      \phs27.1 &      11.9 &    \phn$-$2.2 &       1.9 \\ 
113120 & 58783.7461 &    \phn$-$6.1 &      12.3 &      \phs30.8 &       1.9 \\ 
113120 & 58885.8555 &      \phs27.7 &      12.1 &       $-$30.0 &       1.6 \\ 
137387 & 58781.8320 &      \phs44.7 &   \phn9.3 &       $-$17.4 &       2.1 \\ 
137387 & 58787.8516 &      \phs33.2 &   10.0 &    \phn$-$0.5 &       2.0 \\ 
137387 & 58880.1562 &      \phs19.1 &   10.0 &      \phs41.1 &       2.2 \\ 
152478 & 58709.7188 &      \phs27.7 &   \phn9.1 &      \phs41.0 &       2.3 \\ 
152478 & 58715.5430 &      \phs28.0 &   \phn8.9 &      \phs47.4 &       2.3 \\ 
152478 & 58761.2891 &      \phs27.8 &   \phn9.4 &      \phs58.3 &       2.2 \\ 
157042 & 58731.9102 &    \phn$-$0.7 &   \phn7.2 &      \phs41.5 &       3.0 \\ 
157042 & 58750.6367 &       $-$11.1 &   \phn7.3 &       $-$14.2 &       3.4 \\ 
157042 & 58767.2422 &   \phn\phs2.1 &   \phn7.3 &       $-$26.3 &       3.1 \\ 
157832 & 58731.9766 &      \phs27.4 &   \phn7.3 &        \nodata &    \nodata \\ 
157832 & 58750.7070 &      \phs27.4 &   \phn7.8 &        \nodata &    \nodata \\ 
157832 & 58772.2148 &      \phs19.8 &   \phn7.8 &        \nodata &    \nodata \\ 
191610 & 58718.1445 &       $-$26.1 &   \phn8.7 &        \nodata &    \nodata \\ 
191610 & 58726.9453 &       $-$23.1 &   \phn8.2 &        \nodata &    \nodata \\ 
191610 & 58753.7656 &       $-$23.5 &   \phn7.7 &        \nodata &    \nodata \\ 
194335 & 58718.2695 &   \phn\phs8.4 &      10.3 &      \phs53.3 &       1.5 \\ 
194335 & 58727.0117 &    \phn$-$7.3 &      10.6 &      \phs42.6 &       1.7 \\ 
194335 & 58755.4883 &   \phn\phs4.0 &   \phn8.8 &       $-$88.6 &       1.5 \\ 
214168 & 58734.2344 &    \phn$-$4.0 &      11.0 &        \nodata &    \nodata \\ 
214168 & 58753.7070 &   \phn\phs1.9 &      10.9 &        \nodata &    \nodata \\ 
214168 & 58772.3047 &    \phn$-$2.7 &      11.2 &        \nodata &    \nodata \\  
 \enddata
%\tablecomments{}
\end{deluxetable*}
 
%%%%%%%%%%%%%%%%%%%%%%%%%%%
\section{Detection of \texorpdfstring{\MakeLowercase{sd}O}{sdO} Companion Stars}

In order to confirm the detection of the hot subdwarf companions in these candidate systems, 
we adopted the methodology described in \citet{wang2018} to search for the 
spectral signature of the subdwarf O-type (sdO) star by forming cross-correlation functions (CCFs) 
of the observed FUV spectra with a model spectrum for a very hot star.  
The default model spectrum was derived from the OSTAR2003 grid of solar abundance 
models based upon the non-local thermal equilibrium (non-LTE) atmospheres code TLUSTY and 
the associated radiative transfer code SYNSPEC \citep{lanz2003}. 
The adopted parameters for the model spectrum are $T_{\rm eff} = 45$ kK, 
$\log g = 4.75$, and $V \sin {i} = 0$ km~s$^{-1}$, and these are typical of 
estimates derived for known Be+sdO systems \citep{wang2017}. 
The adopted gravity parameter is the highest value available in the OSTAR2003 grid, 
but it may underestimate the actual value for the subdwarf companions. 

% Appearance of CCFs
The CCFs should ideally reflect the photospheric absorption lines of the components, so we omitted those regions with spectral features that are much broader 
than the line profiles associated with rotational broadening. 
The omitted regions included the deep and broad absorption line profiles of 
Ly$\alpha$ $\lambda 1216$ and \ion{Si}{3} $\lambda 1300$ blend, as well as the 
beginning and ending regions of the spectrum. In addition, we excluded the 
broad wind line features of \ion{Si}{4} $\lambda\lambda 1393, 1402$ in the 
cases of HD~43544, HD~55606, HD~60855, HD~137387, HD~152478, and HD~194335. 
Both the \ion{Si}{4} $\lambda\lambda 1393, 1402$ and \ion{C}{4} $\lambda\lambda 1548, 1550$ 
wind lines were omitted in spectra of HD~29441, HD~113120, and HD~157042.   

% Add CCF of HD128220 as recommended by referee
We caution that the solar abundance model spectrum used to represent the sdO spectrum may not match the abundance patterns accurately. Many low-mass sdO stars have spectra that are He rich with metallic line enhancements. For example, \citet{rauch1993} examined the FUV spectrum of the low-mass sdO star HD~128220 based upon 38 high resolution {\it IUE} spectra. Through a careful spectral analysis using non-LTE models, he found that the star has an effective temperature of $T_{\rm eff} = 40.6 \pm 0.4$ kK, $\log{g} = 4.5 \pm 0.1$, and a mass of $0.54 \pm 0.01\ M_\odot$. Furthermore, the star is helium rich with $n_{He}/n_{H} = 0.30 \pm 0.05$. Because this subdwarf star has measured $T_{\rm eff}$ and $\log{g}$ values that are similar to our model template, but with enhanced helium abundance, it provides a good test case of the applicability of our adopted solar abundance model. We downloaded from MAST\footnote{https://archive.stsci.edu/iue/} a high resolution SWP spectrum of HD~128220 made on HJD 2,444,256.0. The spectrum was reduced and rectified following the procedures reported in \citet{wang2018}. In Figure~3, the spectrum of HD~128220 is illustrated in black in the left panel, and the TLUSTY model spectrum of the subdwarf component is offset by +1.0 and shown in green. In order to bring the observed spectrum into agreement with the rest frame of the model spectrum, we shifted the observed spectrum by adopting a radial velocity of $RV = 15.2 \pm 2.2$ km s$^{-1}$ from \citet{howarth1987} for the date of observations. We then cross correlated the observed spectrum with the TLUSTY model spectrum. The resulting CCF is shown in the right panel in Figure~3. A sharp peak is clearly seen indicating the detection of the subdwarf star in HD~128220 using the model spectrum. Therefore, we can safely assume that the TLUSTY model spectrum provides a sufficiently good match of an sdO spectrum for application in our search for their spectral signature in the observed \emph{HST} spectra. 

\begin{figure*}[ht!]
\gridline{\fig{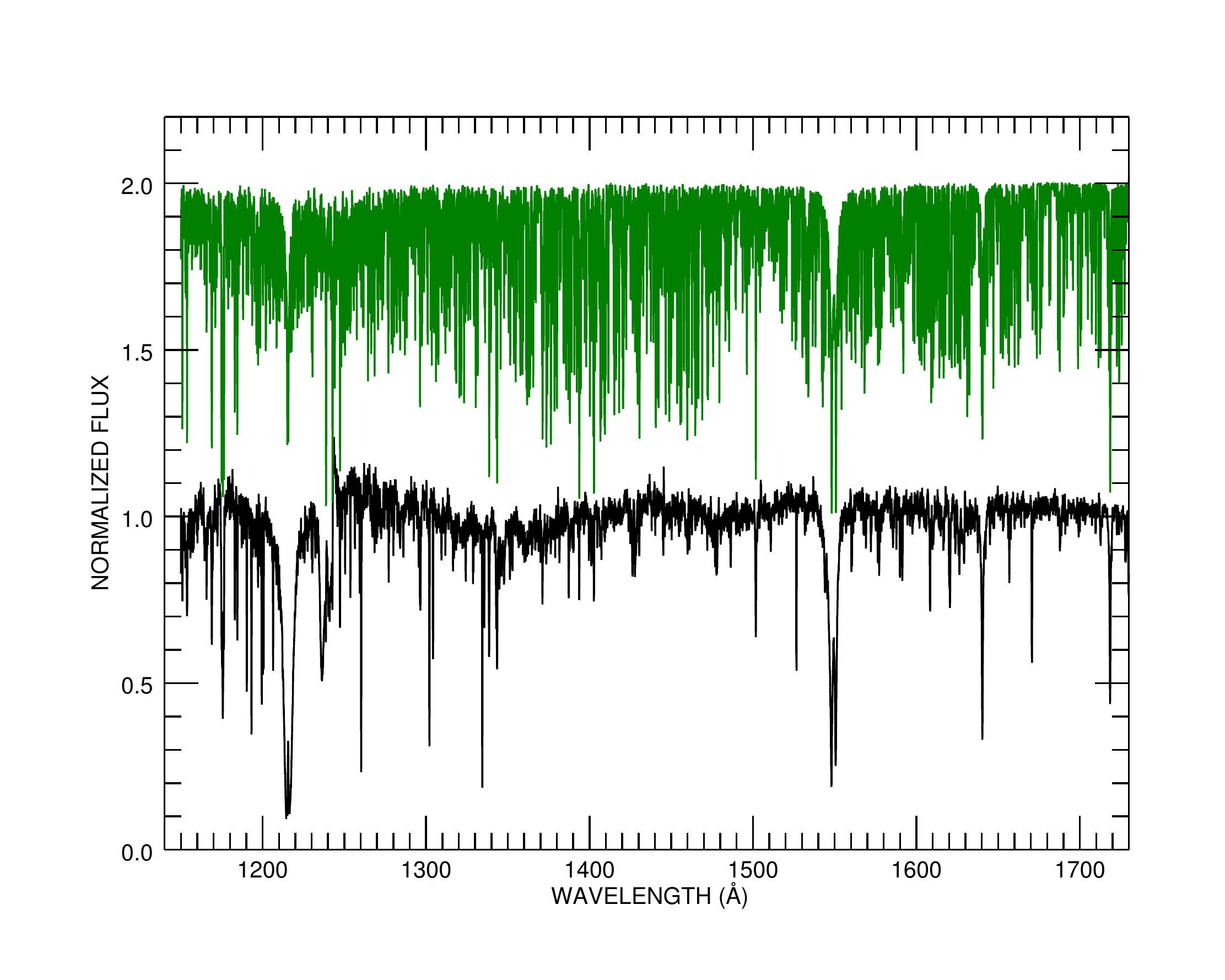}{0.5\textwidth}{}
          \fig{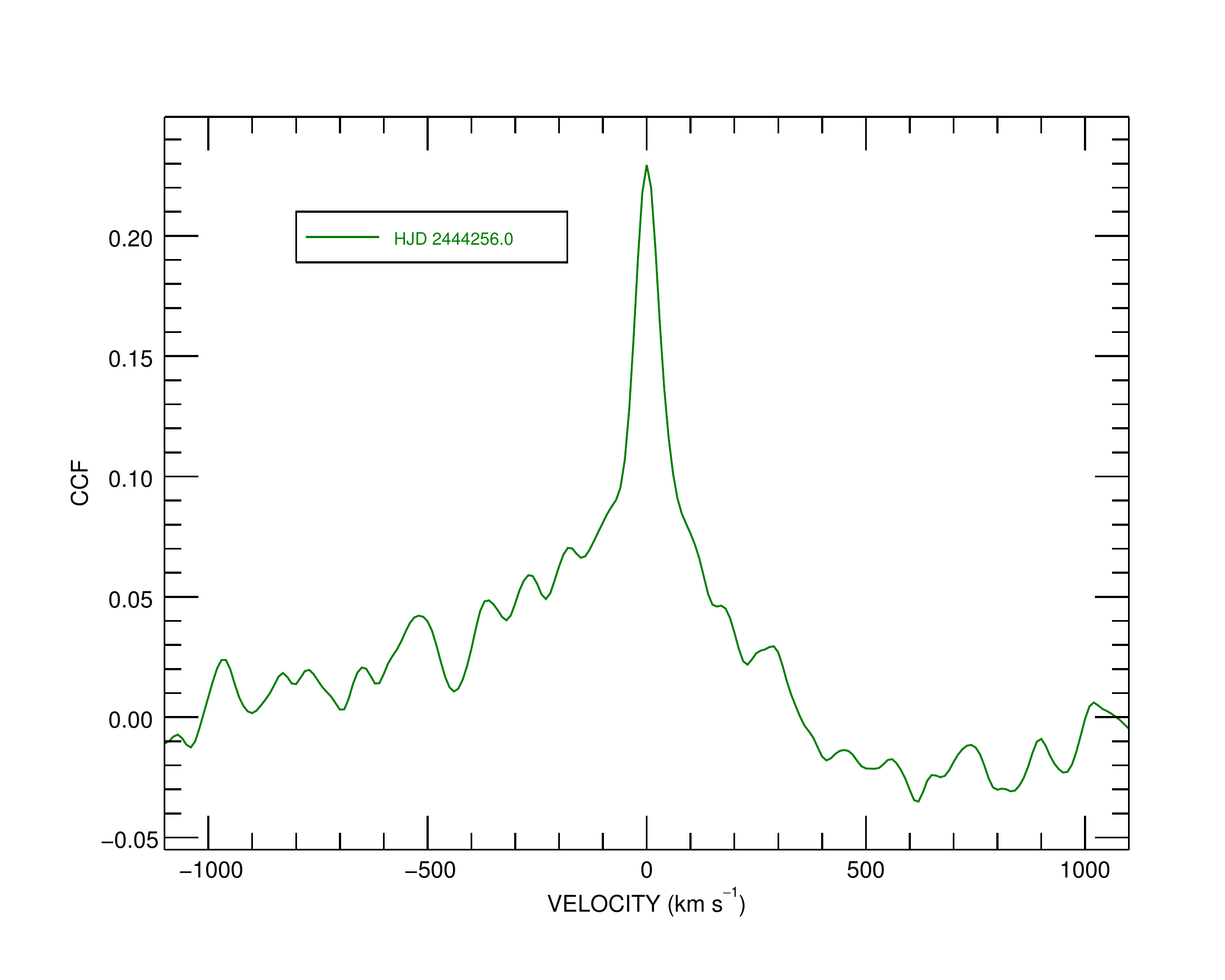}{0.5\textwidth}{}
          }
\caption{Left panel: The \emph{IUE}/SWP spectrum of HD 128220 (black) made on HJD 2444256.0. The TLUSTY model spectrum of the sdO component is offset by +1.0 (green) for $T_{\rm eff} = 45$ kK. Right panel: The CCF of the observed spectrum of HD 128220 cross correlated with the TLUSTY sdO model spectrum.}
\label{fig:CCFs of HD128220}
\end{figure*}

% Add Figure 3: IUE HD128220 spectrum plot
\placefigure{fig:CCFs of HD128220}
%\pagebreak

The resulting CCFs generally are very broad because of correlation of the 
model lines with the rotationally broadened lines of the Be star. 
In those cases of a positive detection, the CCF appears to show 
a sharp peak from the sdO spectral features that is superimposed on top of the 
broad CCF component from the Be star.  We applied the following method 
to remove the Be component from the CCFs and isolate the signal from 
the sdO star.  We first calculated a CCF for model spectra of both the 
Be star (using the parameters in Table 1) and the sdO star.  This simulated 
CCF was shifted from the zero velocity frame to the specific radial 
velocity of the Be star from Section 3 (Table 2), and the CCF amplitude was 
rescaled to match that of the wings in the CCF of the observed spectrum
with the model sdO spectrum.  In many cases, we found that the CCF of the 
model Be and model sdO spectrum did not match well with the observed CCF width, 
so we reset the value of the projected rotational velocity $V \sin i$ for 
the model Be star spectrum in order to attain a better fit of the CCF wings. 
These adjusted $V \sin i$ values are listed in the final column of Table~1, 
and the uncertainties are of order $10\%$ because the widths were set by
visual inspection.  Once the simulated Be model plus sdO model CCF was 
registered in velocity, adjusted in width, and rescaled in CCF amplitude, 
we then subtracted this model from the CCF of the observed spectrum and
model sdO spectrum.  This procedure is illustrated in Figure~4 for the 
case of one observation of HD~137837.  We see that the simulated CCF from 
the model Be star and model sdO star spectrum does match the observed CCF
wings, and by subtracting this component, we form a residual CCF that 
clearly shows the narrow central peak from correlation with the sdO 
spectral lines alone. 

\begin{figure*}
\includegraphics[width=\textwidth]{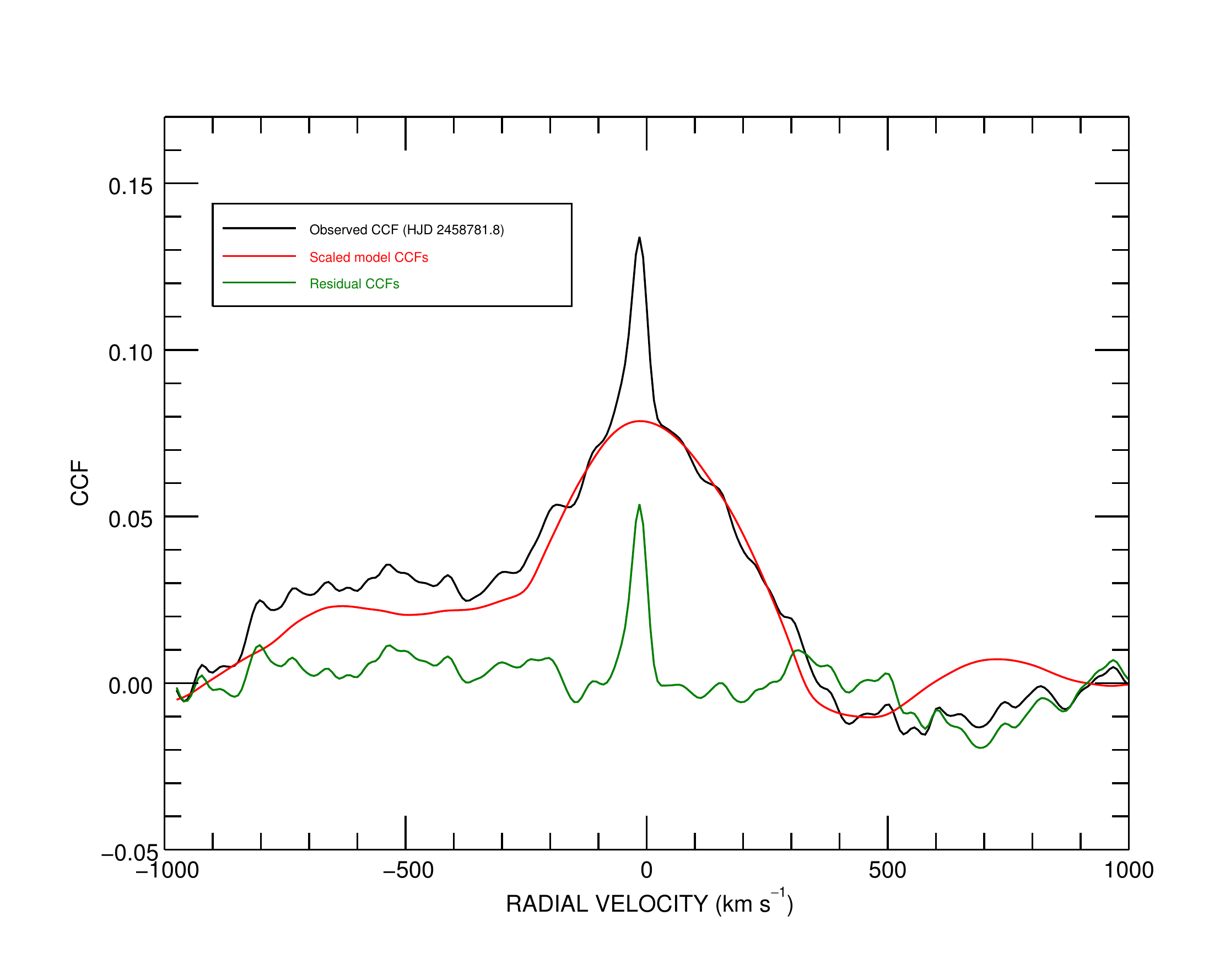}
\caption{The CCF of the observed spectrum of HD 137387 observed on HJD 2458781.8 
from cross correlation with the TLUSTY model spectrum. A scaled Be model spectrum (red) 
was subtracted from the composite feature, and the corresponding residual CCF (green) is 
associated with the sdO companion. }
\label{fig:CCFs of FUV spectra HD137387}
\end{figure*}

% Figure 4 goes here. 
% Figure 4: CCFs of hot target w/ TLUSTY 45 kK, HD 137387
\placefigure{fig:CCFs of FUV spectra HD137387}

We examined the residual peaks formed this way for all the observed spectra, 
and we detected CCF peaks from a hot subdwarf companion in ten Be+sdO targets. 
These positive detections were re-analysed using a grid of test effective 
temperatures for the sdO component (see Section 5), and we adopted the temperature
that maximized the residual CCF peaks as the default for the model sdO spectrum. 
We then repeated the procedure of subtracting the Be component in the CCF 
using a model with the adopted sdO temperature instead of the default 
$T_{\rm eff} = 45$~kK value.  The final residual CCFs are plotted for the 
ten positive detection cases in Figures 5, 6, and 7.  No residual peaks 
were found that were signficantly larger than the variation in the CCFs 
far from zero velocity for HD~157832, HD~191610, and HD~214168.  
The residual CCFs for these three non-detections are shown in Figure~8 
and are based on the default $T_{\rm eff} = 45$~kK model for an sdO component. 
It is puzzling that no obvious CCF peaks were found for these three cases, 
given that clear peaks were detected in CCFs constructed in a similar 
way for the lower S/N {\it IUE} spectra (and often in multiple observations)
\citep{wang2018}.  This may result from temporal variations in the overall 
visibility of the spectral lines of the sdO component (Section 8). 

\begin{figure*}[ht!]
\gridline{\fig{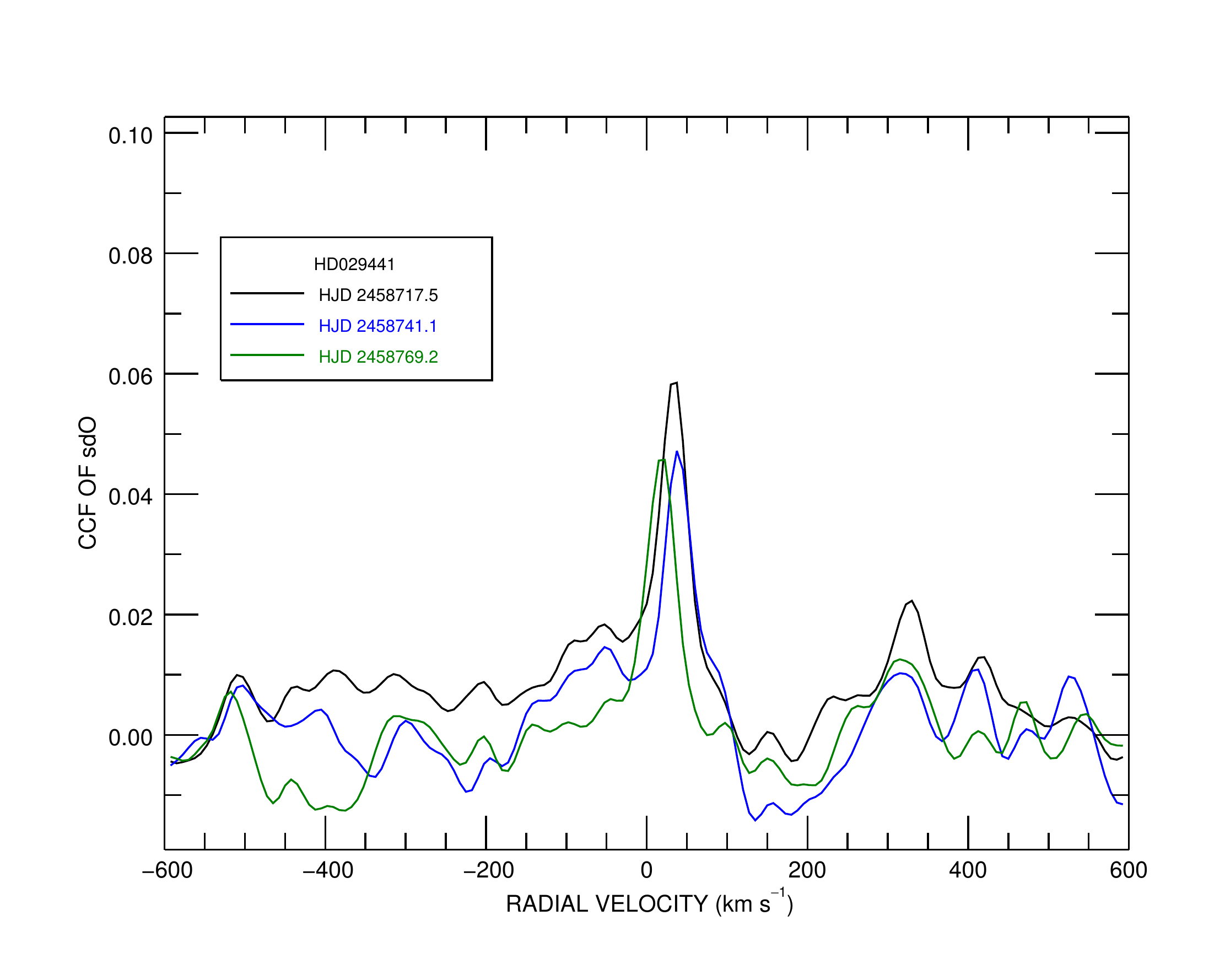}{0.5\textwidth}{(a) HD 29441}
          \fig{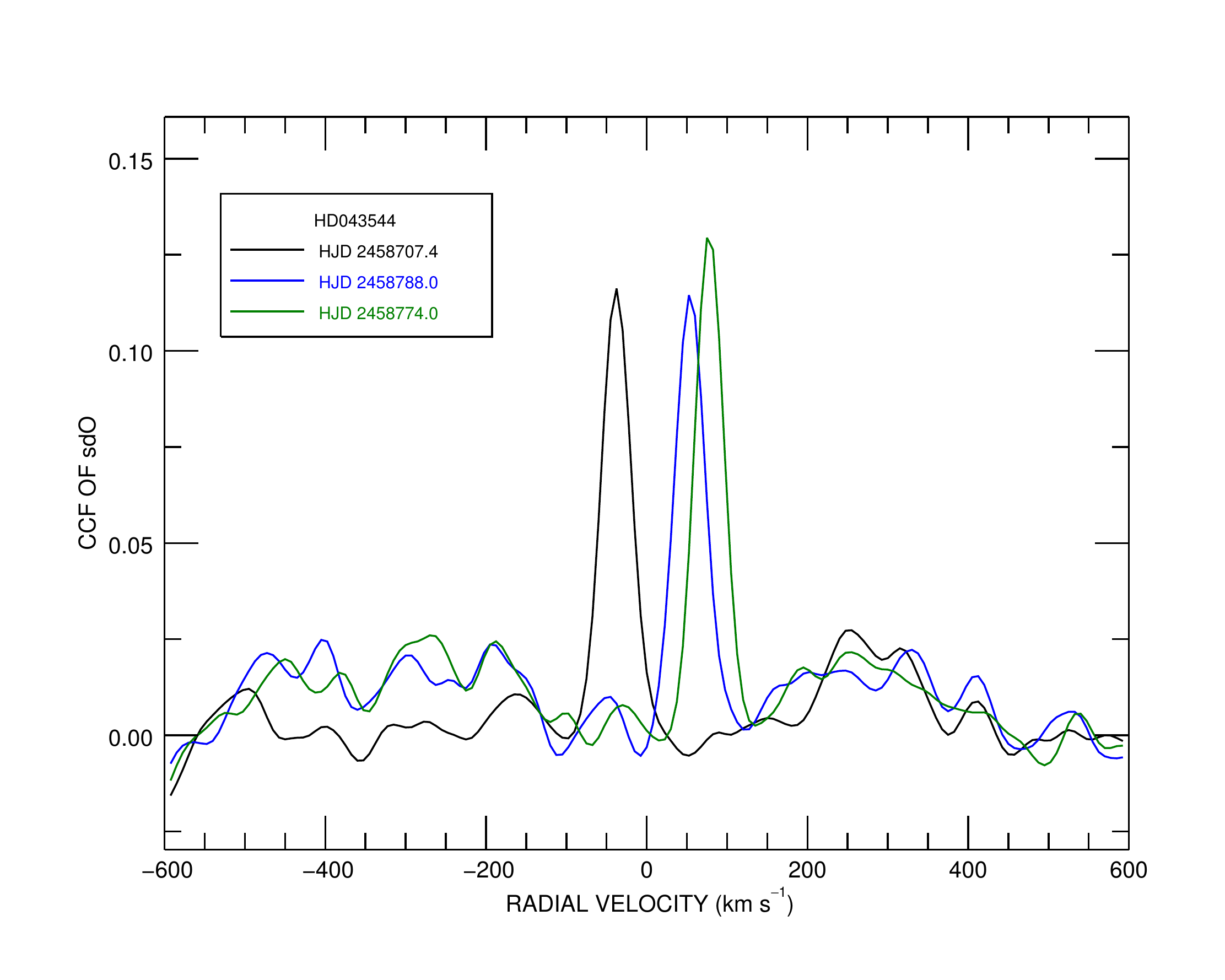}{0.5\textwidth}{(b) HD 43544}
          }
\gridline{\fig{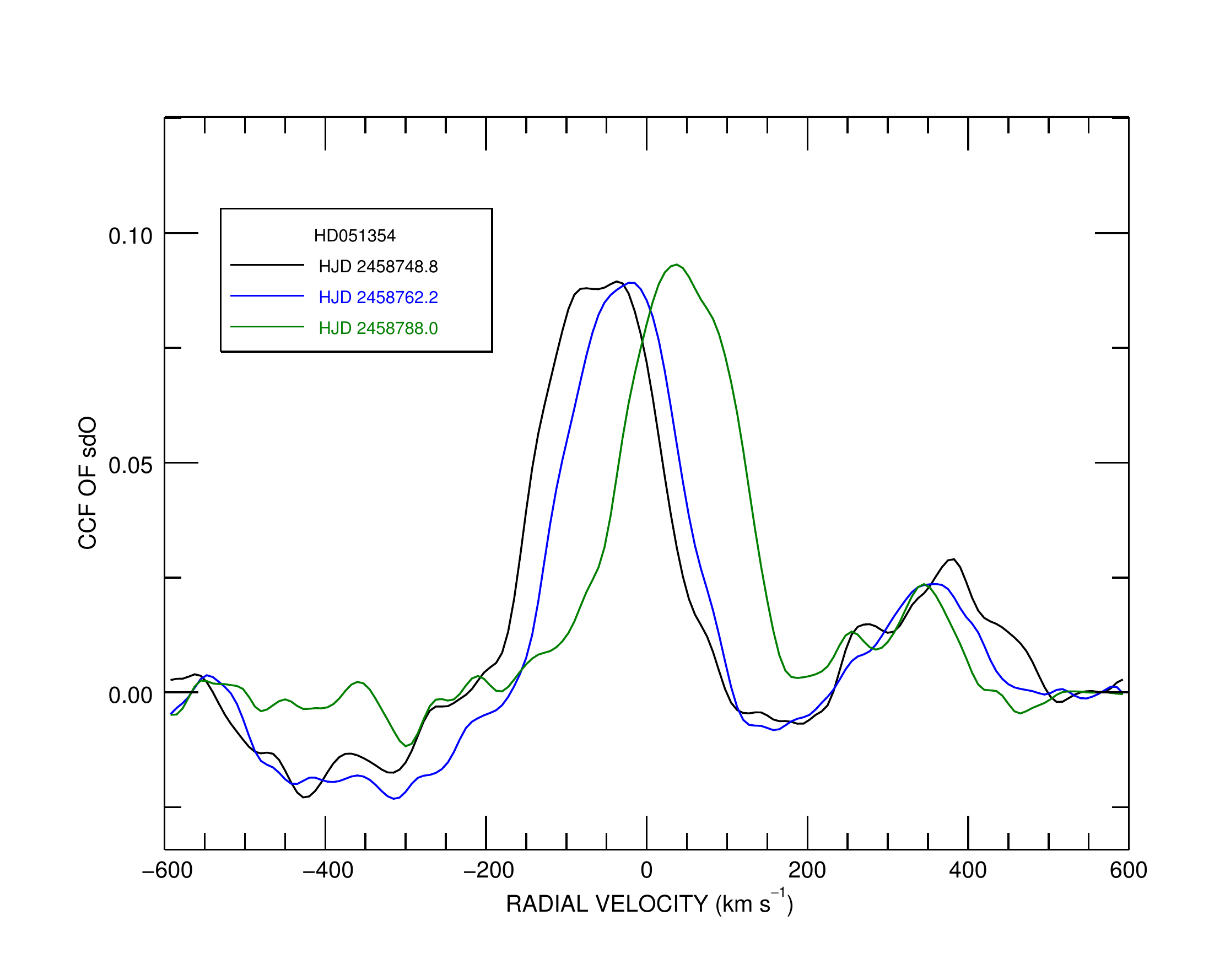}{0.5\textwidth}{(c) HD 51354}
          \fig{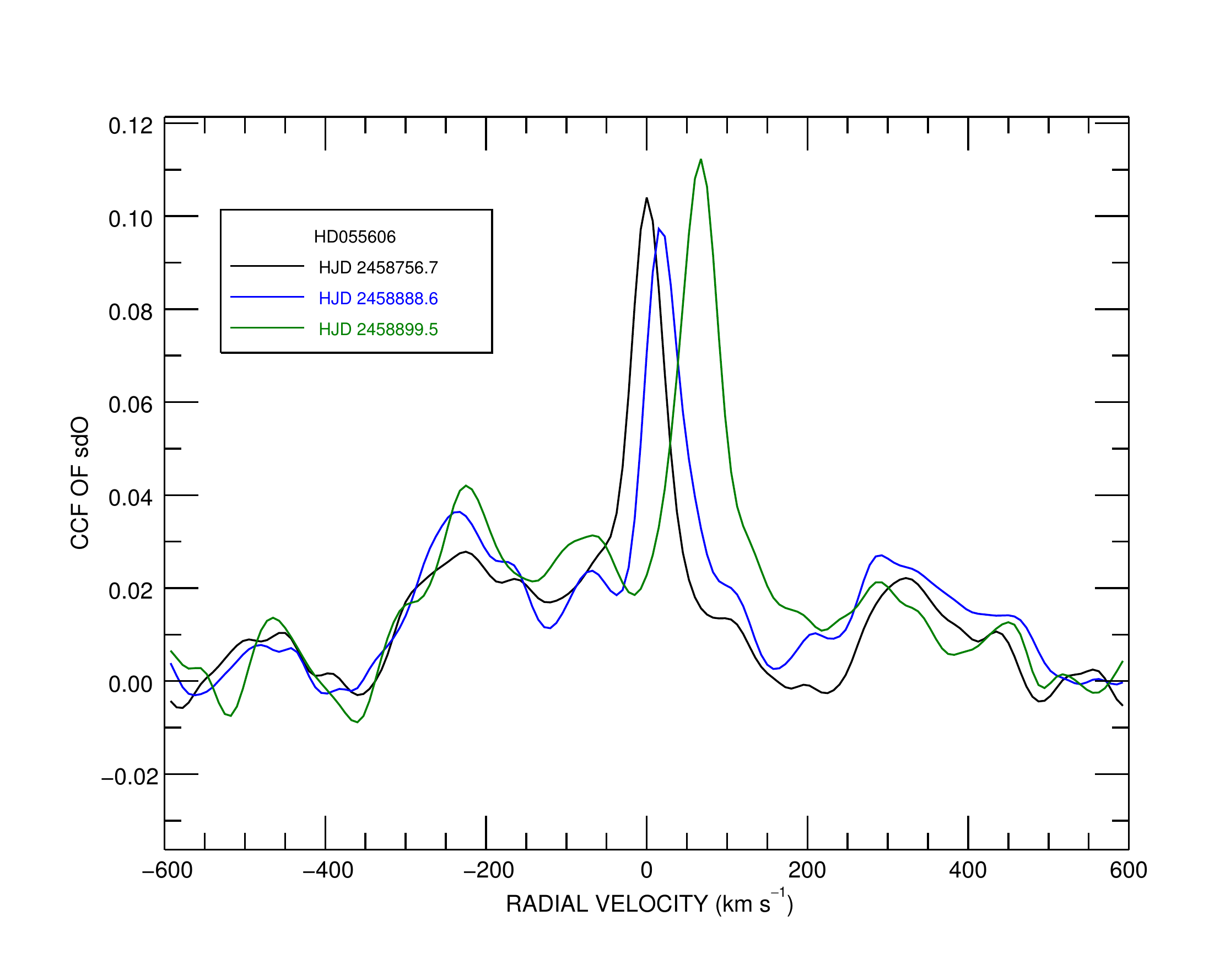}{0.5\textwidth}{(d) HD 55606}
          }
\caption{Residual CCFs of sdO companion stars in confirmed Be+sdO binary systems.}
\label{fig:CCFs1}
\end{figure*}

% Figures 5, 6, and 7, here.
% Figure 5: CCFs of sdO 
\placefigure{fig:CCFs1}
\pagebreak

\begin{figure*}  
\gridline{\fig{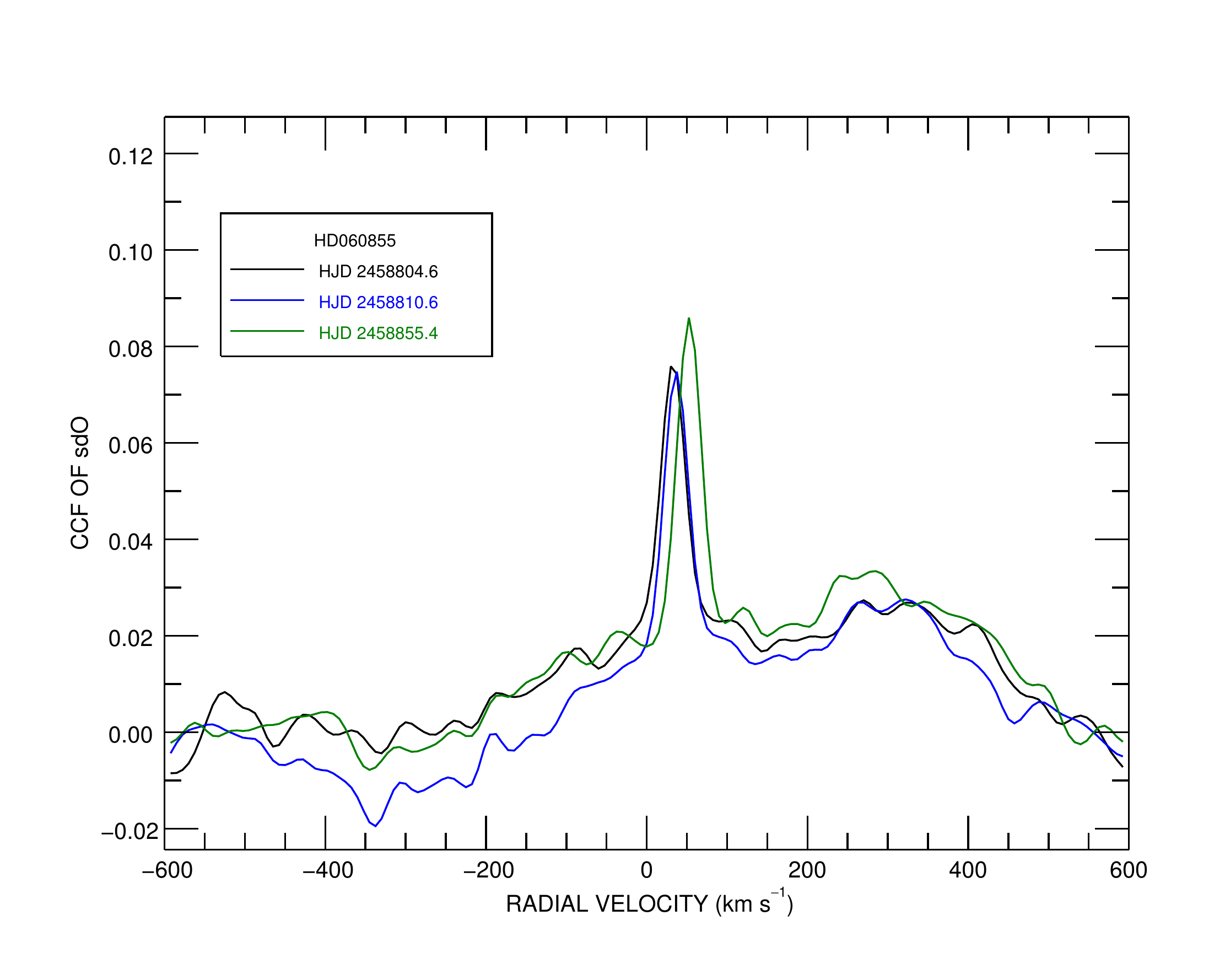}{0.5\textwidth}{(a) HD 60855}
	\fig{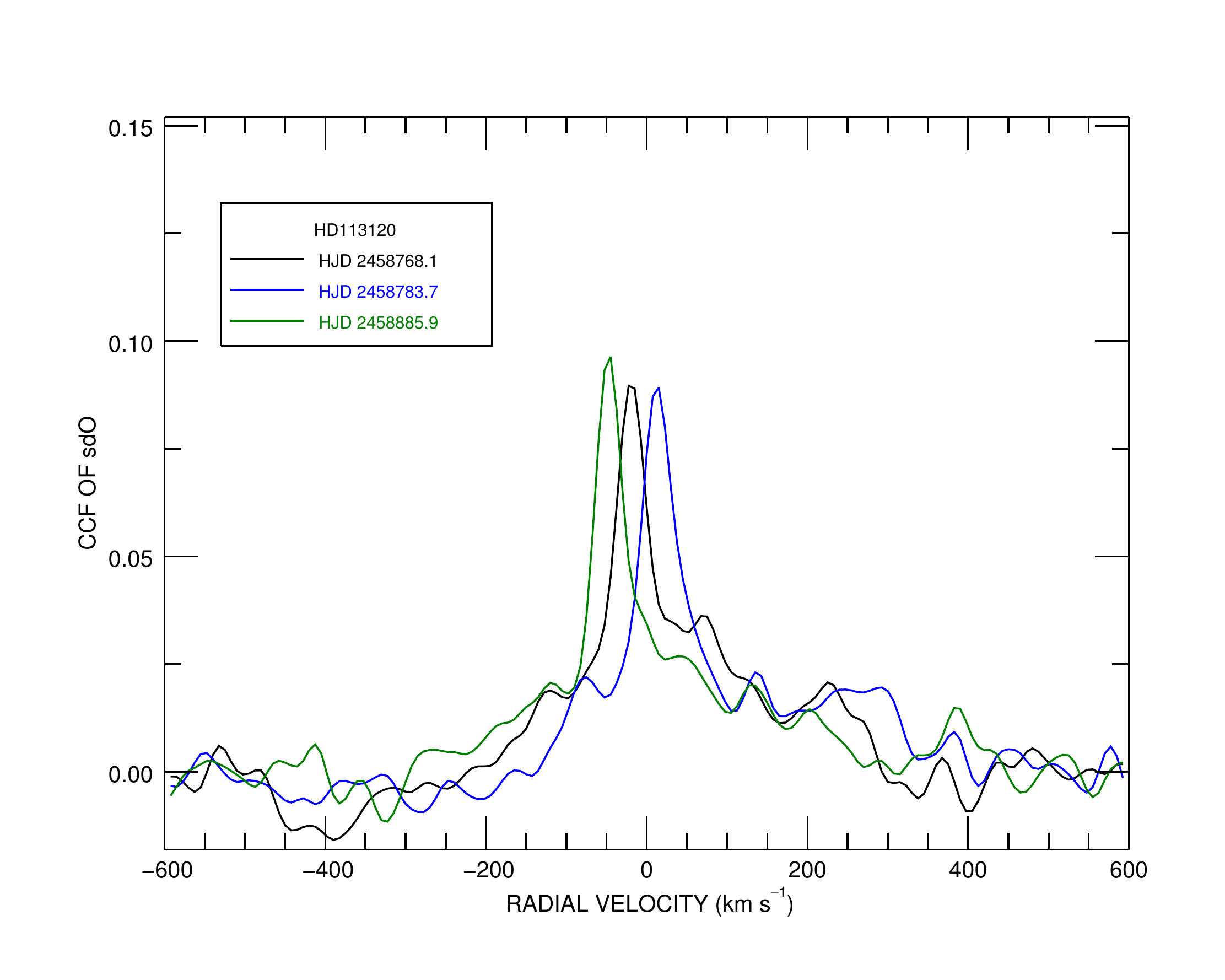}{0.5\textwidth}{(b) HD 113120}
	}      
 \gridline{\fig{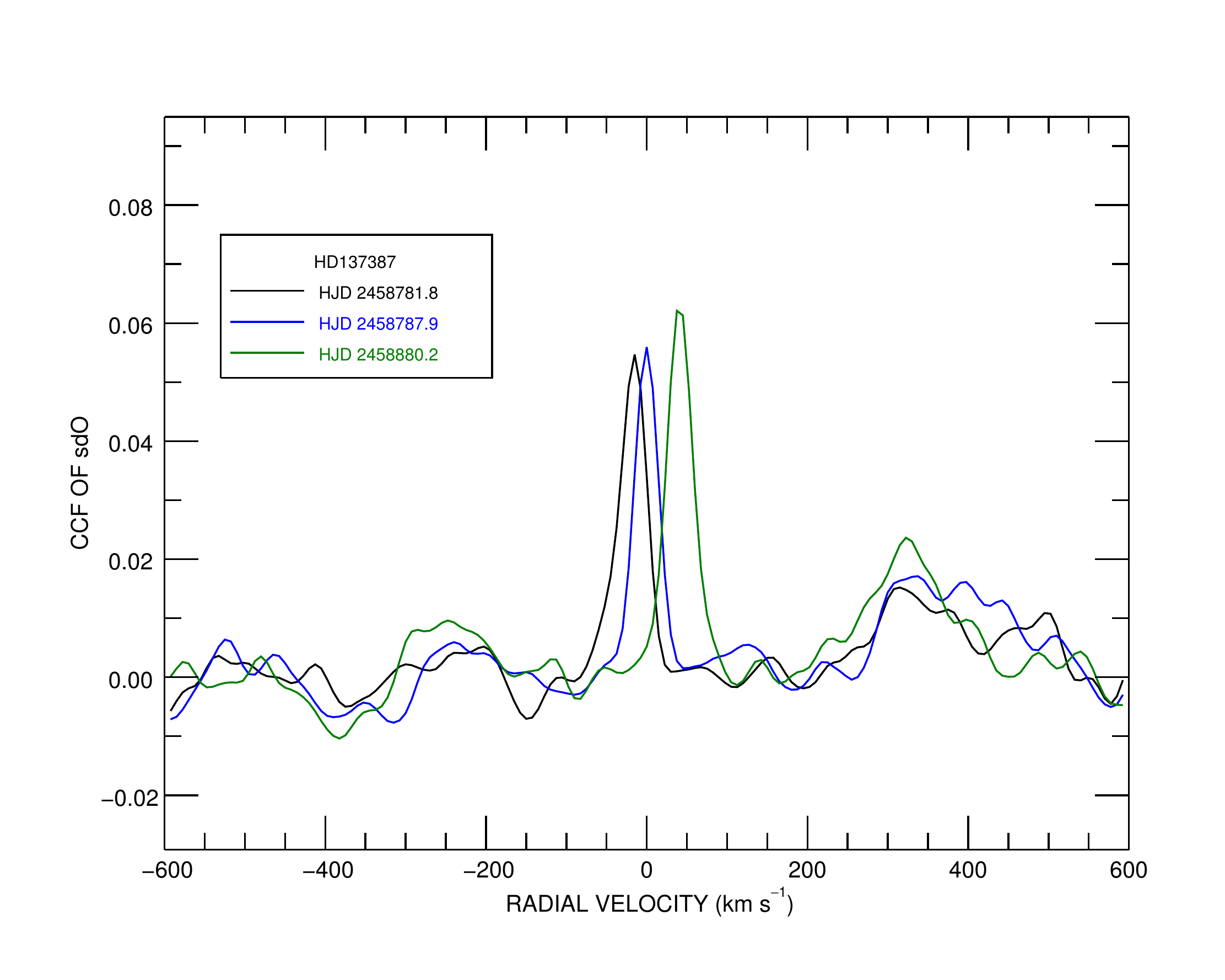}{0.5\textwidth}{(c) HD 137387}
	\fig{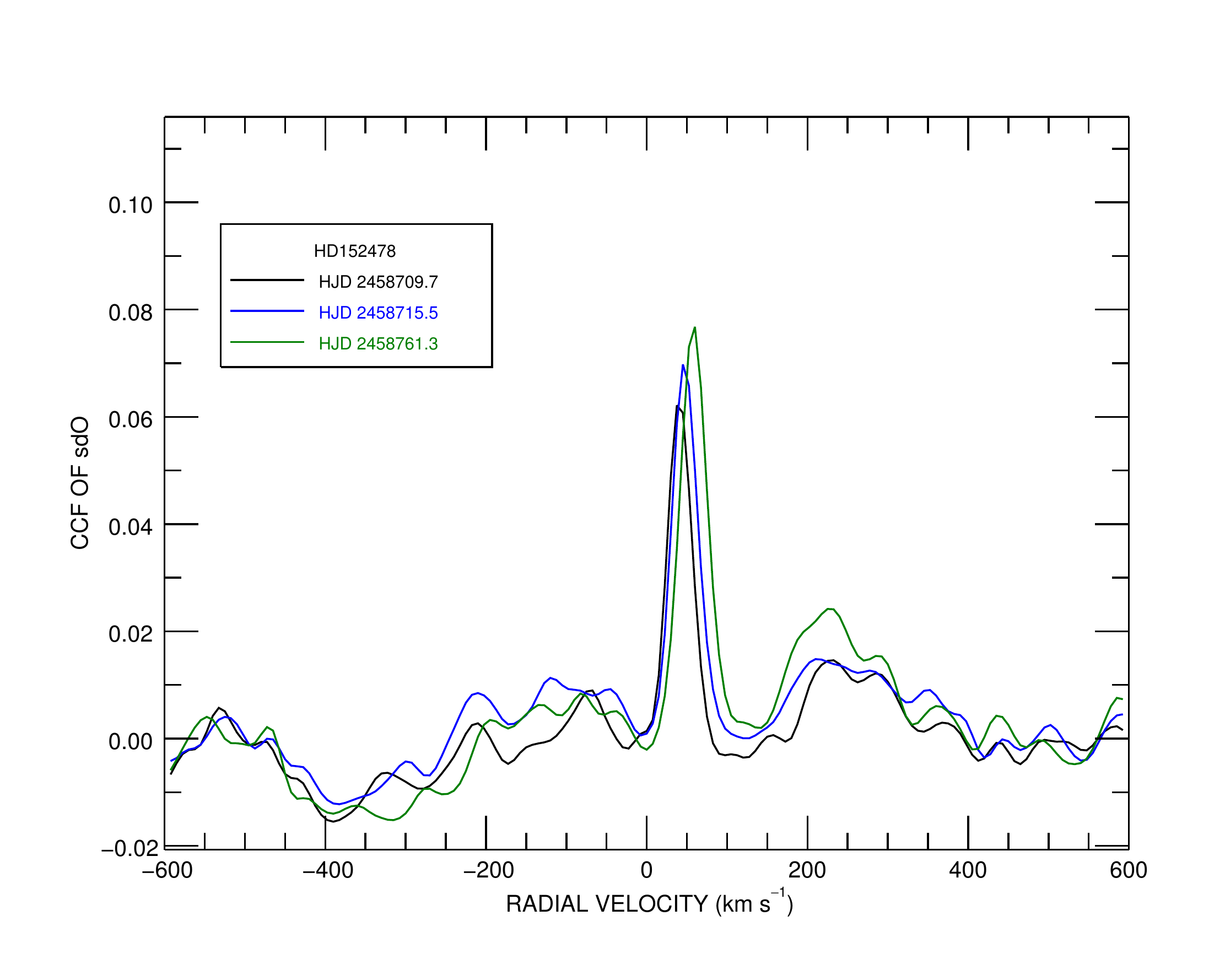}{0.5\textwidth}{(d) HD 152478 }
	}
\caption{Residual CCFs of sdO companion stars in confirmed Be+sdO binary systems.}
\label{fig:CCFs2}
\end{figure*}

% Figure 6: CCFs of sdO 
\placefigure{fig:CCFs2}

\begin{figure*}  
\gridline{\fig{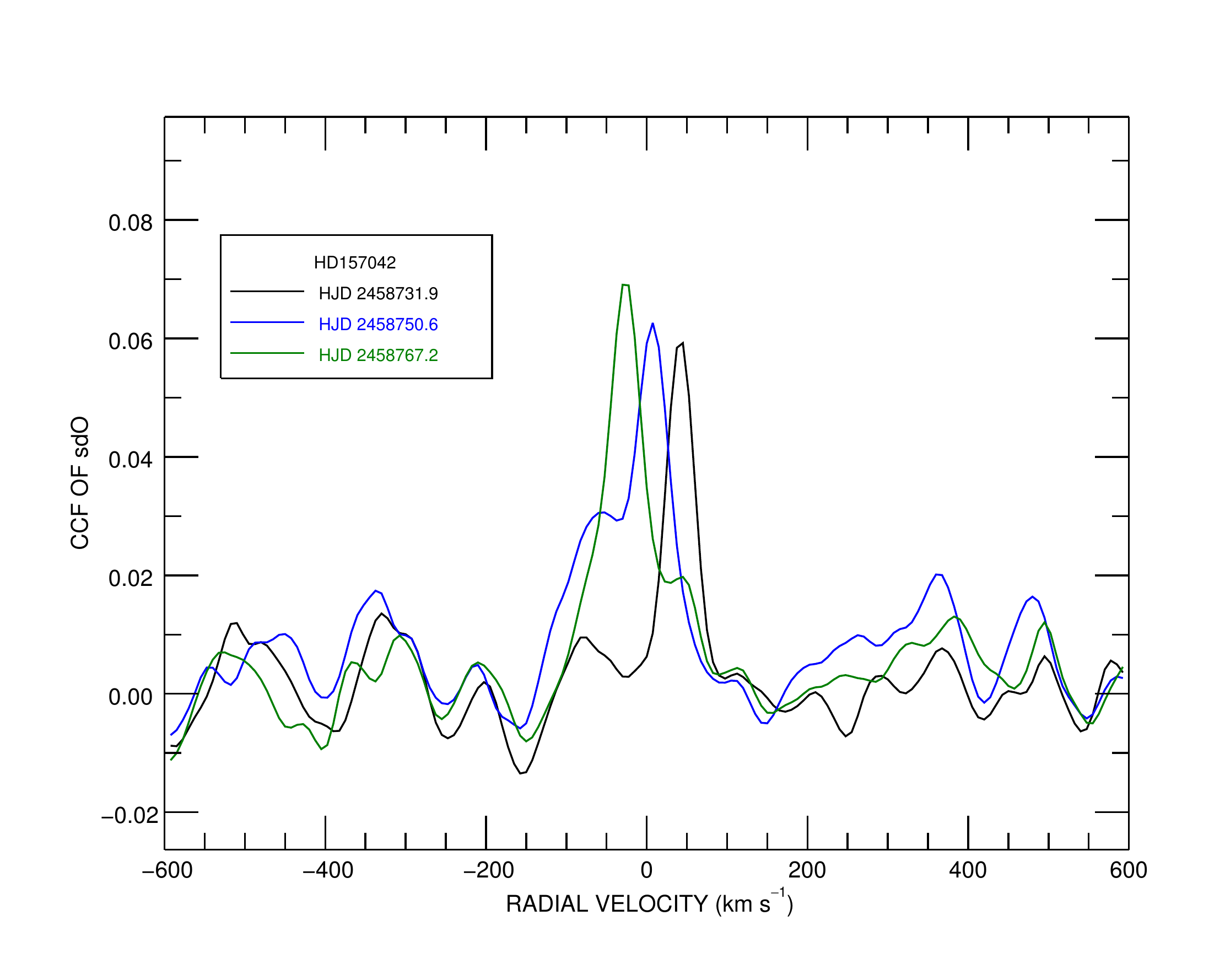}{0.5\textwidth}{(a) HD 157042}
	\fig{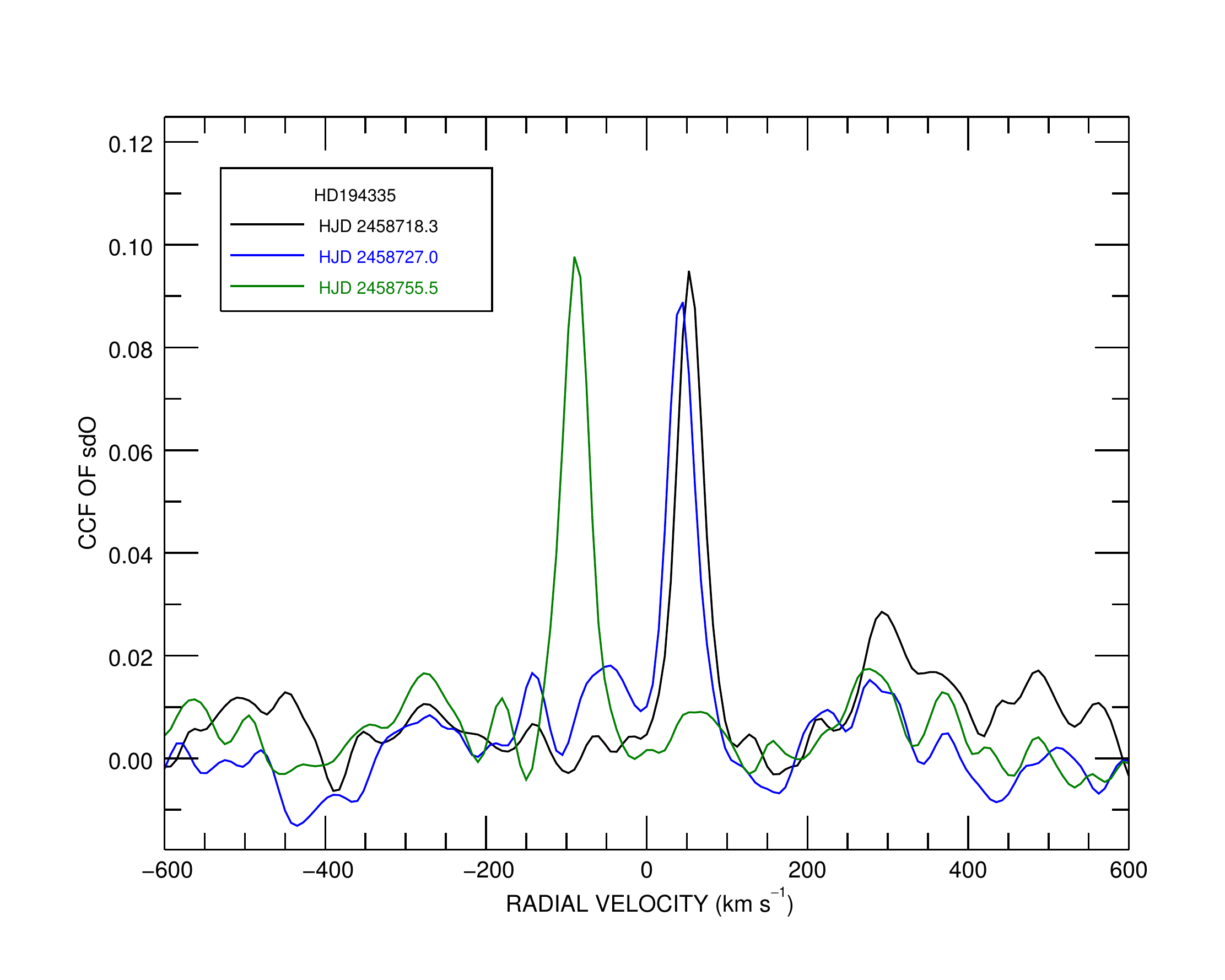}{0.5\textwidth}{(b) HD 194335 }
	}
\caption{Residual CCFs of sdO companion stars in confirmed Be+sdO binary systems.}
\label{fig:CCFs3}
\end{figure*}

% Figure 7: CCFs of sdO 
\placefigure{fig:CCFs3}

\begin{figure*}  
\gridline{\fig{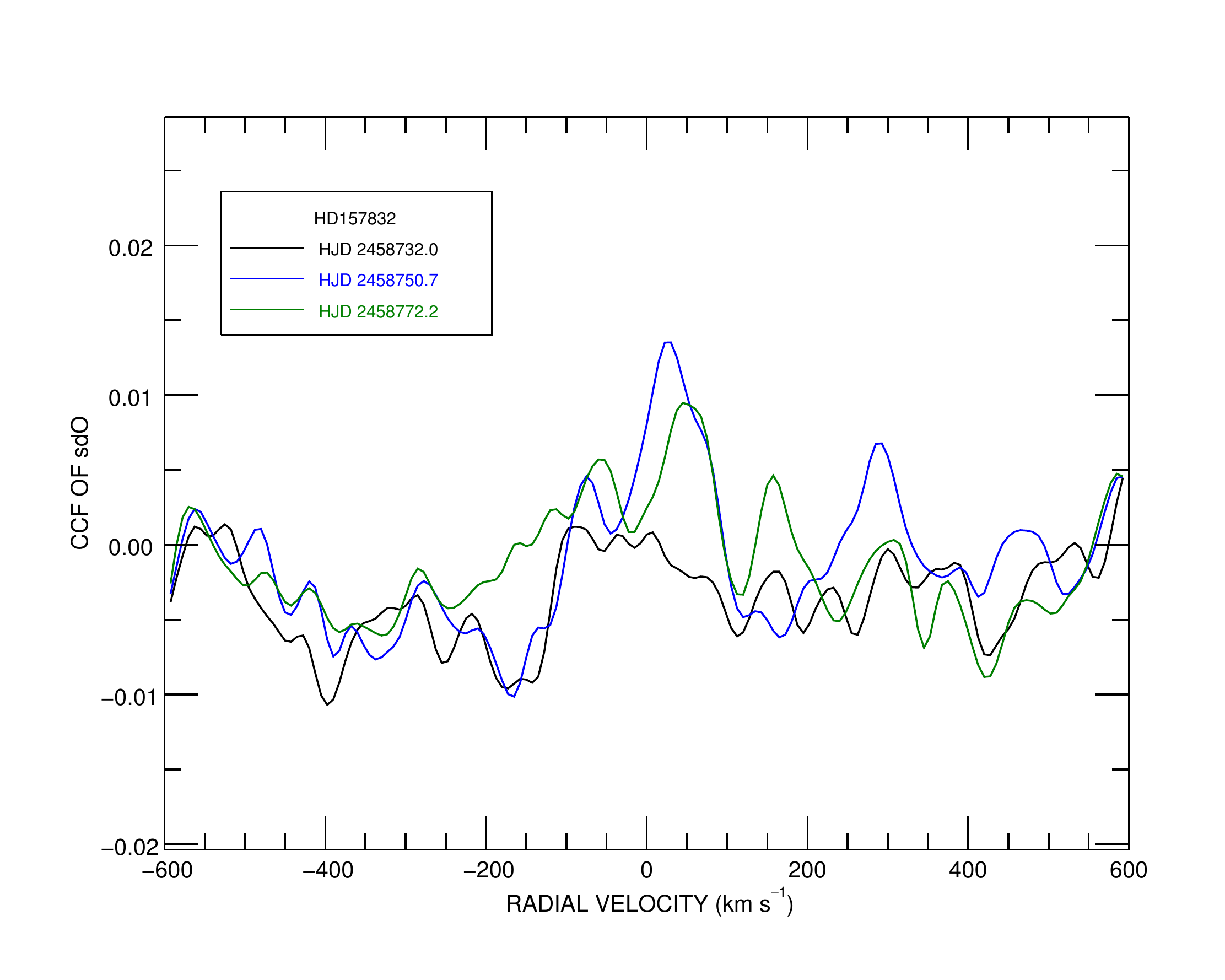}{0.5\textwidth}{(a) HD 157832}
	\fig{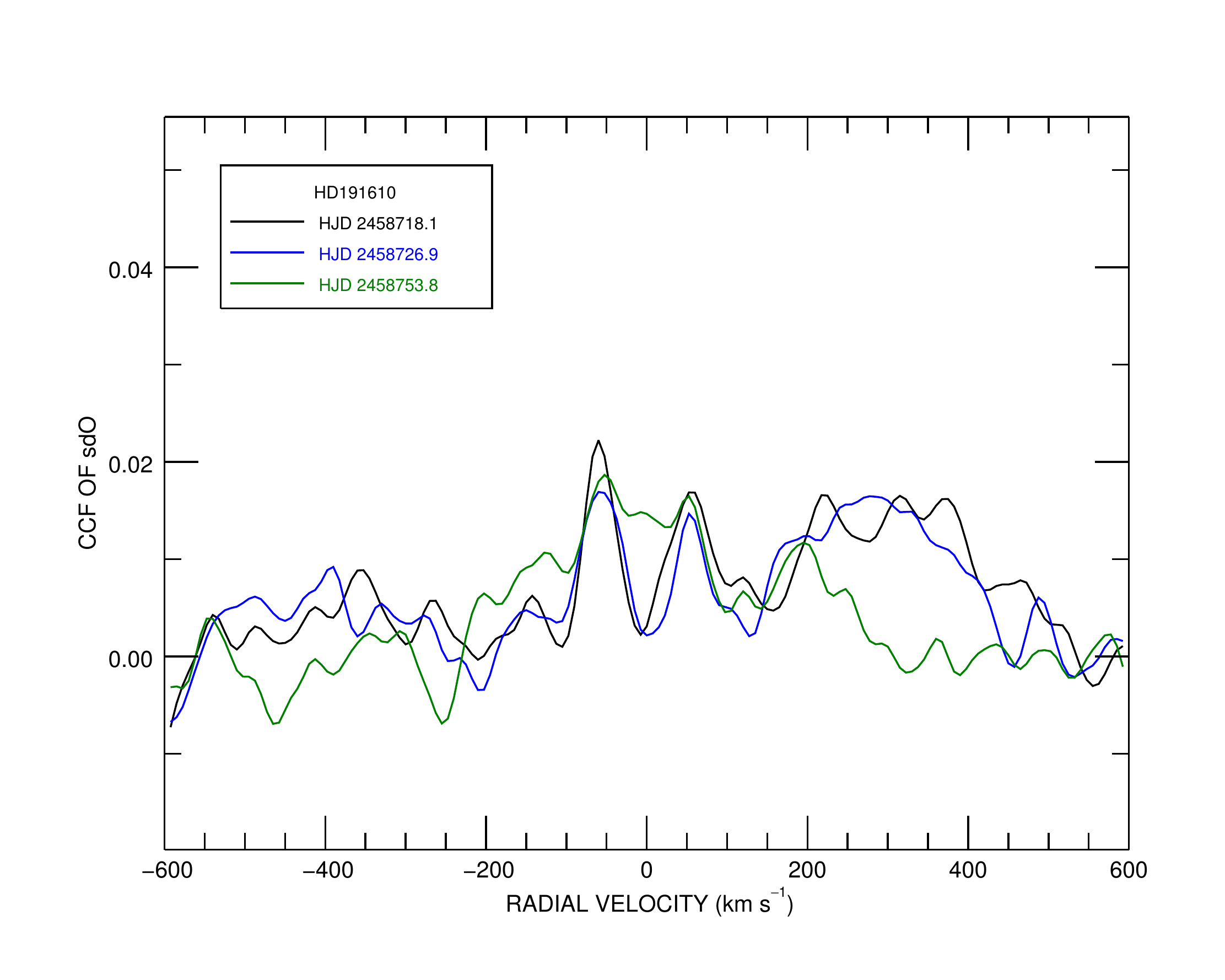}{0.5\textwidth}{(b) HD 191610}
	}      
 \gridline{\fig{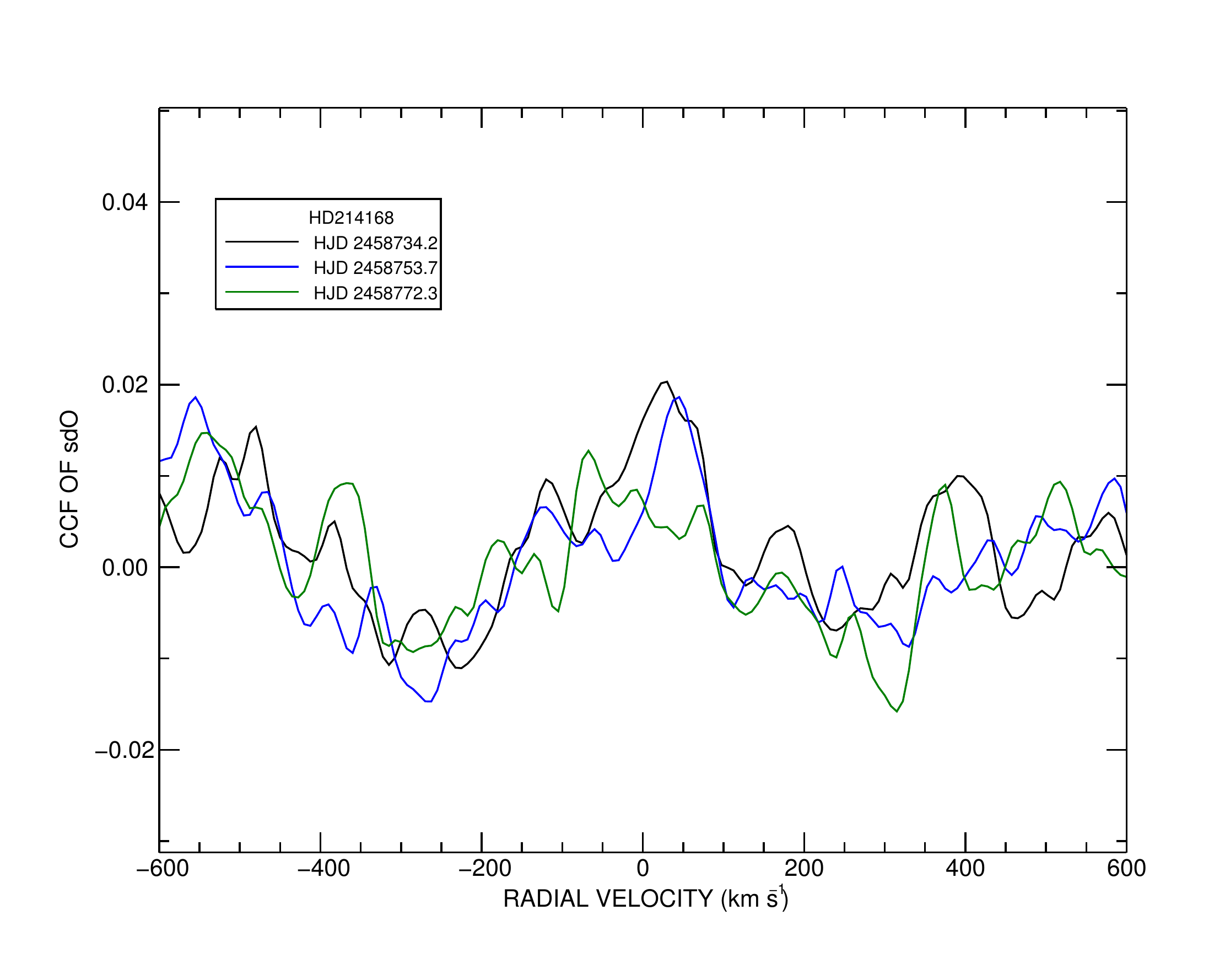}{0.5\textwidth}{(c) HD 214168}
	}
\caption{Residual CCF plots for the null detections.}
\label{fig:CCFs4}
\end{figure*}

% Figure 8 here. 
% Figure 8: CCFs of null detections, HD157832
\placefigure{fig:CCFs4}

%%%%%%%%%%%%%%%%%%%%%%%%%%%%%%%%%
\newpage
\section{Physical Properties of the \texorpdfstring{\MakeLowercase{sd}O}{sdO}  Stars \label{subsec: sdO features}}

%% sdO RVs
\subsection{Radial Velocities \label{tab: RVs of sdO}} % 

Inspection of the residual CCFs formed by subtracting the Be star component 
shows the presence of narrow peaks that are associated with the sdO spectral 
lines (Figs.\ 5, 6, 7).  We find that the peaks show distinct changes in  
Doppler shift that reflect the orbital motion of the sdO component. 
Most of the residual peaks are narrow with a broadening dominated by
the instrumental profile and intrinsic line widths in the CCF (Section 5.2). 
Consequently, we made simple Gaussian fits of the residual peaks in 
order to determine the Gaussian central position and its uncertainty. 
These radial velocities and their uncertainties are presented in 
columns 5 and 6, respectively, of Table 2 for the ten systems with 
positive detections.  These velocity measurements will be valuable for 
future work on the determination of the orbital elements.  Note that 
in principle it is possible to estimate the mass ratio from the co-variations 
of the Be star and sdO star radial velocities, but we found that the 
uncertainties in the Be star velocities were too large to establish 
any reliable limits on the mass ratio. 

%%%
\subsection{Projected Rotational Velocity \label{tab: Vsini}} 

The width of the peak in the residual CCF is directly related to the projected 
rotational velocity $V \sin i$ of the sdO line profiles.  We made a simulation 
of the dependence of the CCF peak FWHM (from Gaussian fits) as a function of 
assumed $V \sin i$ by calculating model spectra on the observed grid that 
were convolved with a rotational broadening function for a range in $V \sin i$.
We then formed CCFs of these model spectra with the adopted model sdO spectrum, 
and we made Gaussian fits of the resulting CCF peaks to build a 
functional relation between $V \sin i$ and CCF FWHM.  An example of this 
relation from the numerical tests is shown in Figure 9.  Comparison of the 
observed FWHM results with these relations indicates that the sdO star spectral lines 
are mainly unresolved in the $R=20000$ versions of {\it HST}/STIS spectra.  
The only exception is the case of HD~51354 that shows significantly broader 
CCF peaks than those in any of the other detected sdO targets.   
The derived value of $V \sin i$ for HD~51354 is given in column 3 of Table~3, 
and upper limits are reported for the other nine cases where the sdO signal is detected. 
The uncertainty in $V \sin i$ is the standard deviation of the derived 
values from the FWHM measurements in the residual CCFs of the three observations. The upper limits are set by the largest estimate from the three spectra or by the limit associated with the spectral resolution.  

\begin{figure*}
\includegraphics[width=\textwidth]{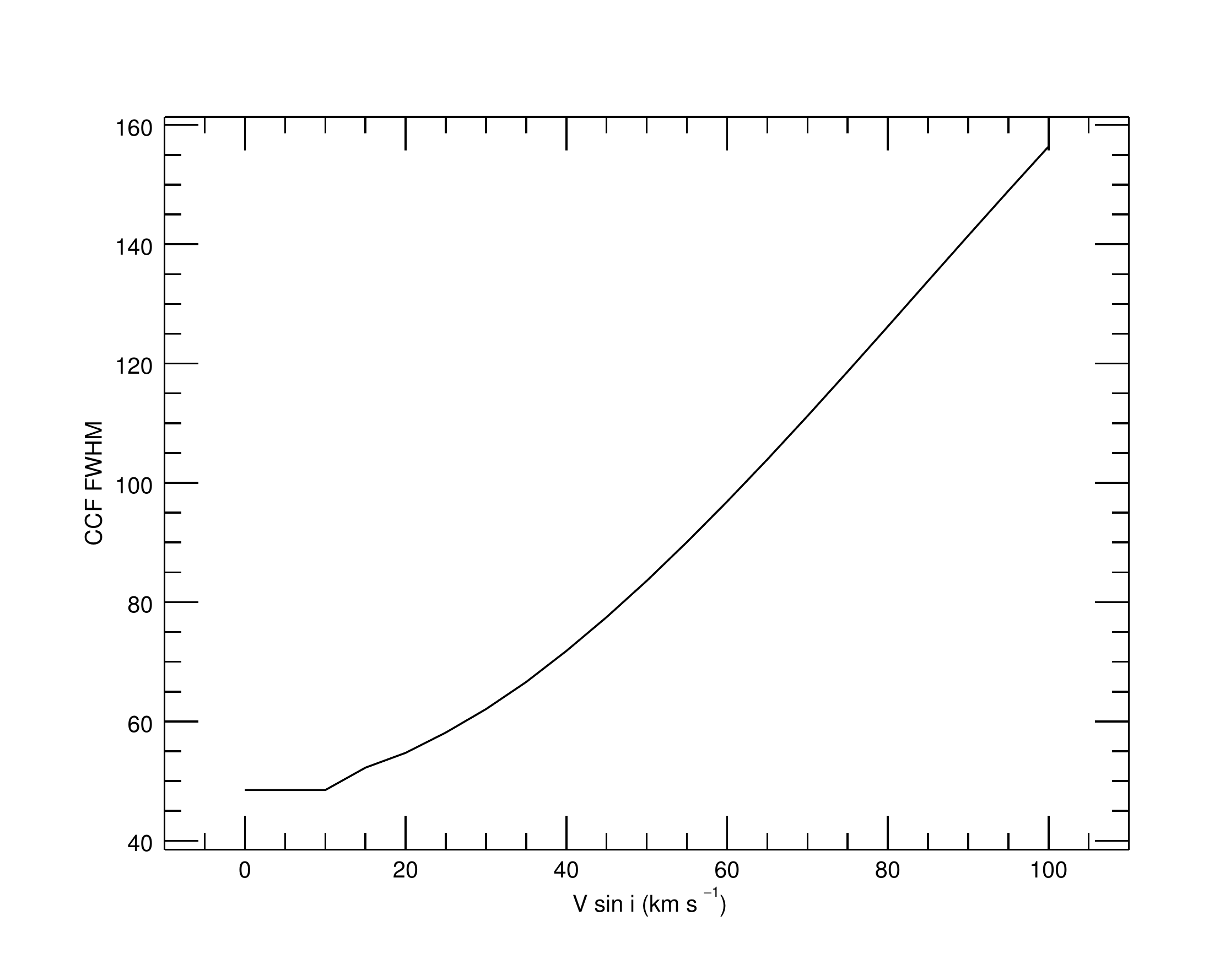}
\caption{The distribution of CCF full width at half maximum (FWHM) as a function of assumed model sdO $V \sin{i}$ values in the binary system HD 43544. Note that no rotational broadening is applied in the models for $V\sin i \leq 10$ km~s$^{-1}$.}
\label{fig:CCFvsini}
\end{figure*}

% Figure 9: CCF FWHM VS Vsini 
\placefigure{fig:CCFvsini}

% Table 3: Physical parameters of sdO
\begin{deluxetable*}{llccccc}
%\tabletypesize{\scriptsize}
\tablenum{3}
\tablecaption{Physical properties of sdO stars \label{tab: sdO parameters}}
\tablewidth{0pt}
\tablehead{
\colhead{HD} &  \colhead{$T_{\rm eff}$}  & \colhead{$V\sin{i}$}    &\colhead{$f_2/f_1$} & \colhead{$F_2/F_1$} & \colhead{$R_2$} & \colhead{$\log{L}$}  \\
\colhead{Number} & \colhead{(K)} & \colhead{(km s$^{-1}$)} & \colhead{} & \colhead{} & \colhead{($R_\odot$)} & \colhead{($L_\odot$)}
}
\startdata
 29441  & 40000  & $<$ 15   & 0.027 $\pm$ 0.003 & \phn9.9 $\pm$ 2.4   &  0.28 $\pm$ 0.04  & $2.26^{+0.14}_{-0.21}$  \\ 
 43544  & 38200 & $<$ 15  &  0.090 $\pm$ 0.009 & \phn6.8 $\pm$ 1.5    &   0.51 $\pm$ 0.08  & $2.70^{+0.15}_{-0.23}$   \\ 
 51354  & 43500 &  102 $\pm$ 4 &  0.099 $\pm$ 0.027  & 11.7 $\pm$ 2.7 &  0.47 $\pm$ 0.09  & $2.85^{+0.13}_{-0.18}$ \\ 
 55606  & 40900 &  $<$ 24 &  0.041 $\pm$ 0.002 & \phn8.4 $\pm$ 1.9    &  0.31 $\pm$ 0.04  & $2.38^{+0.13}_{-0.19}$ \\ 
 60855  & 42000 &  $<$ 27 &  0.041 $\pm$ 0.003  & 10.9 $\pm$ 2.6 &  0.50 $\pm$ 0.07  &$2.85^{+0.14}_{-0.20}$    \\ 
113120  & 45000  & $<$ 36 &  0.041 $\pm$ 0.009  & \phn7.6 $\pm$ 3.6 & 0.30 $\pm$ 0.10   &$2.52^{+0.23}_{-0.53}$  \\ 
137387  & 40000 & $<$ 17  & 0.032 $\pm$ 0.003  & \phn4.8 $\pm$ 1.0  &  0.44 $\pm$ 0.06  &$2.65^{+0.14}_{-0.20}$  \\ 
152478  & 42000 &  $<$ 15 & 0.049 $\pm$ 0.003  & 11.5 $\pm$ 2.7   &  0.27 $\pm$ 0.04 &$2.31^{+0.14}_{-0.21}$   \\ 
157042  & 33800  & $<$ 36  & 0.026 $\pm$ 0.003  & \phn2.5 $\pm$ 0.6 &  0.61 $\pm$ 0.09  & $2.64^{+0.15}_{-0.23}$    \\ 
157832  & 45000\tablenotemark{a} & \nodata  &   $<$ 0.007  &  \phn5.2   &  $<$ 0.37 &  $<$ 2.70  \\ 
191610  & 45000\tablenotemark{a}  & \nodata  &  $<$ 0.024  &  11.5   & $<$ 0.19   &    $<$ 2.13 \\ 
194335  & 43500 & $<$ 15  & 0.047 $\pm$ 0.007  & \phn4.8 $\pm$ 0.9   &  0.52 $\pm$ 0.07 &$2.94^{+0.15}_{-0.23}$  \\  
214168  & 45000\tablenotemark{a} & \nodata   & $<$ 0.019  & \phn3.6   & $<$ 0.42   &  $<$ 2.82   \\ 
\enddata
%\tablecomments{}
\tablenotetext{a}{Assumed sdO temperature value.}
\end{deluxetable*}

%%%
\subsection{Effective Temperature \label{tab: Teff}} 

The residual CCF peak heights are sensitive to the degree of agreement between 
the model sdO spectrum and the sdO line contribution to the observed spectrum. 
The model properties are primarily set by the assumed effective temperature, 
so for each positive detection, we calculated residual CCFs over a range 
in assumed sdO temperature in the same way as described above (Section 4)
and then we measured the resulting peak height.
Figure~10 shows the variation in peak height with assumed effective temperature
$T_{\rm eff}$ for one observation of HD~43544 plus a spline fit of the relation. 
We took $T_{\rm eff}$ at the position of the maximum of the spline fit 
as the best estimate of temperature of the sdO star, and these are listed for 
the ten detected Be+sdO systems in column 2 of Table~3.  The range in the 
resulting $T_{\rm eff}$ values between observations is generally small, but given the uncertainty 
in the other model parameters (such as how much the actual gravity differs 
from the assumed $\log g = 4.75$), we estimate that temperature uncertainties
are comparable to the grid step size in the OSTAR2003 models, 
$\triangle T_{\rm eff} = 2.5$~kK.

\begin{figure*}
\includegraphics[width=\textwidth]{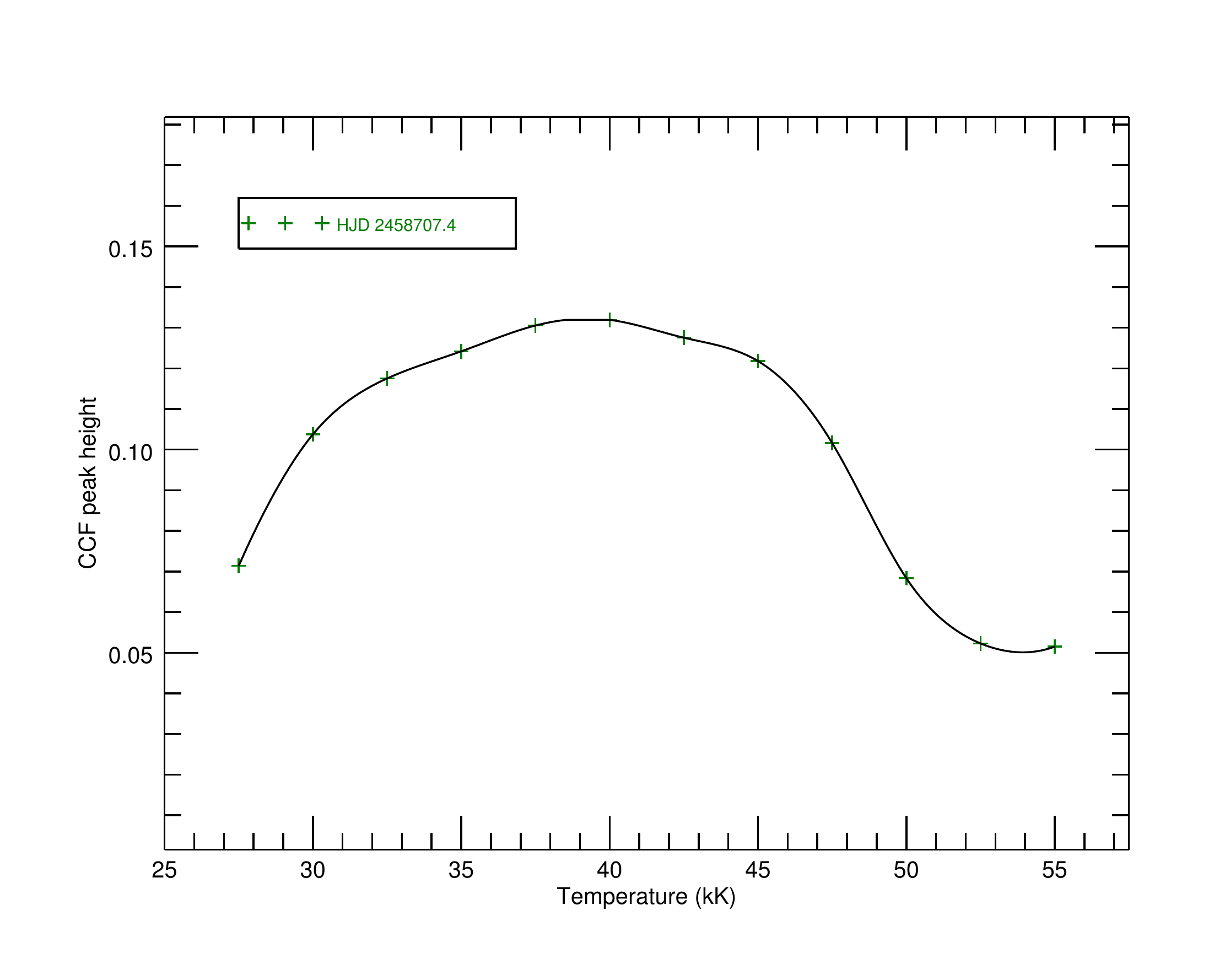}
\caption{Distribution of CCF peak heights as a function of $T_{\rm eff}$ of 
the sdO companion in the binary system HD 43544 calculated for 
the spectrum obtained on HJD 2458707.4. A spline fit was applied to the 
distribution to estimate the effective temperature of the sdO companion 
from the value where CCF reaches a local maximum.} 
\label{fig:CCFteff}
\end{figure*}

% Figure 10 goes here. 
% Figure 10: CCFs peak heights VS sdO Teff
\placefigure{fig:CCFteff}

%%%
\subsection{Monochromatic Flux Ratio \label{tab: flux ratio}}

The height of the residual CCF peak is closely related to the flux contribution 
of the sdO star to the combined FUV flux.  We relied upon a simulation of 
the residual CCFs for models of the spectra of the Be star and sdO star that 
were made over a range of test values of monochromatic flux ratio $r=(f_2/f_1)$.
Each composite model was created using the best fit parameters for the Be star 
from Table~1 (from the BSTAR2006 grid) and for the sdO star from Table~3
(from the OSTAR2003 grid), and then the two rectified spectra were added by
\begin{equation}
f({\rm model}) = (f({\rm Be}) + r f({\rm sdO}))/(1 + r).
\end{equation}
This approach assumes a constant flux ratio across the observed wavelength 
range, which is a reasonable assumption given the high temperatures of 
both components.   We then calculated the CCF of the model spectrum 
with an sdO model for the best fit $T_{\rm eff}$ and $V \sin i = 0$ km s$^{-1}$ 
in the same way as we did for the observed spectra (Section 4). 
The Be contribution to the resulting CCF was then removed as done previously,
and the residual CCF peak height for the sdO component was measured. 
The result is a relation between assumed flux ratio $r$ and residual 
CCF peak height, and an example is shown in Figure~11.   
We interpolated within each of these target specific curves to 
determine the flux ratio $r$ from the measured residual CCF peak heights. 
These flux ratio estimates and their standard deviations are listed 
in column 4 of Table~3. 

\begin{figure*}[ht!]
\includegraphics[width=\textwidth]{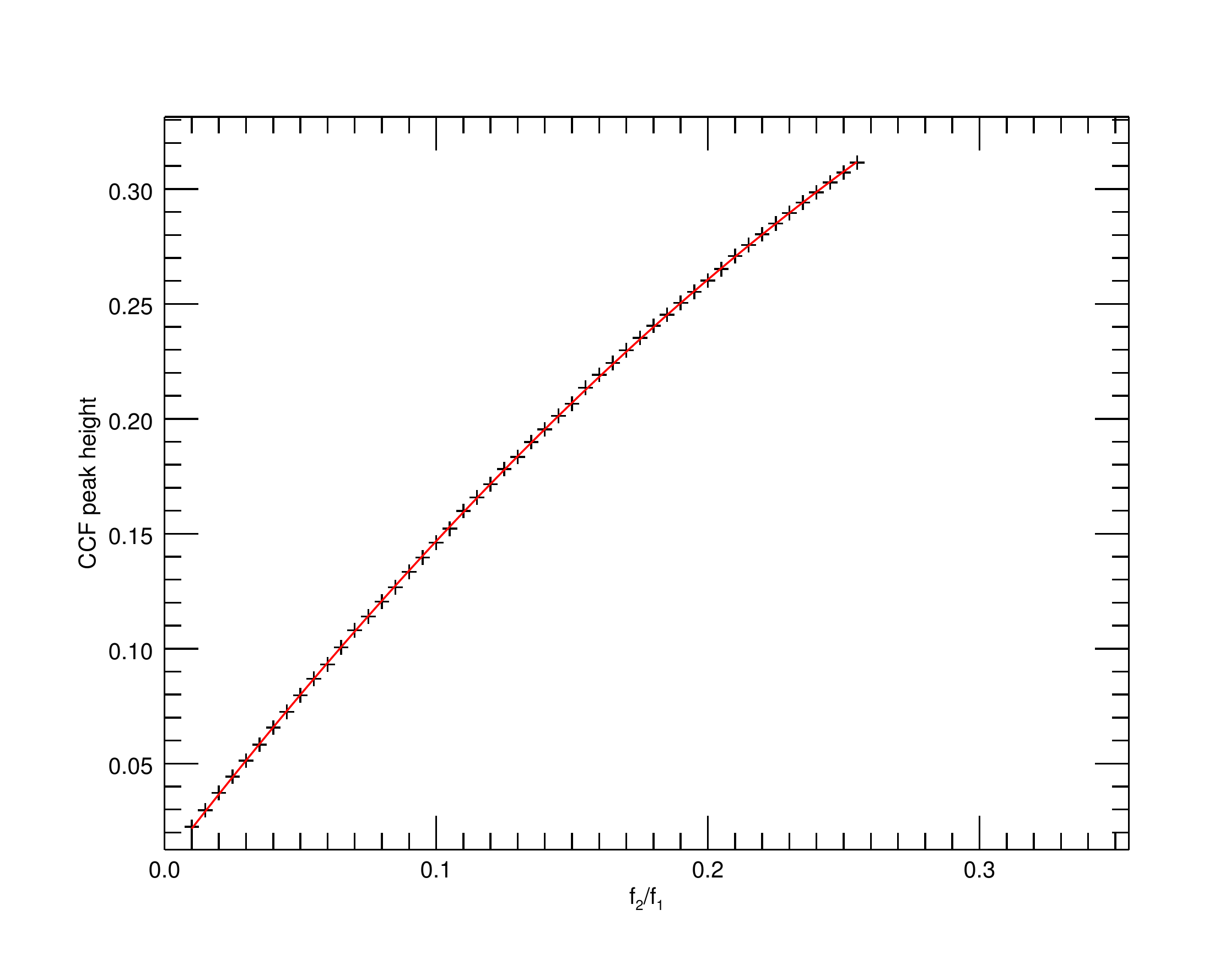}
\caption{Distribution of the CCF peak heights as a function of assumed model flux ratio $f_2/f_1$ for the binary system HD 43544 calculated from TLUSTY model spectra. The final estimation of $f_2/f_1$ value of the sdO companion in the system was obtained through interpolating the observed CCF peak height to the distribution. }
\label{fig:CCFfluxratio}
\end{figure*}

% Figure 11 goes here. 
% Figure 11: CCF peak heights VS f2/f1 
\placefigure{fig:CCFfluxratio}

The observed flux ratio $f_2/f_1$ is related to surface flux ratios $F_2/F_1$ 
and area ratio $(R_2/R_1)^2$ by 
\begin{equation}
{{f_2}\over {f_1}} = {{F_2}\over {F_1}} \left({{R_2}\over {R_1}}\right) ^2.
\end{equation}
The surface fluxes were taken as the average fluxes from the TLUSTY/SYNSPEC 
flux models over the range 1450 -- 1460 \AA, which corresponds to the near center 
part of the observed wavelength range.  These fluxes depend mainly upon the 
adopted temperatures of the Be and sdO stars, and the surface flux ratio $F_2/F_1$ 
is given in column 5 of Table~3.  The uncertainty in this ratio was determined 
by the errors in the effective temperatures (assumed as $\sigma (T_{\rm eff}) = 2.5$~kK).
Thus, we can use the observed flux ratio and surface flux ratio to arrive at 
the radius ratio.  In Section 6 below, we derive estimates for the Be star 
radius $R_1$ from the spectral energy distribution and distance.  Then, we solve
for the sdO star radius $R_2$ from the radius ratio, and these radii are given 
in column 6 of Table~3.   Finally, we estimate the luminosity of the sdO star in 
column 7 using 
\begin{equation}
\log L/L_\odot = 4 \log T/T_\odot + 2 \log R/R_\odot
\end{equation}
in which the nominal solar temperature is $T_\odot = 5772$~K \citep{prsa2016}.

%%%%%%%%%%%%%%%%%%%%%%%%%%%%%%
\section{Spectral Energy Distributions of \texorpdfstring{B\MakeLowercase{e}}{Be}  stars \label{tab: SED fitting}}

The radius of the sdO star can be found from the radius ratio (Section 5.4) 
and the radius of the Be star.  In this section, we estimate the Be star 
radius by comparing the observed spectral energy distribution (SED) with models 
for the composite Be+sdO system in order to determine the Be star angular 
diameter.  We then use a distance estimate to find the Be star physical radius,
which in turn leads to the sdO star radius (Table~3).

The angular diameter of the limb darkened disk $\theta$ (in units of radians) 
of a single star is found by the inverse-square law:
\begin{equation}
{ {f_\lambda ({\rm observed})}\over
  {F_\lambda ({\rm emitted}) } } =
  ( {R_\star / d} )^2 ~10^{-0.4 A_\lambda} =
  {1\over 4} \theta^2~10^{-0.4 A_\lambda}
\end{equation}
where the ratio of the observed and emitted fluxes
depends on the square of the ratio of stellar radius $R_\star$
to distance $d$ and the interstellar extinction $A_\lambda$.
Here we create a model spectrum for the Be+sdO system $F_\lambda$ that is 
scaled to the Be star flux, and then we fit this to the observed flux $f_\lambda$
to find the angular diameter $\theta$ and reddening $E(B-V)$. 
We adopted an extinction curve law $A_\lambda$ from \citet{fitzpatrick1999}
that is a function of the reddening $E(B-V)$ and the ratio of
total-to-selective extinction $R=A_V/E(B-V)$ (set at a value of 3.1).   

Low resolution ($R=500$) model spectra were formed 
from surface fluxes from BSTAR2006 for the Be star 
and from OSTAR2003 for the sdO star using the atmospheric parameters listed 
in Tables 1 and 3, respectively.  Then the combined spectrum is given by 
$F_\lambda ({\rm Be}) + F_\lambda ({\rm sdO}) (R_2/R_1)^2$, using the radius 
ratio determined in Section 5.4.  In the cases of the sdO non-detections 
(HD~157832, HD~191610, HD~214168), only the Be star flux is included in the model. 

The flux versions of the {\it HST}/STIS spectra were averaged and rebinned into
nine wavelength segments (effective resolution of $\approx 14$) to represent 
the flux in the FUV.  We then extended the wavelength range into the NUV 
by adding fluxes measured by the TD-1 mission \citep{thompson1978} and 
fluxes derived from the Johnson $UBV$ magnitudes \citep{mermilliod1987} 
using the flux calibration of \citet{colina1996}. 
Longer wavelength fluxes were omitted because the Be star disks become important flux 
contributors beyond the optical range.  
There were two exceptions to this general scheme.  The target HD~113120 has a 
close visual companion that is excluded in the small aperture for the 
{\it HST}/STIS observations, but is present in the longer wavelength measurements, 
so the latter were corrected to correspond to the Be+sdO system alone (see Appendix).
The other target with a close companion is HD~214168, and the three {\it HST}/STIS 
spectra record different amounts of companion flux (see Appendix).  In this case, 
we selected the spectrum with lowest flux as most representative of the Be star 
(no sdO detection) and then adjusted the longer wavelength fluxes for the Be star alone.

The composite model spectra were fit to the observed SED using the relation above
and the non-linear, least-squares fitting code MPFIT \citep{markwardt2009} to find the best-fit values of $E(B-V)$ and $\theta$ that are listed in columns 2 and 5 of Table~4, respectively. The observed and model fit SEDs are plotted in Figures 12, 13, and 14 for systems with sdO detections (Be+sdO model fluxes) and in Figure 15 for the sdO non-detections (Be model fluxes only).  The fluxes increase towards the shorter 
FUV wavelengths because of the hot temperatures of the targets and their companions, 
and the extinction and reddening values are relatively small in all cases. Column 3 of Table~4 lists published estimates of $E(B-V)$ from references cited in column 4. These published values agree with our measurements within 0.1 mag. We include in column 6 of Table~4 the Be star angular size estimates from the JMMC Stellar 
Diameters Catalog \citep{bourges2014} for comparison with our result $\theta$.  In most cases 
there is satisfactory agreement, but we caution that the JMMC results do not account for 
the flux of the sdO and other companions.    
 
\begin{figure*}
\gridline{\fig{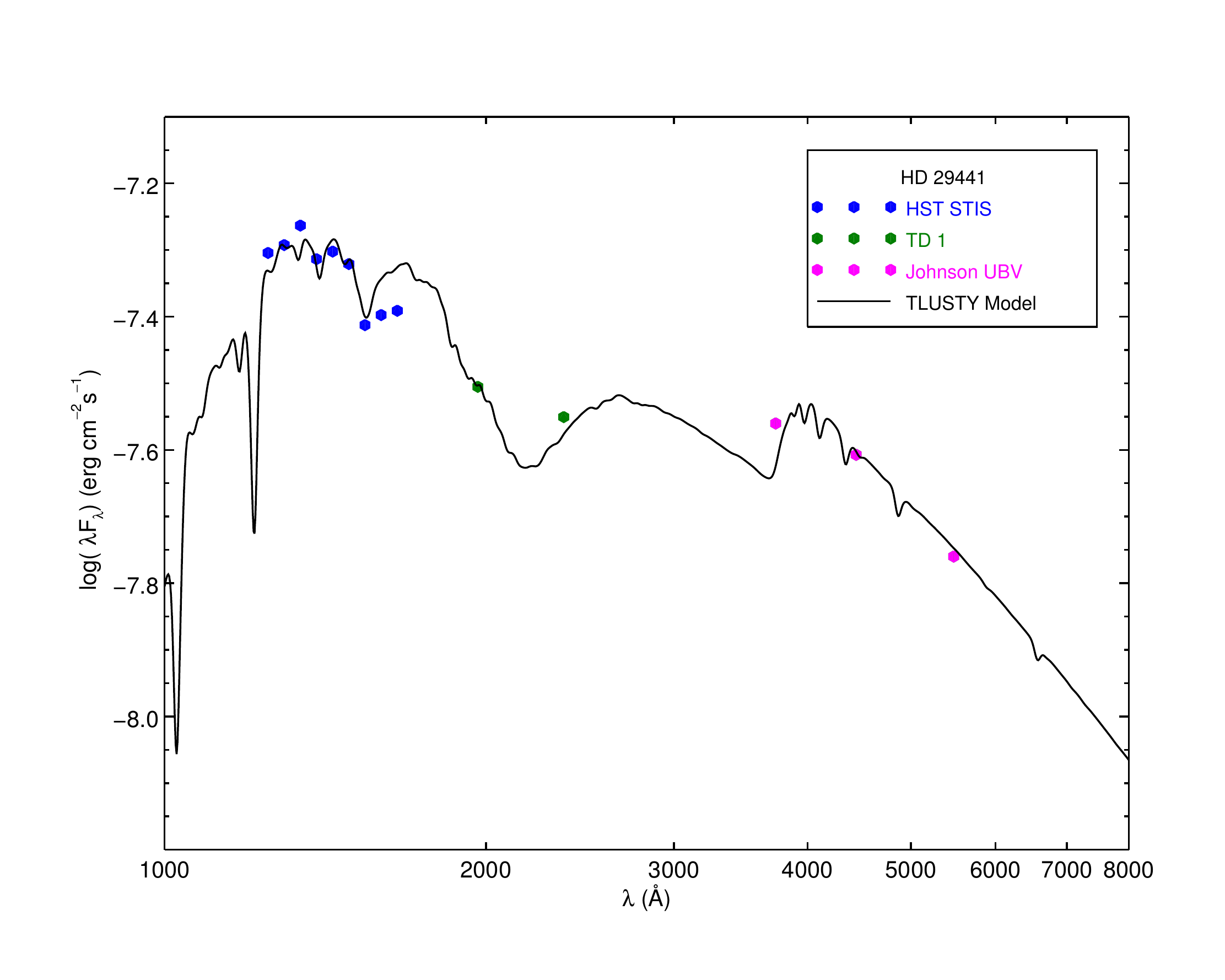}{0.5\textwidth}{(a) HD 29441}
          \fig{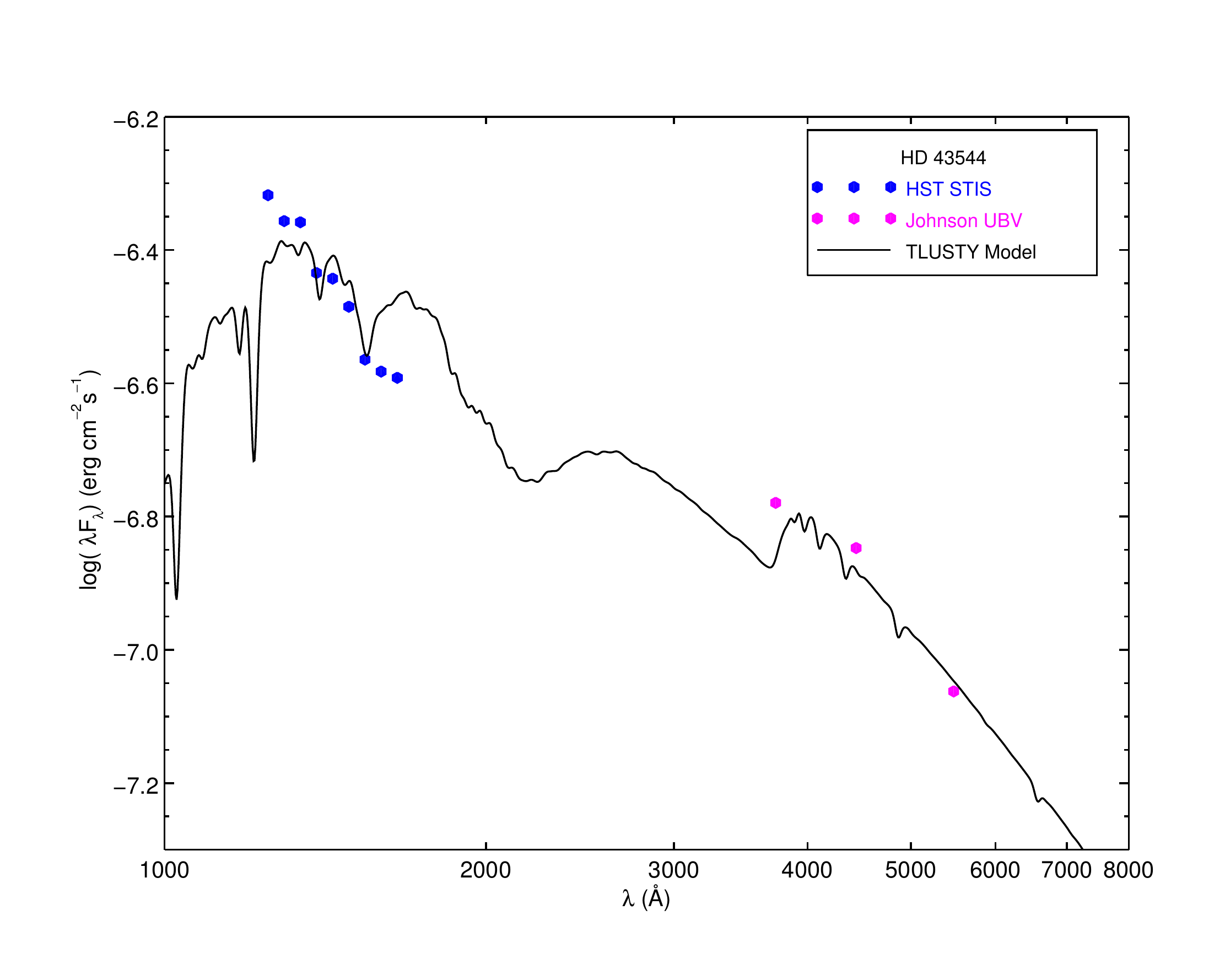}{0.5\textwidth}{(b) HD 43544}
          }
\gridline{\fig{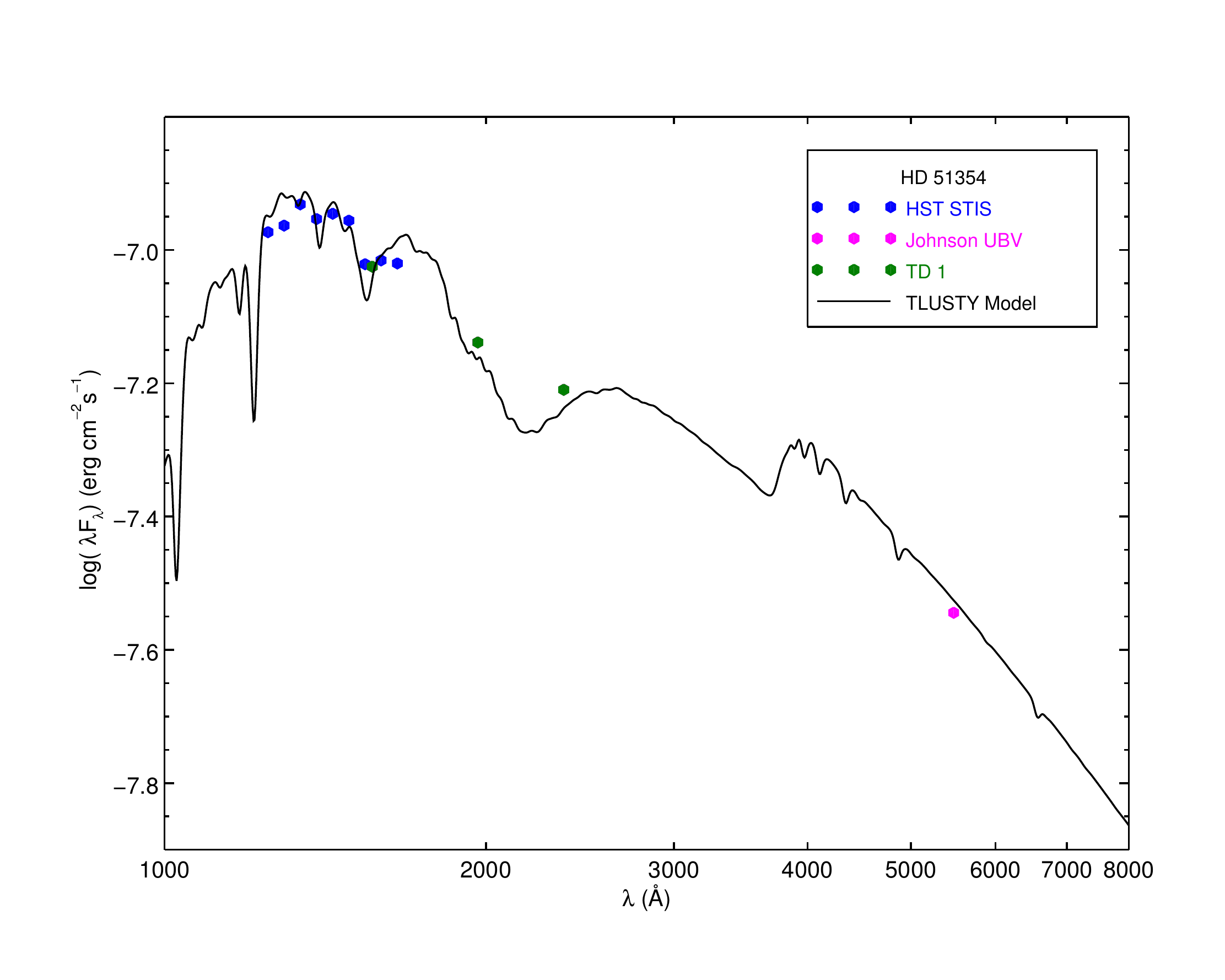}{0.5\textwidth}{(c) HD 51354}
          \fig{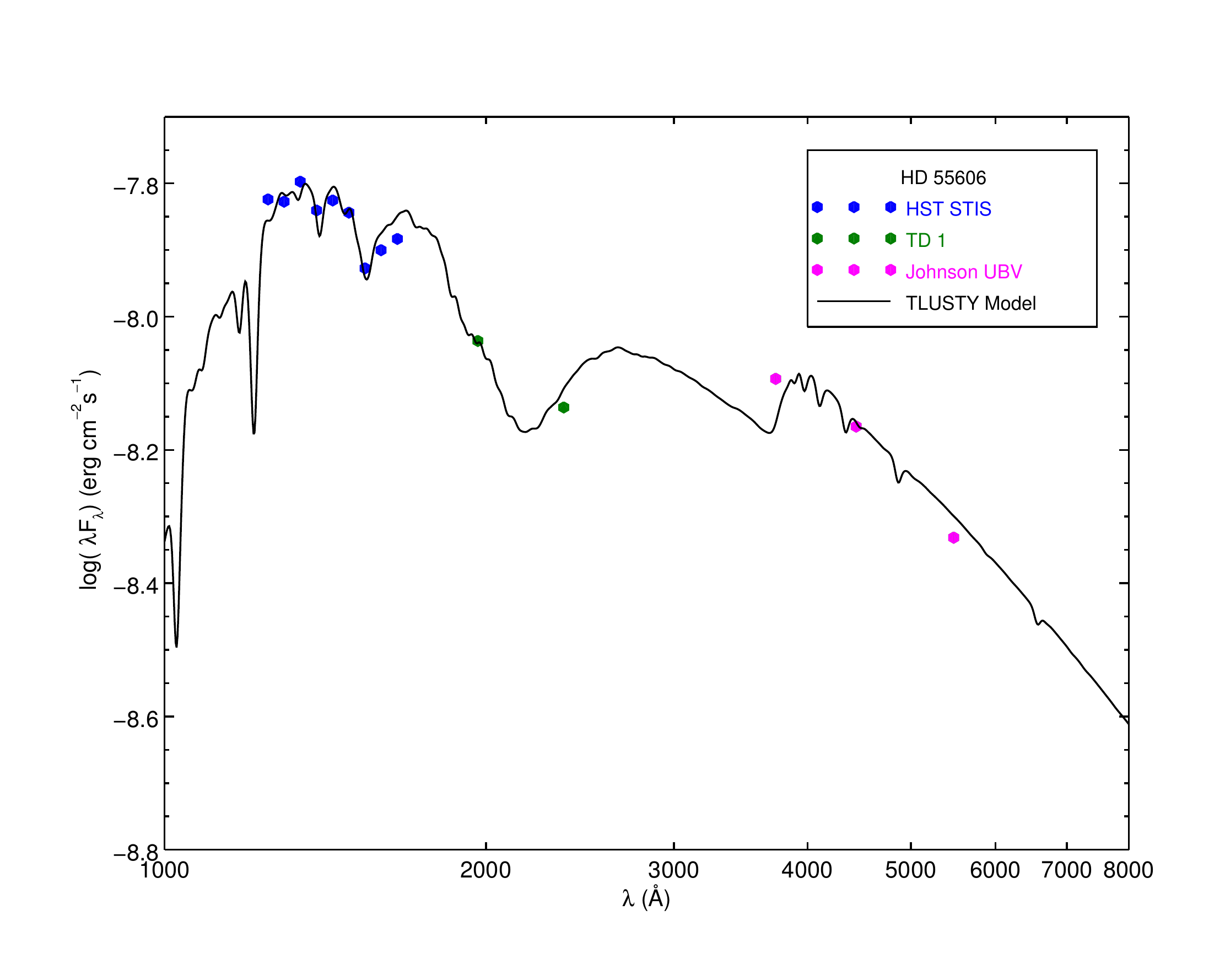}{0.5\textwidth}{(d) HD 55606}
          }
\caption{Spectral energy distribution of the Be stars: blue points are FUV \emph{HST}/STIS observations, 
green points are UV observations from \cite{thompson1978}, and magenta points are optical 
Johnson $UBV$ observations from \cite{mermilliod2006}.}
\label{fig:SED1}
\end{figure*}

% Figures 12, 13, 14 here
% Figure 12: SED fits 
\placefigure{fig:SED1}

\begin{figure*}  
\gridline{\fig{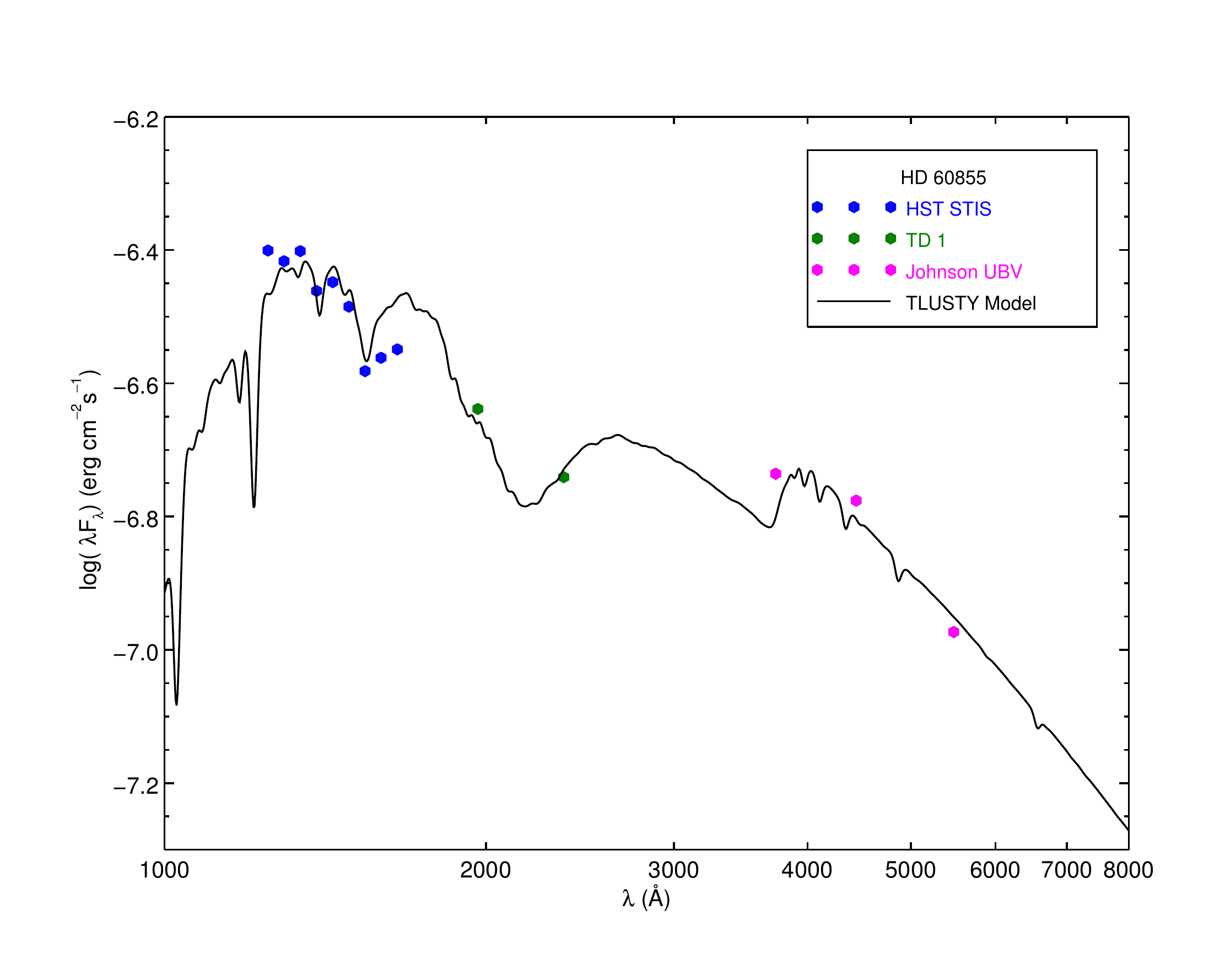}{0.5\textwidth}{(a) HD 60855}
	\fig{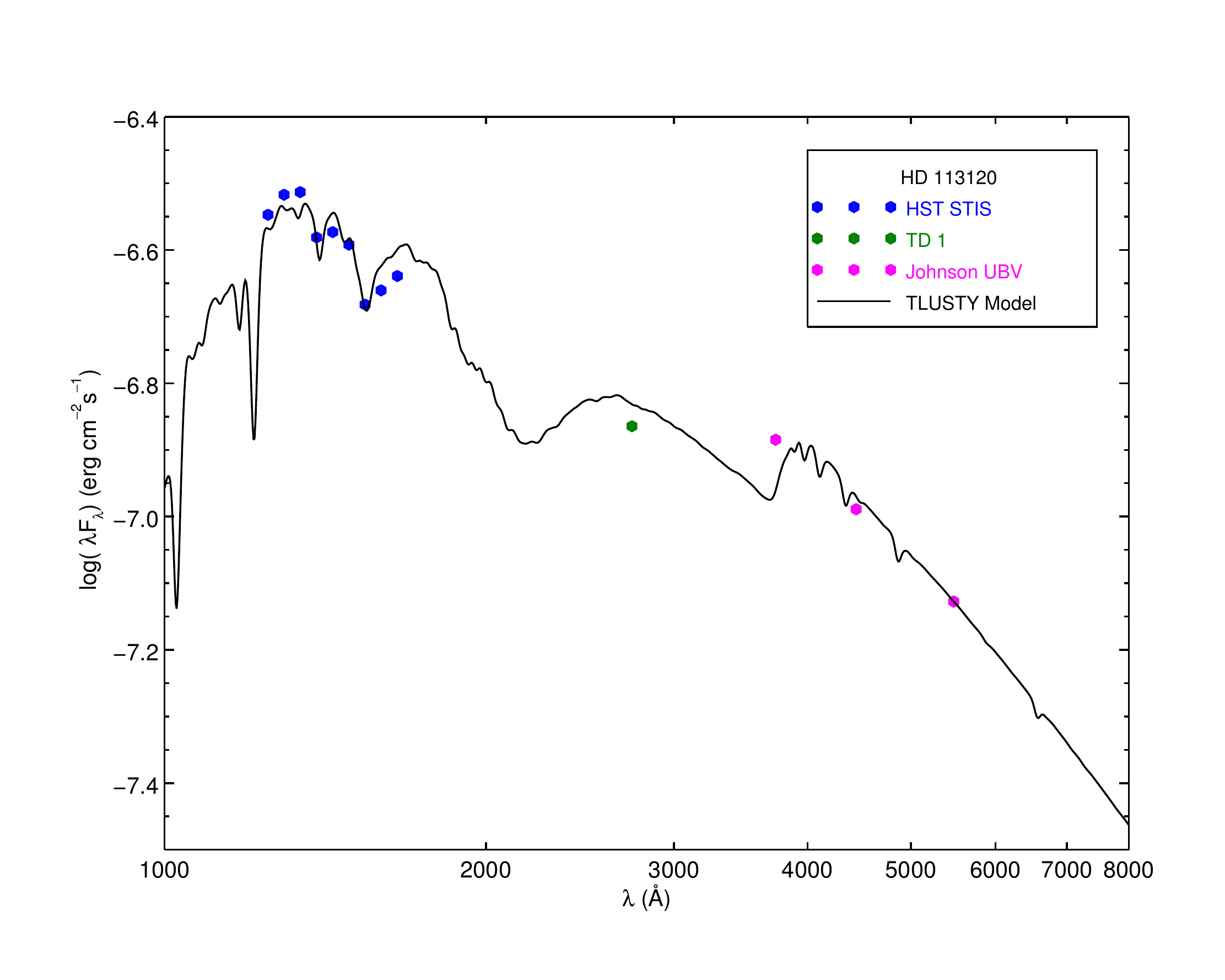}{0.5\textwidth}{(b) HD 113120}
	}      
 \gridline{\fig{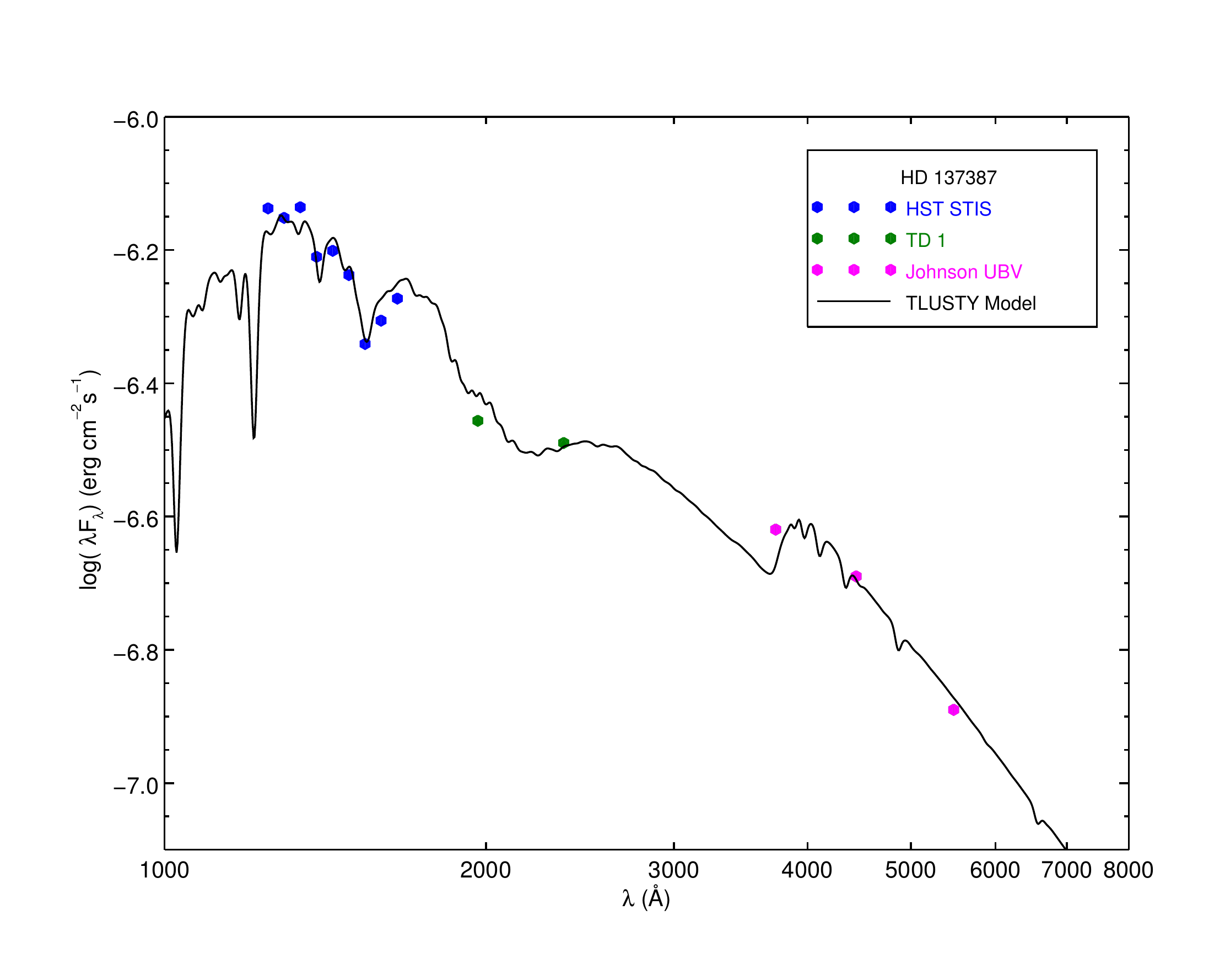}{0.5\textwidth}{(c) HD 137387}
	\fig{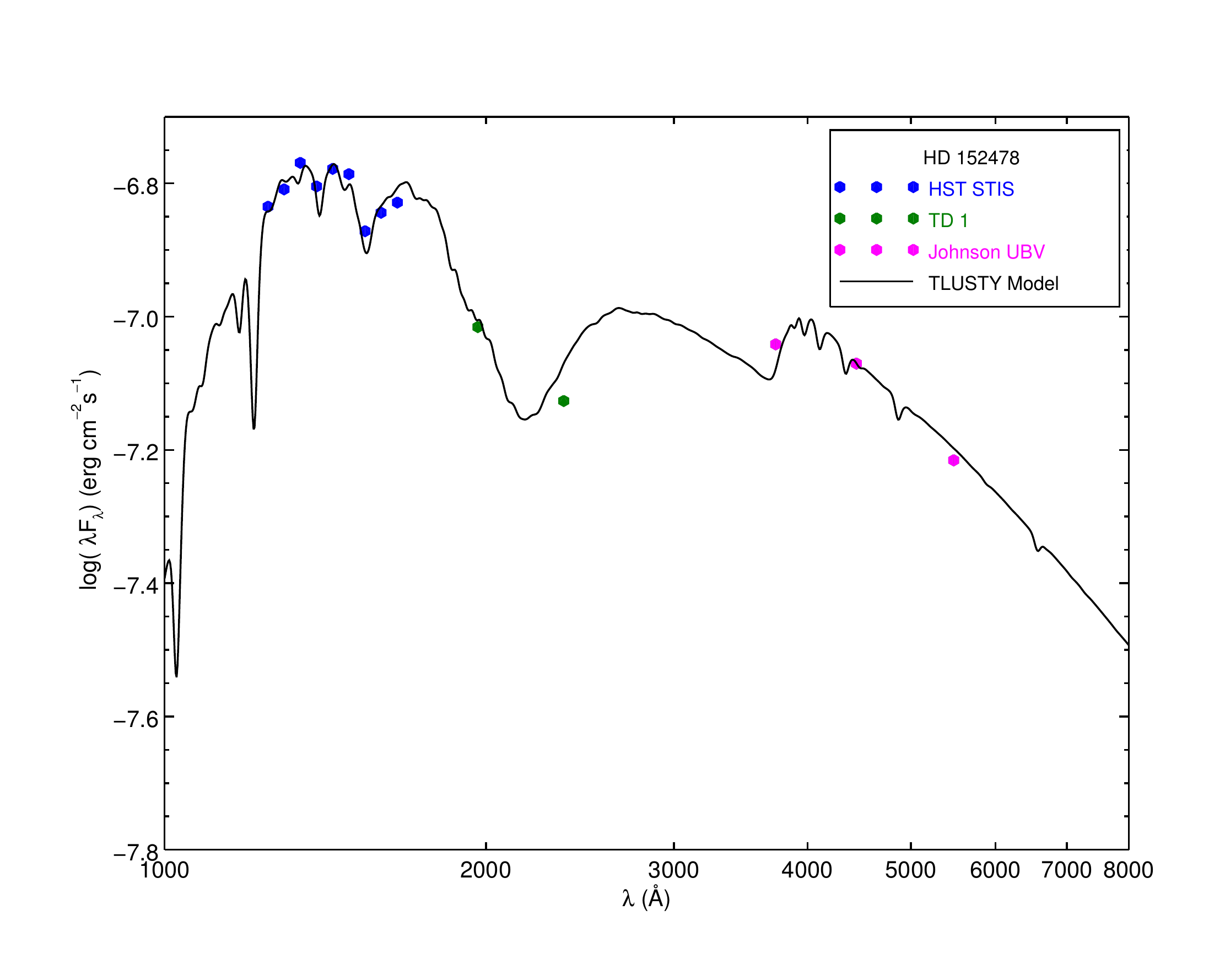}{0.5\textwidth}{(d) HD 152478 }
	}
\caption{Spectral energy distribution of the Be stars in the same format as Fig.\ 12.}
\label{fig:SED2}
\end{figure*}

% Figure 13: SED fit 
\placefigure{fig:SED2}
     
\begin{figure*}  
\gridline{\fig{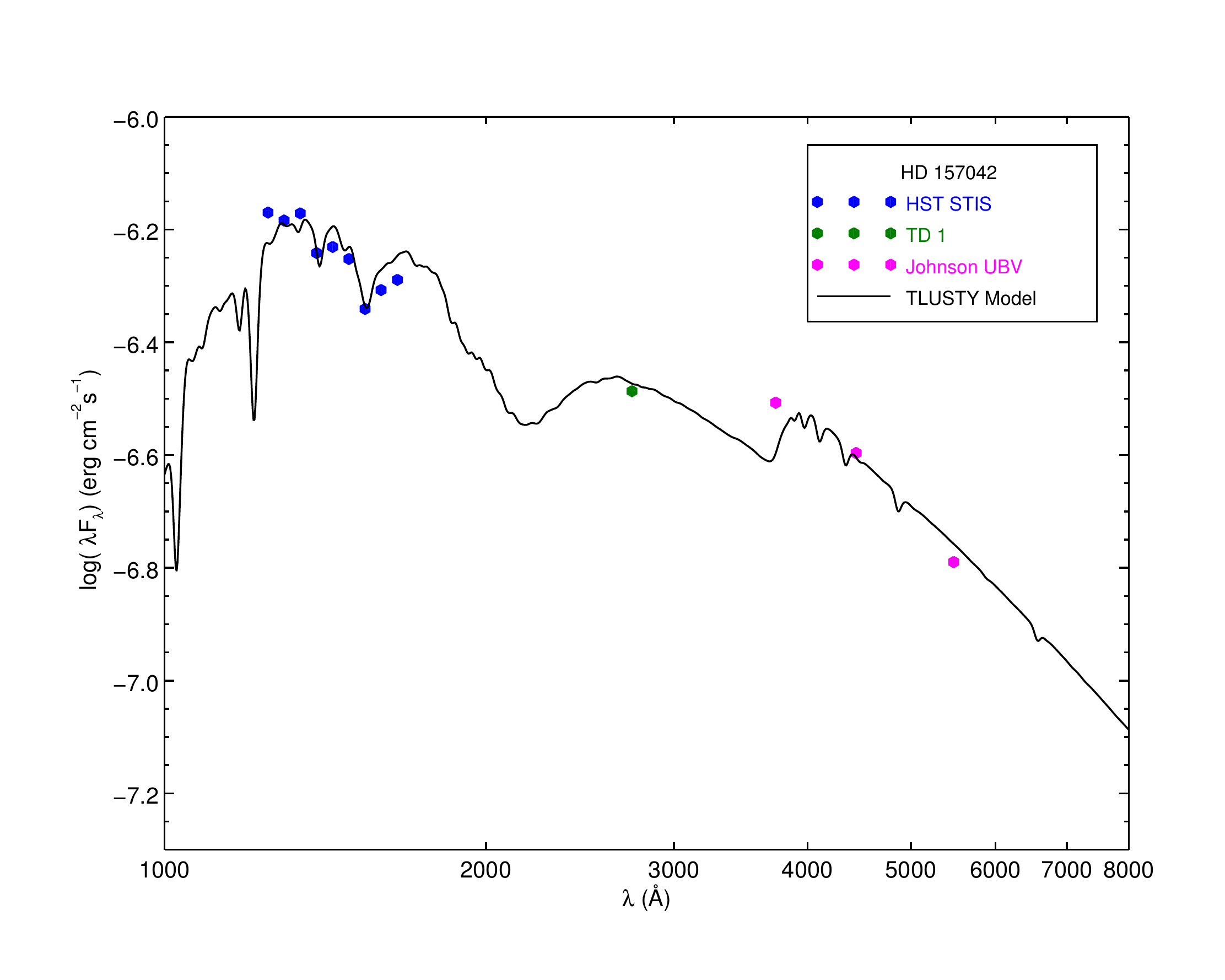}{0.5\textwidth}{(a) HD 157042}
	\fig{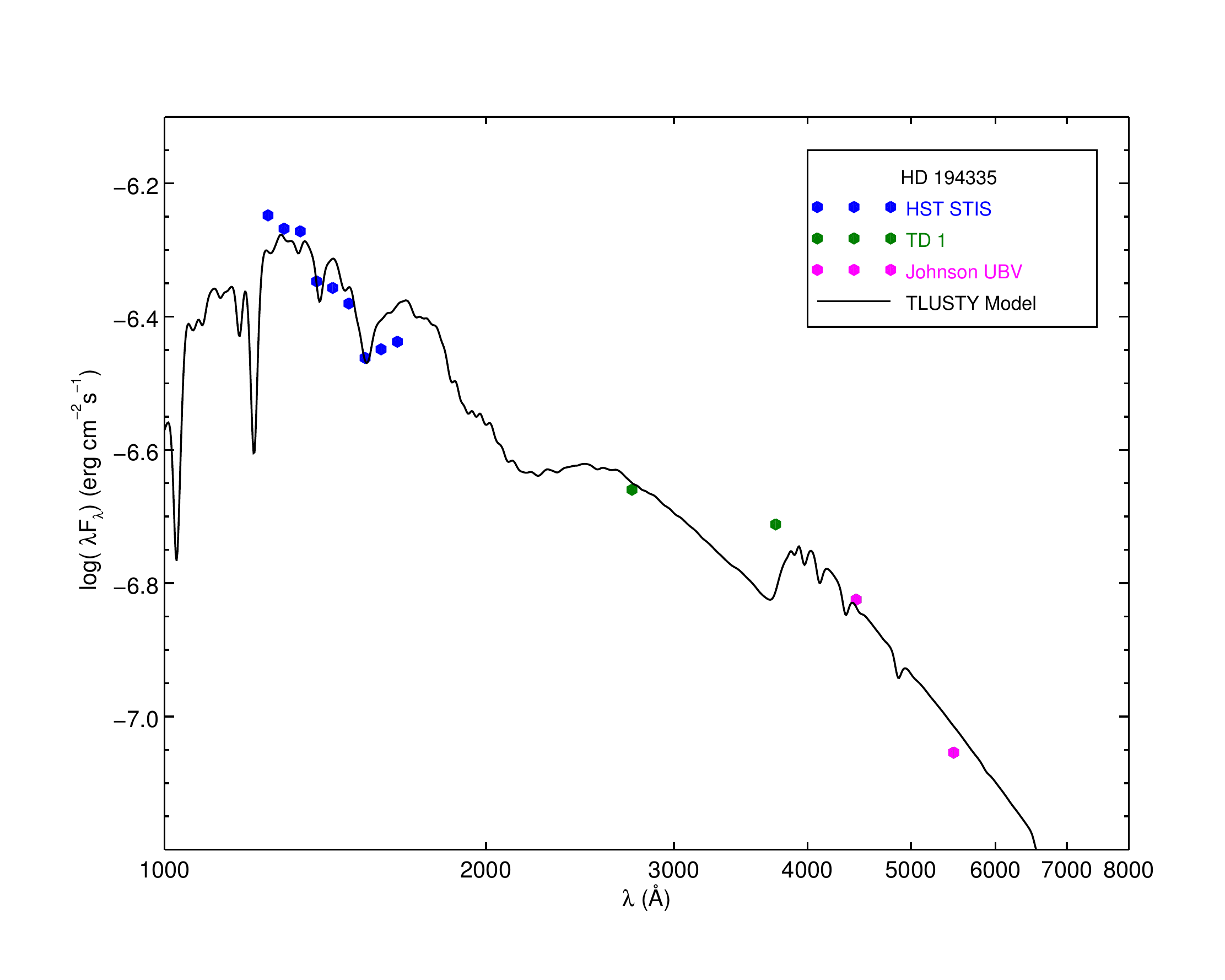}{0.5\textwidth}{(b) HD 194335 }
	}
\caption{Spectral energy distribution of the Be stars in the same format as Fig.\ 12.}
\label{fig:SED3}
\end{figure*}

% Figure 14: SED fit 
\placefigure{fig:SED3}
     
\begin{figure*}  
\gridline{\fig{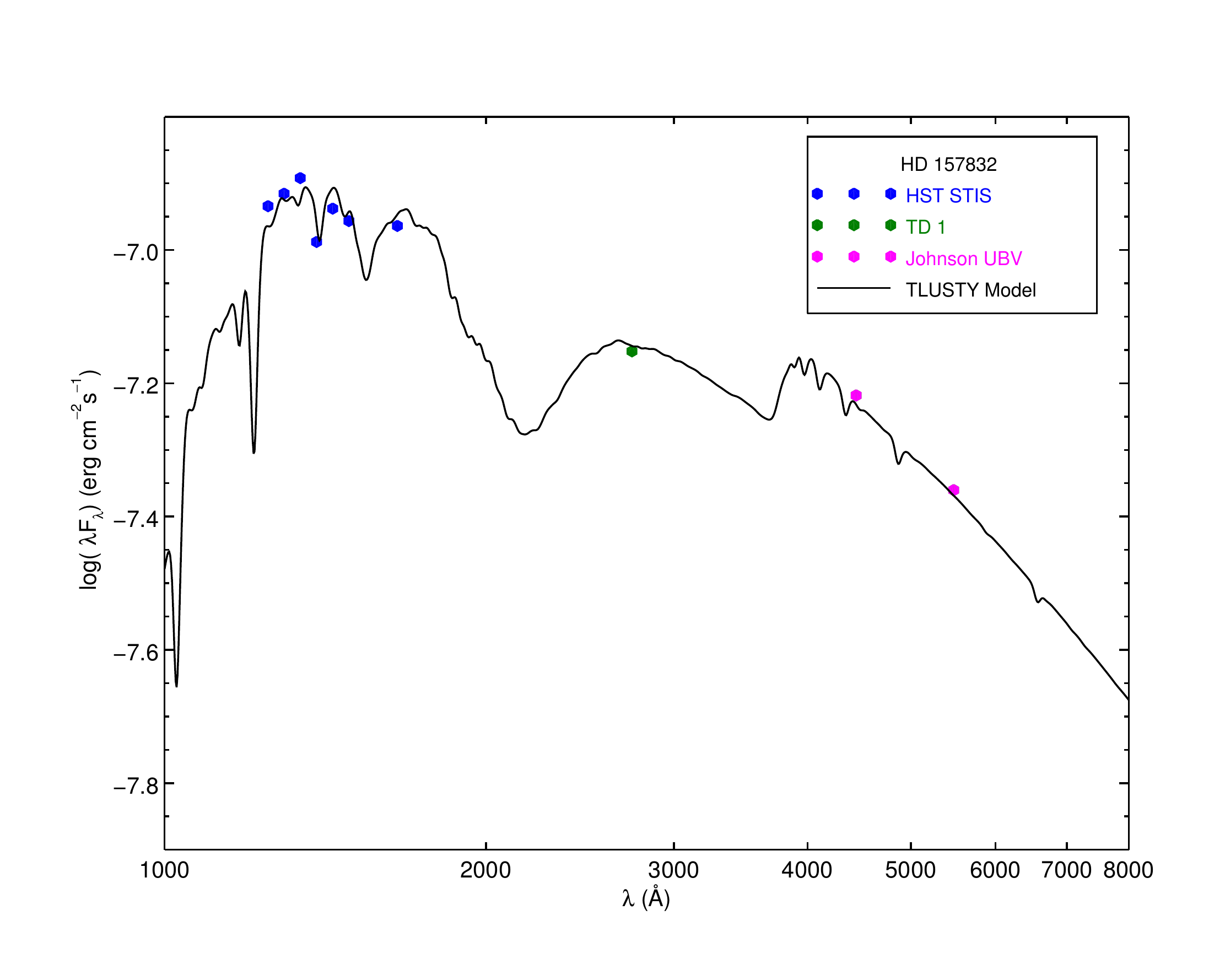}{0.5\textwidth}{(a) HD 157832}
	\fig{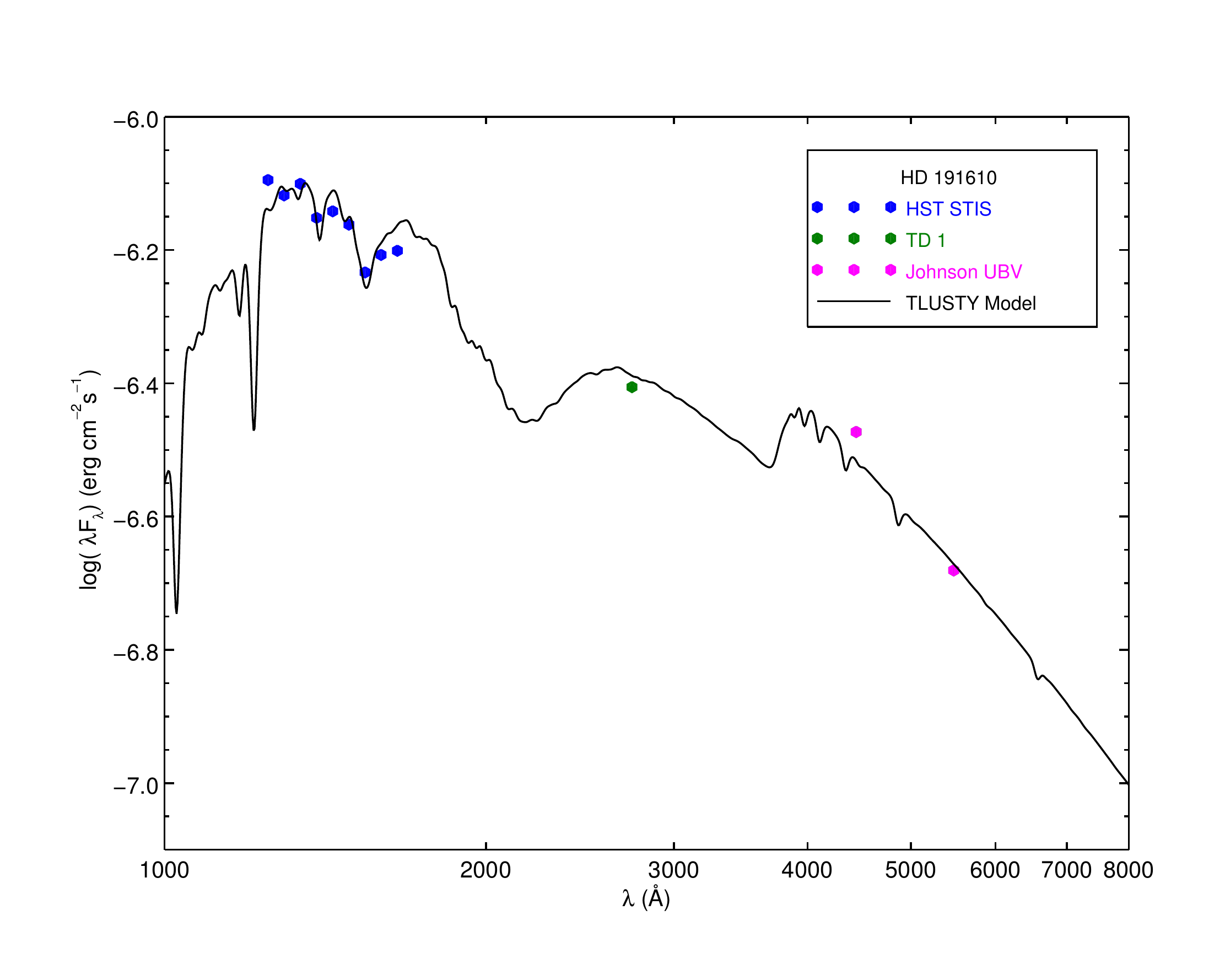}{0.5\textwidth}{(b) HD 191610}
	}      
 \gridline{\fig{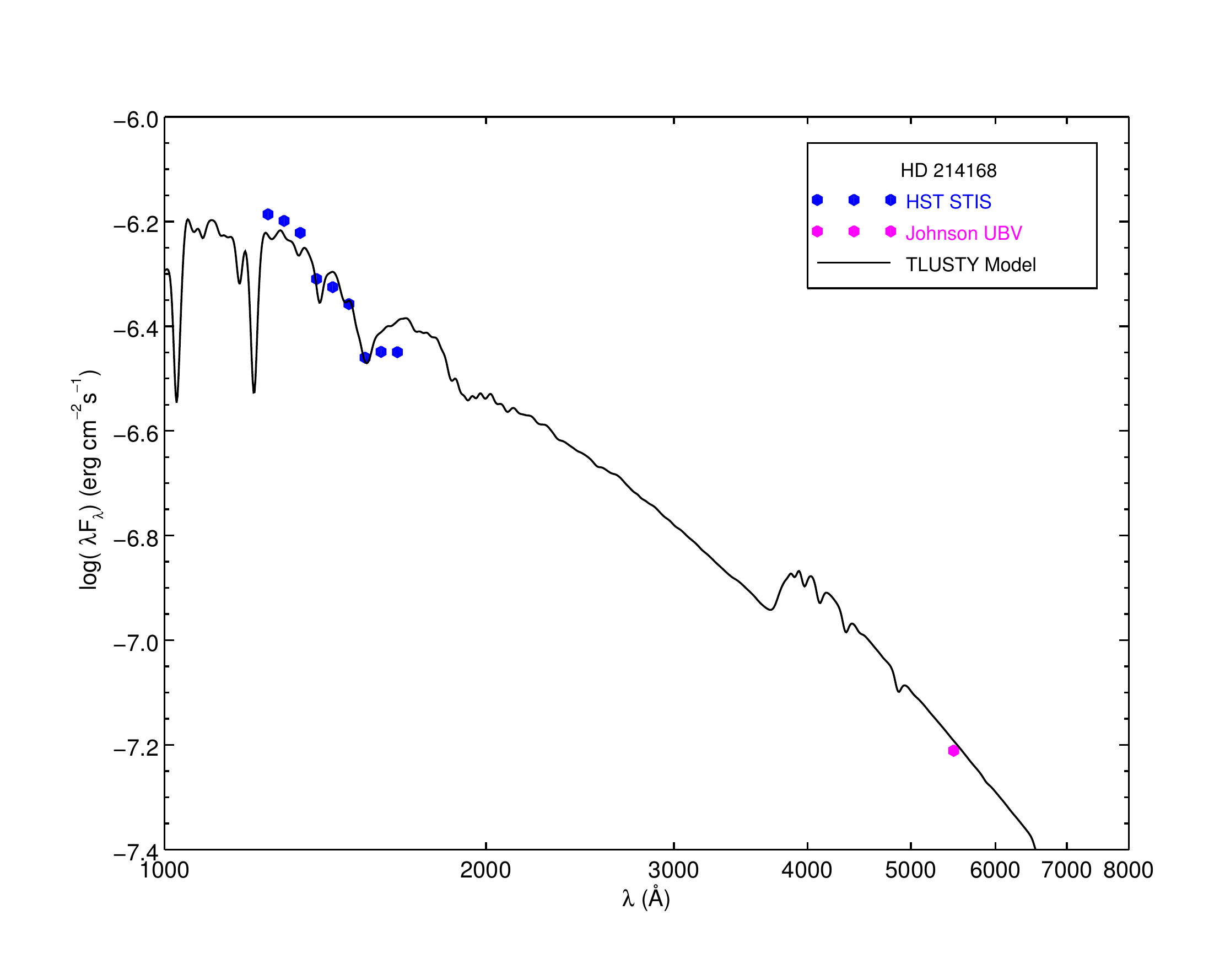}{0.5\textwidth}{(c) HD 214168}
	}
\caption{Spectral energy distribution of the Be stars in the same format as Fig.\ 12.}
\label{fig:SED4}
\end{figure*}

% Figure 15 here  
% Figure 15: SED fit for null detection targets
\placefigure{fig:SED4}

% Table 4: SED fitting parameters
\begin{deluxetable*}{lccccccc}
\rotate
%\tabletypesize{\scriptsize}
\tablenum{4}
\tablecaption{Spectral energy distribution fitting parameters \label{tab: SED parameters}}
\tablewidth{0pt}
\tablehead{
\colhead{HD} &  \colhead{$E(B-V)_{HST}$}  & \colhead{$E(B-V)$} & \colhead{Reference} & \colhead{$\theta$} & \colhead{$\theta_{JMCC}$} & \colhead{$d$} & \colhead{$R_{Be}$}  \\
\colhead{Number} &  \colhead{(mag)} &  \colhead{(mag)} &  &  \colhead{(mas)} & \colhead{(mas)} & \colhead{(pc)} & \colhead{($R_\odot$)} 
}
\startdata
 29441  & 0.198 $\pm$ 0.014 & 0.202 & 1 & 0.0673 $\pm$ 0.0033 & 0.0763 $\pm$ 0.0030 &  \phn737 $\pm$ 38  &  \phn5.34 $\pm$ 0.38 \\ 
 43544  & 0.155 $\pm$ 0.024 &  0.054 $\pm$ 0.013 & 2 & 0.1315 $\pm$ 0.0113 & 0.1169 $\pm$ 0.0042 &  \phn314 $\pm$  6\phn  &  \phn4.44 $\pm$ 0.39 \\ 
 51354  & 0.172 $\pm$ 0.015  &  0.07 & 3 & 0.0779 $\pm$ 0.0042 & 0.0876 $\pm$ 0.0030 &  \phn608 $\pm$ 21  &  \phn5.09 $\pm$ 0.33 \\ 
 55606  & 0.220 $\pm$ 0.012 & 0.151 $\pm$ 0.100 & 2 & 0.0343 $\pm$ 0.0014 & \nodata             & 1090 $\pm$ 45  &  \phn4.01 $\pm$ 0.23 \\ 
 60855  & 0.204 $\pm$ 0.016 & 0.295 $\pm$ 0.054  & 2 & 0.1586 $\pm$ 0.0092 & 0.1877 $\pm$ 0.0078 &  \phn483 $\pm$ 25  &  \phn8.24 $\pm$ 0.64 \\ 
113120  & 0.176 $\pm$ 0.014 &  0.295 $\pm$ 0.054 & 2 & 0.1246 $\pm$ 0.0059 & \nodata             &  \phn307 $\pm$ 57  &  \phn4.11 $\pm$ 0.79 \\ 
137387  & 0.125 $\pm$ 0.012 & 0.107 $\pm$ 0.009 & 2 & 0.1558 $\pm$ 0.0066 & 0.1599 $\pm$ 0.0060 &  \phn326 $\pm$ 11  &  \phn5.45 $\pm$ 0.29 \\ 
152478  & 0.252 $\pm$ 0.011 & 0.241 $\pm$ 0.037 & 2 & 0.1274 $\pm$ 0.0051 & 0.1576 $\pm$ 0.0055 &  \phn308 $\pm$  7\phn  &  \phn4.21 $\pm$ 0.20 \\ 
157042  & 0.187 $\pm$ 0.015 &  0.111 $\pm$ 0.027 & 2 & 0.1931 $\pm$ 0.0103 & 0.1845 $\pm$ 0.0164 &  \phn291 $\pm$ 15  &  \phn6.04 $\pm$ 0.44 \\ 
157832  & 0.229 $\pm$ 0.011 & 0.138  & 1 & 0.1025 $\pm$ 0.0041 & 0.1341 $\pm$ 0.0052 & 1078 $\pm$ 85  & 11.88 $\pm$ 1.05 \\ 
191610  & 0.182 $\pm$ 0.012 & 0.064 $\pm$ 0.019 & 2 & 0.2144 $\pm$ 0.0095 & 0.2140 $\pm$ 0.0079 &  \phn188 $\pm$  8\phn  &  \phn4.34 $\pm$ 0.27 \\ 
194335  & 0.122 $\pm$ 0.017  & 0.045 $\pm$ 0.019 & 2 & 0.1311 $\pm$ 0.0078 & 0.1225 $\pm$ 0.0046 &  \phn373 $\pm$ 11  &  \phn5.26 $\pm$ 0.35 \\  
214168  & 0.025 $\pm$ 0.051  & 0.093 $\pm$ 0.024 & 2 & 0.0943 $\pm$ 0.0170 & 0.0899 $\pm$ 0.0035 &  \phn590 $\pm$ 25  &  \phn5.98 $\pm$ 1.11 \\ 
\enddata
\tablecomments{Indices of references: (1) \citet{kervella2019} (2) \citet{zorec2016} (3) \citet{gontcharov2018b} }
%\tablenotetext{d.}{ Estimated values based upon spectral type.}
\end{deluxetable*}

We multiply the angular sizes by the distance to arrive at the physical radii of the Be stars. 
The distances from the Gaia DR2 survey \citep{bailer-jones2018} are given in column 5 of Table~4, 
with the exception of the case of HD~113120 where the {\it Hipparcos} distance is adopted 
(see Appendix).  The Be star radii appear in the final column of Table~4, and they are close 
to expectations for B-type main sequence stars (except perhaps for HD~157832; see Appendix). 
We used these Be star radii estimates together with the ratio of the radii derived from the 
flux ratios to find the sdO star radii that are given in column 6 of Table~3 (Section 5.4).

%%%%%%%%%%%%%%%%%%%%%
\section{Evolutionary Tracks for the Subdwarf Stars}\label{sec:disc}

\citet{gotberg2018} explored the properties of the stripped stars in binaries by
calculating MESA evolutionary tracks for the mass donor stars.  These models 
assume a Case B scenario in which mass transfer occurs as the initially 
more massive star expands as it evolves towards the red giant phase. 
After losing the outer envelope, the remnant spends about $10\% \sim 20\%$ of its 
lifetime as a He-core burning, hot subdwarf.  We show examples of the 
evolutionary tracks from \citet{gotberg2018} in $(\log T_{\rm eff}, \log R/R_\odot)$
and $(\log T_{\rm eff}, \log L/L_\odot)$ diagrams in Figures 16 and 17, respectively.
These particular tracks assume an initial solar composition for the star. 
Four tracks are illustrated in each case for initial masses of 3.30, 3.65,
4.04, and $7.37\ M_\odot$ for the mass donor star (sdO progenitor). 
In both diagrams, we see how the star evolves from the main sequence 
to cooler temperature and larger size. In the case of the low mass sdO progenitor
($3.30\ M_\odot$), the star grows to fill its Roche limit after a period of 300 Myr. Once large scale Roche lobe overflow commences, the donor star is quickly transformed to a hotter and smaller (less luminous) object, a stage that may last only 3 Myr. The star then continues He-core burning for another 60 Myr in the portion of the tracks just beyond the first 
minimum in radius and luminosity (see also Fig.~2 in \citealt{gotberg2018}).
The masses of sdO remnant at this stage are 0.67, 0.74, 0.85, and $1.85 M_\odot$
for the four mass tracks, respectively.

\begin{figure*}
\includegraphics[width=\textwidth]{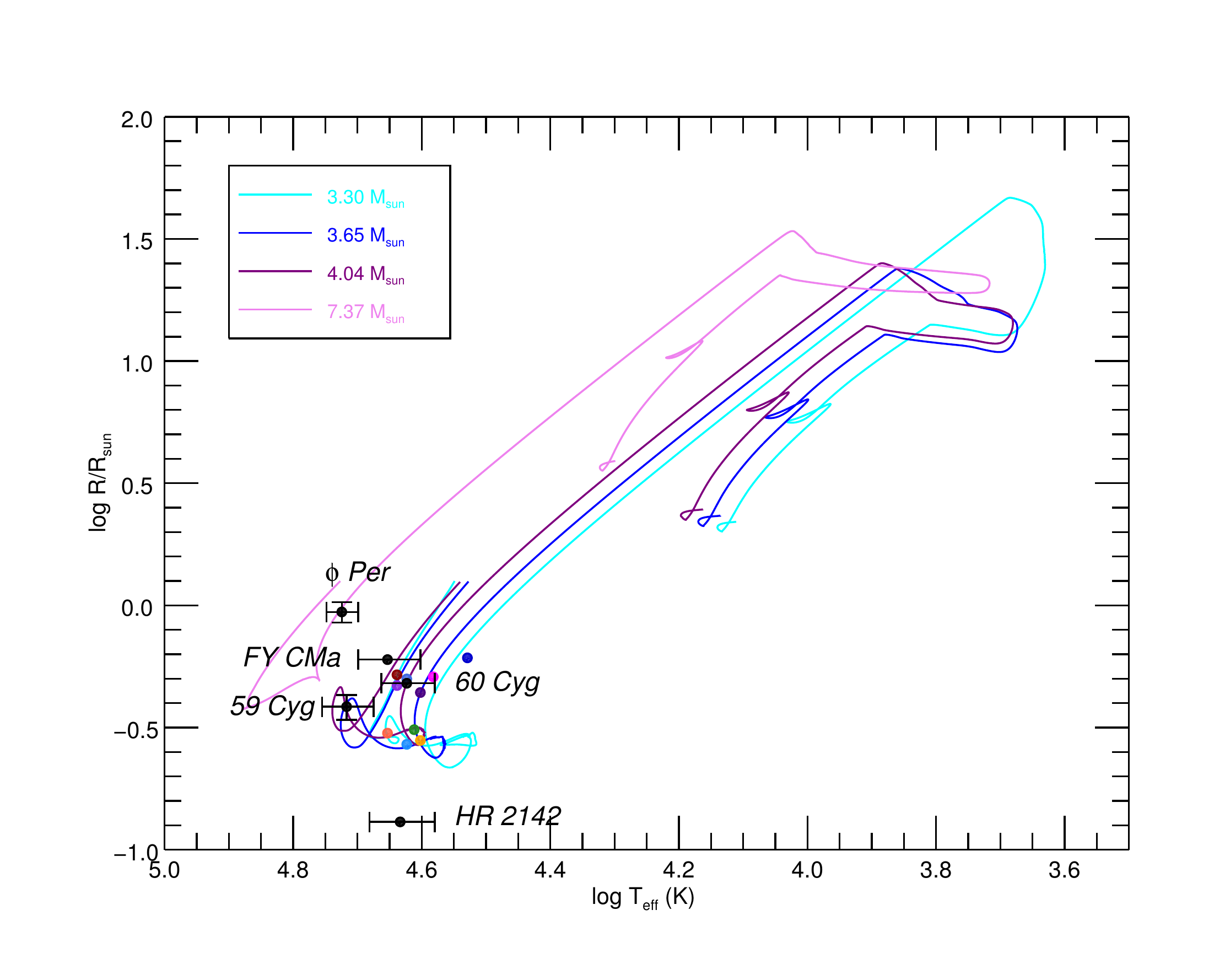}
\caption{The evolutionary tracks in the ($\log{T_{\rm eff}}$, $\log{R_\odot}$) plane 
of stripped stars with masses in a range of $3.30 - 7.37\ M_\odot$ adopted from \citet{gotberg2018}.  
The corresponding parameters of detected sdO stars are marked by filled circles and color coded as follows: HD 29441 (orange), HD 43544 (fuchsia), HD  51354 (blue violet), HD 55606 (forest green), HD 60855 (royal blue), HD 113120 (tomato), HD 137387 (indigo), HD 152478 (dodger blue), HD 157042 (medium blue), HD194335 (dark red). The five prior known Be+sdO binaries are shown in black circles.}
\label{fig:et1}
\end{figure*}

% Fig. 16 here
% Figure 16: Evolutionary track, logTeff VS logR 
\placefigure{fig:et1}

\begin{figure*}
\includegraphics[width=\textwidth]{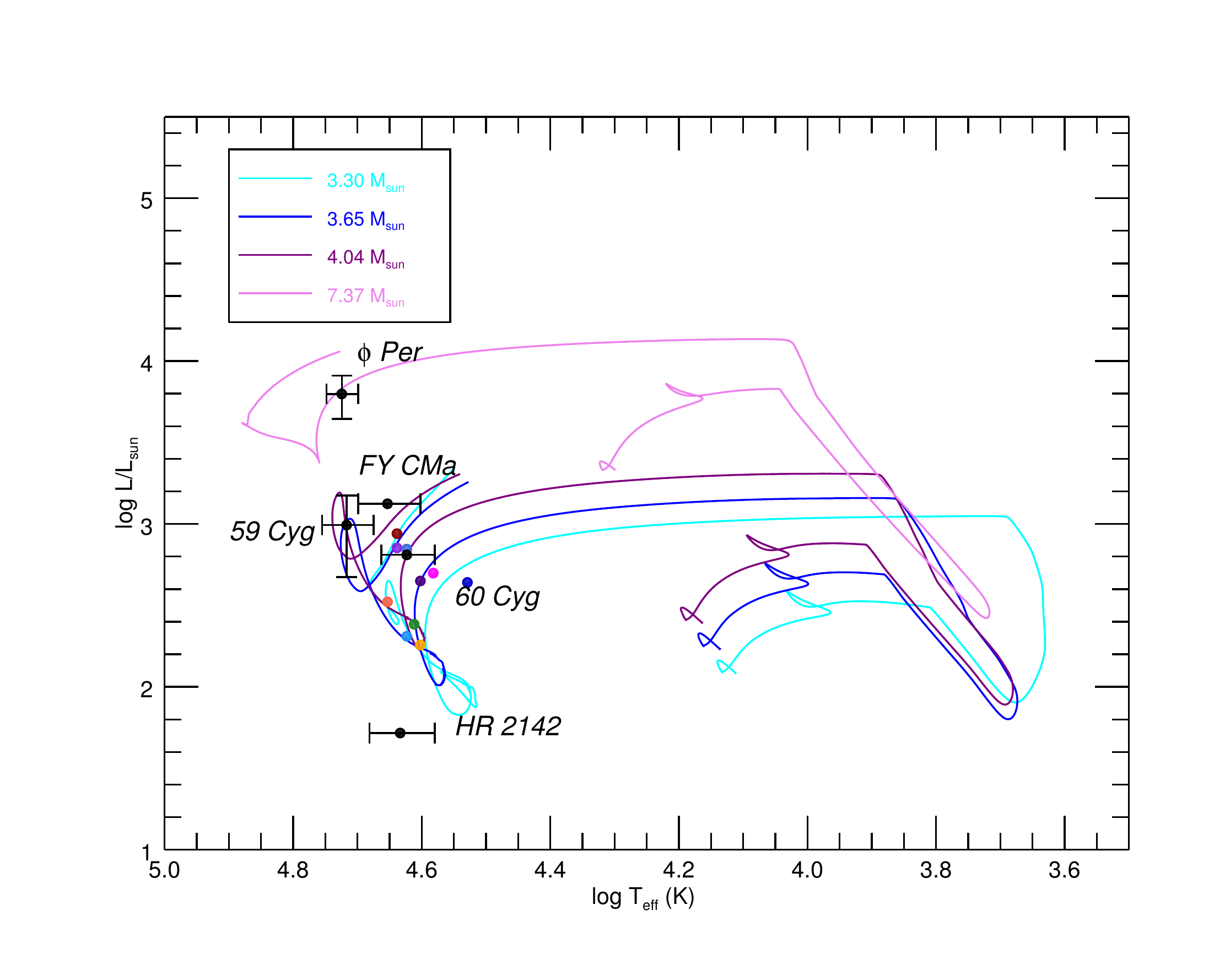}
\caption{The evolutionary tracks in the ($\log{T_{\rm eff}}$, $\log{L_\odot}$) plane 
of stripped stars with masses in a range of $3.3-7.37\ M_\odot$ adopted from \citet{gotberg2018}.
The corresponding parameters of detected sdO stars are marked by filled circles and color coded as follows: HD 29441 (orange), HD 43544 (fuchsia), HD  51354 (blue violet), HD 55606 (forest green), HD 60855 (royal blue), HD 113120 (tomato), HD 137387 (indigo), HD 152478 (dodger blue), HD 157042 (medium blue), HD194335 (dark red). The five prior known Be+sdO binaries are shown in black circles.}
\label{fig:et2}
\end{figure*}

% Fig. 17 here 
% Figure 17: Evolutionary track, logTeff VS logL
\placefigure{fig:et2}

We include in Figures 16 and 17 our estimates for the sdO stellar parameters as summarized in Table~3. We also plot the parameters for the 
five Be+sdO systems that were known prior to our study.  We see that most 
of the sdO companions do indeed have temperatures, radii, and luminosities 
that are comparable to the predictions from \citet{gotberg2018} for 
progenitors with masses in the range of 3 to $5 M_\odot$. The sdO companions in four of our stars (HD 29441, HD 55606, HD 113120, and HD 152478)  are shown at the He-core burning stage on the evolutionary track, and the rest of the sdO stars are at an inflated stage either from an early contraction phase or the later inflation phase at the He shell burning stage. If the masses of the sample Be stars are typical of those of single B-stars ($\approx 4$ to $8 M_\odot$), then these systems probably began in 
near equal mass binaries, and systemic mass loss during the mass transfer 
phase was probably minimal.  

The only system in the sample with a known orbit 
and mass estimates is HD~55606, and \citet{chojnowski2018} find a 
mass for the sdO star in the range 0.83 to $0.90 M_\odot$.  
The evolutionary tracks suggest there is an approximately linear 
mass -- luminosity relation during the He-core burning stage 
(see Fig.\ 3 in \citealt{gotberg2018}).  Using the luminosity estimate 
for the sdO star in HD~55606 from Table~3, the tracks predict a 
mass in the range of 0.77 to $0.91 M_\odot$, in agreement with 
observed estimate for the sdO star from \citet{chojnowski2018}. 

\citet{schootemeijer2018} created similar evolutionary tracks to study 
the current stage of the Be+sdO system $\phi$~Per.  They concluded that 
the sdO star's luminosity was large for its mass ($1.2 M_\odot$), and 
consequently the star may have entered the subsequent and short-lived 
He-shell burning stage of evolution.  This also appears to be the 
conclusion from the tracks from \citet{gotberg2018} that show that
the parameters of $\phi$~Per fall on a track for a more massive 
remnant ($1.85 M_\odot$).   \citet{schootemeijer2018} argued that 
most of the Be+sdO binaries would have fainter sdO components found 
in the longer lasting He-core burning stage.  The sdO companions 
discovered in the {\it HST}/STIS sample all have lower luminosities 
than that of $\phi$~Per, and this supports the idea that most 
such systems will be found in the fainter, He-core burning phase. 

The faintest of the sdO companions plotted in Figures 16 and 17 is 
that in the HR~2142 system.  \citet{peters2016} described how the 
CCF peak strength from the sdO companion has an orbital phase 
dependence in the case of HR~2142, and they argued that the 
sdO star was obscured by varying amounts by circumstellar gas. 
Thus, the implied radius and luminosity of the sdO component 
in HR~2142 are both lower limits to the actual values. 

%%%%%%%%%%%%%%%%%%%%% 
\newpage
\section{Conclusions}\label{sec:conc}

The primary goal of this investigation was to verify the presence
of the spectral lines of hot subdwarf stars through a CCF analysis
of FUV spectra from {\it HST}/STIS.  We succeeded in detecting the
CCF signal from the sdO star in nine of twelve candidates 
identified from lower quality {\it IUE} spectroscopy and in 
one case, HD~55606, where the sdO spectral features were first 
seen in optical spectra \citep{chojnowski2018}.  We can be confident 
that these detections are reliable, because the CCF peaks of the 
sdO component in these Be+sdO binaries are much narrower than 
the rotationally broadened component of the Be star, the peaks 
attain a maximum using a model CCF template for a star much 
hotter than the Be star, and the peaks display radial velocity 
variations indicative of Doppler shifts from orbital motion. 
On the other hand, the peak heights indicate that the sdO companions
are relatively faint compared to their Be star primaries, which 
makes their detection difficult by other means. In fact, in our sample of 10 confirmed Be+sdO systems, only HD~55606 was known to be a binary prior to this work.

The radial velocity measurements of the sdO components are an 
important first step in determining the orbital elements of the 
binaries and placing limits on the masses.  The only system in 
the sample with a known orbit is HD~55606 with an orbital period
of 93.76 days derived by  \citet{chojnowski2018}, and the velocities from 
the {\it HST}/STIS spectra are consistent with their radial 
velocity curve for the sdO star.  We combined the {\it HST}/STIS 
and {\it IUE} velocity measurements to make a preliminary 
orbit for HD~194335 that indicates an orbital period of 60.3 days
(see Appendix).  We can make a rough estimate of the orbital 
period $P$ of the other systems by assuming a probable Be star mass $M_1$, 
system mass ratio $q=M_2/M_1$, and inclination $i$.  
For a circular orbit, the semiamplitude of the sdO star will be
\begin{equation}
K_2 = {{\sin i} \over {1+q}} {{2\pi }\over {P}} M_1^{1/3} (1+q)^{1/3} P^{2/3}.
\end{equation}
If we assume $M_1 = 6 M_\odot$, $q=0.1$, and $i=60^\circ$, then 
the predicted semiamplitude is $K_2 = 315 ~{\rm km~s}^{-1} ~P^{-1/3}$,
where the period is measured in days.  

We can then make a numerical 
fit of the three observed radial velocities from {\it HST}/STIS 
by solving for three parameters: 
the period $P$, the epoch of maximum velocity $T_0$, 
and the systemic velocity $\gamma$.  The results range from $P=84$ days 
for HD~137837 to $P=346$ days for HD~60855.  These are nominally upper 
limits on the period because we make the simplifying assumption that 
the {\it HST}/STIS spectra cover only part of one orbit. 
Nevertheless, orbital periods on the order of months are what is 
found in the six other systems with known orbits and are consistent with the 
expectations for the enlargement of the orbit with mass transfer 
following mass ratio reversal \citep{wellstein2001}.

The CCF peaks from the sdO components are generally narrow, 
and the derived projected rotational velocity $V \sin i$ is 
small (unresolved) except in the case of HD~51354 where 
$V \sin i = 102$ km~s$^{-1}$.  We expect that the progenitors
of the sdO stars attained synchronous rotation due to tidal 
forcing when they filled their Roche lobe, and consequently, 
their projected rotational velocities are small because of 
the long orbital periods and very small radii.  The large 
$V \sin i$ of HD~51354 may be the result of spin up by 
reverse mass transfer.  The Be primary star loses mass into 
its outflowing equatorial disk, and some of this gas may 
end up being accreted by the sdO companion. The angular 
momentum carried by the accreted gas could lead to spin up. 
However, it is unknown why this process would occur in 
this Be+sdO system and not the others. 

The peak heights of the sdO CCF component vary with the 
assumed temperature of the model spectrum used to form the 
cross-correlation function.  We adopt a temperature that 
maximizes the CCF peak strength, and this yields effective 
temperatures in the range of 34 to 45~kK (Table~3). 
These are all much hotter than those of the Be stars and 
are close to model expectations for stripped stars (Section 7).

The CCF peak height also depends on the fractional flux 
contribution of the sdO spectrum to the total combined flux, 
and we used numerical simulations of model spectra for the
Be and sdO stars to derive the monochromatic flux ratio 
in each case (Table~3).  We can then compare the observed 
flux ratio with the surface flux ratio from the estimated
temperatures of the components to find the radius ratio. 
We then determined the Be star radius from a fit of the 
spectral energy distribution and the distance (Section 6)
to finally arrive at estimates of the sdO stellar radii
(Table~3).  Although the sdO stars contribute a minor 
fraction of the overall flux, it should be possible to 
find evidence of their spectral lines in high 
S/N optical spectra and to determine visual orbits for 
the closer systems through long baseline optical 
interferometry \citep{mourard2015}.

The question remains about why we did not detect any clear 
sdO CCF peak in the {\it HST}/STIS spectra of HD~157832, 
HD~191610, and HD~214168, even though such peaks were 
detected in CCFs from lower quality {\it IUE} spectra
\citep{wang2018}.  We note that in the latter two cases, 
there are many archival {\it IUE} spectra, but \citet{wang2018}
were only able to find significant CCF peaks for about 
half of those spectra (see Appendix).  This suggests that 
there is an inherent temporal variability in the contribution
of the sdO component to the combined FUV spectrum.  
A similar situation was found by \citet{peters2016} for the 
Be+sdO system HR~2142, in which the detection of the peak 
was limited to certain parts of the orbit.  We suspect 
that this variability is due changes in obscuration of the
sdO star by nearby circumstellar gas.  Thus, the 
non-detection of the sdO CCF peaks in these three cases 
does not necessarily imply that these are single Be stars
that lack sdO companions. 

We found that the derived temperatures and radii of the 
sdO components are broadly consistent with model predictions
for stars stripped of their envelopes by binary interaction 
\citep{gotberg2018}. These parameters agree with the
long-lived stage of He-core burning in the remnants that 
lasts about $10\% \sim 20\%$ of the star's lifetime.  There is a 
predicted relation between mass and luminosity at this stage, 
and our derived luminosity for the sdO in HD~55606 corresponds
to a mass that agrees within uncertainties with that derived 
from the observations presented by \citet{chojnowski2018}.

The {\it HST}/STIS spectra of this sample of Be stars have 
given us a remarkable picture of how stars are transformed 
through binary interactions to create hot, stripped down subdwarfs.
These spectra reveal the temperatures, radii, and hence, luminosities
of the sdO stars, but what remains is to determine their masses 
by determining the full orbital elements.  This is a difficult 
task because the sdO spectral lines in the optical part of the
spectrum are weak and few in number, and the lines of the 
Be star are very shallow and broad ($V \sin i \approx 300$ km~s$^{-1}$)
compared to the orbital semiamplitude ($K_1 < 10$ km~s$^{-1}$). Nevertheless, this observational work is essential in order 
to derive the component masses and to make detailed comparisons 
with model predictions \citep{gotberg2018}.  Searches for 
Be+sdO systems among Be stars in clusters are particularly 
important to compare their ages to model predictions. 
The fact that Be stars lack main sequence companions compared
to normal B stars is evidence that many Be stars are the 
descendants of pairs of interacting binaries \citep{bodensteiner2020a},
so there should be many examples of this hidden stage of evolution 
that remain to be discovered.

%%%%%%%%%%%%%%%%%%%%%%%%%%%
\acknowledgments
We thank Denise Taylor of STScI for her help in planning the observations with \emph{HST}. This work was supported by NASA through a grant from the Space Telescope Science Institute, under program GO-15659. Institutional support was provided from the GSU College of Arts and Science.  

%% To help institutions obtain information on the effectiveness of their 
%% telescopes the AAS Journals has created a group of keywords for telescope 
%% facilities.
%
%% Following the acknowledgments section, use the following syntax and the
%% \facility{} or \facilities{} macros to list the keywords of facilities used 
%% in the research for the paper.  Each keyword is check against the master 
%% list during copy editing.  Individual instruments can be provided in 
%% parentheses, after the keyword, but they are not verified.

\vspace{5.5mm}
\facilities{HST (STIS), Gemini:Gillett (\`Alopeke)}

%% Similar to \facility{}, there is the optional \software command to allow 
%% authors a place to specify which programs were used during the creation of 
%% the manuscript. Authors should list each code and include either a
%% citation or url to the code inside ()s when available.

%\software{
 %         }

%% Appendix material should be preceded with a single \appendix command.
%% There should be a \section command for each appendix. Mark appendix
%% subsections with the same markup you use in the main body of the paper.

%% Each Appendix (indicated with \section) will be lettered A, B, C, etc.
%% The equation counter will reset when it encounters the \appendix
%% command and will number appendix equations (A1), (A2), etc. The
%% Figure and Table counter will not reset.

%%%%%%%%%%%%%%%%%%%%%%%%%%%%%%
\appendix

% Notes on Individual Stars 
\section{Notes on Individual Stars}\label{sec:notes}

{\sl HD~29441 (V1150 Tau)}. There is only one {\it IUE} high dispersion 
spectrum for this target, and the CCF peak of the hot component appears
with a radial velocity of $V_r=-60.7 \pm 2.5$ km~s$^{-1}$ \citep{wang2018},
compared to the positive radial velocities found here (Table 2). 
\citet{Slettebak1997} note that the star is distant from the Galactic plane, 
with $z=-308$ pc for the Gaia DR2 distance (Table 4).  \citet{Slettebak1997} 
describe the broad lines in the optical spectrum, which they classify as 
B2~III/IVe. 

{\sl HD~43544 (HR 2249)}. \citet{levenhagen2006} determined parameters for
this star that they classify as B2/3~Ve, and they estimate an evolutionary 
mass of $8.5 M_\odot$ for the Be star.  The {\it HST}/STIS spectra CCF peaks show a 
large velocity range that spans the single {\it IUE} measurement \citep{wang2018}.
\citet{huang2010} find a small velocity variation for the Be component 
between two observations. 

{\sl HD~51354 (QY Gem)}. The CCFs of the hot companion are remarkably broad compared 
to all the other detections, and this was also found in the CCFs from two {\it IUE} 
spectra \citep{wang2018}.  These CCF peaks show large velocity variations, 
while velocity measurements of the Be star display relatively little scatter
\citep{chojnowski2017}.

{\sl HD~55606}. The orbital variations of the hot, narrow-lined component 
were discovered in optical spectra by \citet{chojnowski2018} who determined a double-lined
orbital solution.  The {\it HST}/STIS spectra CCF velocities are consistent with their orbit. 
There are no {\it IUE} observations for this target. 

{\sl HD~60855 (HR 2921; V378 Pup)}. This star is a candidate blue straggler 
in the open cluster NGC~2422 \citep{pols1991}, and it has several wide and 
faint companions listed in the Washington Double Star catalog.  There are 
six {\it IUE} CCF velocity measurements, all of which are lower than the three 
measurements presented in Table 2.  

{\sl HD~113120 (HR 4930; LS Mus)}. There are large velocity variations in both
the three {\it IUE} \citep{wang2018} and {\it HST}/STIS spectra CCF measurements for the hot component.  
This star has a nearby companion at a separation of 0.56 arcsec \citep{Hartkopf1996}.
It was also resolved by {\it Hipparcos} with a magnitude difference of 
$\triangle H_p = 2.84 \pm 0.04$ mag \citep{esa1997}. 
This companion falls outside the {\it HST}/STIS 
aperture, but its flux does contribute to the TD-1 and Johnson $UBV$ measurements. 
Consequently, we used $\triangle H_p$ to add $0.077 \pm 0.003$ mag to correct the TD-1 
and $UBV$ measurements (assuming similar colors for both components) to estimate the 
magnitudes and fluxes without the companion's contribution.  The SED fitting was 
made using these adjusted supplementary fluxes.  The distance from Gaia DR2 is very 
uncertain (756 to 1496 pc; \citealt{bailer-jones2018}) probably due to complications
from the companion's flux. \citet{Krelowski2017} list three distance estimates:
250 pc from \ion{Ca}{2} interstellar line strength, 307 pc from {\it Hipparcos} 
\citep{leeuwen2007}, and 351 pc from spectrophotometric fits.  We adopt the 
{\it Hipparcos} result in Table 4. 

{\sl HD~137387 ($\kappa^1$ Aps; HR 5730)}. 
\citet{Boubert2018} suggest that the object is a runaway star with a peculiar 
space velocity of 57 km~s$^{-1}$, but \citet{Jilinski2010} derive a much lower
value of 32 km~s$^{-1}$. \citet{Jilinski2010} obtained seven radial velocities 
for the Be star and concluded that it is a binary.  There are large velocity 
variations for the hot component in both the four {\it IUE} and three {\it HST}/STIS 
spectra CCF measurements. 

{\sl HD~152478 (HR 6274; V846 Ara)}. There are two {\it IUE} velocity measurements
with a large difference \citep{wang2018}, while all three {\it HST}/STIS spectra CCF measurements
are similar and redshifted. \citet{Jilinski2010} found that the Be star absorption lines
are too broad ($V\sin i = 370$ km~s$^{-1}$) for reliable radial velocity measurements. 

{\sl HD~157042 ($\iota$ Ara; HR 6451)}. This star displays relatively large velocity
variations in both the four {\it IUE} and three {\it HST}/STIS spectra CCF measurements for the 
hot companion.  The violet and red peaks of the H$\alpha$ emission show 
relative strength variations \citep{Dachs1986,Dachs1992,Mennickent1991} that 
may be related to orbital phase.

{\sl HD~157832 (V750 Ara)}. This candidate binary was identified by \citet{wang2018} 
from CCF weak peaks observed in two {\it IUE} spectra.  The residual peaks from
the {\it HST}/STIS spectra CCFs are also weak (Fig.\ 7), which suggests that any hot component
in this system is faint.  The Gaia DR2 distance of 1078 pc \citep{bailer-jones2018} 
may be too large.  For example, \citet{kozok1985} gives a lower distance of 780 pc, 
and \citet{lopes2011} suggest an even smaller value of 530 pc. 
Thus, the Be star radius given in Table~4 may be an overestimate. 
\citet{lopes2011} discuss the hard and intense X-ray 
flux from this target, making this star an analog of the Be star X-ray emitter $\gamma$~Cas.
\citet{Langer2020} argue that such X-ray emission may originate where the wind 
from a hot sdO star strikes the outer region of the Be star disk. 

{\sl HD~191610 (28 Cyg; V1624 Cyg)}. The CCF residuals shown in Figure 7 
indicate little or no evidence of a hot component.  \citet{wang2018} 
examined 46 high dispersion SWP spectra from {\it IUE} to search for 
evidence of a CCF peak from a hot companion, and a peak was measured 
for only 25 of the 46 spectra (with CCF peak amplitude of $\approx 0.025$). 
This suggests that there is some kind of temporal variability that makes
detection more favorable at some epochs rather than others.  Thus, we still 
regard HD~191610 as a viable Be+sdO candidate system, despite the non-detection
in the HST/STIS CCFs.  \citet{Becker2015} obtained 64 measurements of the 
Be star radial velocity from high signal-to-noise and high dispersion optical
spectra, and they measured a residual ``jitter'' of 13.4 km~s$^{-1}$, which is 
probably comparable to the orbital semiamplitude of the Be star. 

{\sl HD~194335 (HR 7807; V2119 Cyg)}. The CCF peaks show large velocity variations
in both the four {\it IUE} measurements \citep{wang2018} and in the three {\it HST}/STIS 
measurements (Table 2).  We combined these to make a preliminary circular orbital fit 
with period $P= 60.286 \pm 0.010$~d, epoch of maximum velocity 
$T_0=$ HJD~2,458,721.2 $\pm 0.6$, systemic velocity $\gamma = -18.9 \pm 1.2$ km~s$^{-1}$, 
and semiamplitude $K_1= 75.5 \pm 2.4$ km~s$^{-1}$.  The derived systemic 
velocity is comparable to the median from four measurements of the 
Be star, $-30$ km~s$^{-1}$, made by \citet{Costado2017}.

{\sl HD~214168 (8~Lac~A; HR 8603)}.  This star is a member of the Lac~OB1 association
\citep{Kaltcheva2009} and is the northern and brighter component of a pair 
separated by 22 arcsec.  The target is also a close binary resolved by 
speckle interferometry (CHR~112 Aa,Ab; \citealt{McAlister1987}). 
Recently \citet{Tokovinin2019} presented an orbital solution based upon 
speckle measurements that yields an orbital period of 42 years and an 
angular semimajor axis of 0.057 arcsec.  The derived total mass is 
$17 \pm 8 M_\odot$ based upon the distance from Gaia DR2 ($542 \pm 30$ pc;
\citealt{bailer-jones2018}).  We obtained speckle observations with the \`Alopeke 
instrument on the Gemini North Telescope on 2019 October 12 and with the 
562 nm and 832 nm filters.  The measured separation was 0.041 arcsec and 
position angle 133.4 degrees east from north, and this position is consistent
with the orbit from \citet{Tokovinin2019}.  The average magnitude difference in 
the optical range from the \`Alopeke and Tokovinin et al.\ measurements is 
$\Delta m = 0.43 \pm 0.10$ mag.  

The {\it HST}/STIS observations are centered on the brighter object within an aperture 
with a projected size of $0.2\times0.05$ arcsec on the sky, so some flux from
the Ab component may be recorded in the spectra.  We found that the measured 
flux in the 1300 - 1400 \AA\ range was $(4.27, 4.72, 3.71)\times 10^{-10}$ 
erg~cm$^{-2}$~s$^{-1}$~\AA $^{-1}$ for the three observations consecutively, 
a much larger relative variation than observed in other targets, so we assume 
that the spectrum recorded more (less) flux from component Ab in the second (third)
observation. However, there is no discernible difference between the three spectra,
which suggests that the Ab component has a similar spectral appearance 
with broad lines and a temperature like that of the Be star. 

We performed a CCF analysis to search for the spectral lines of the Ab component by
constructing CCFs for a grid of lower temperature models and then removing 
the associated contribution to the CCF from the Be star in the same way as we did in
the search for the signal of a hot companion.  We detected a weak and narrow 
signal in the residual CCFs that is shown in Figure 18.  The residual peaks 
attained maximum strength for CCF model spectra with $T_{\rm eff} = 23 \pm 4$ kK, 
and the width and height of the peak suggest $V \sin i = 31 \pm 5$ km~s$^{-1}$
and $f_2/f_1 \approx 0.07$, respectively.  The measured velocities from 
Gaussian fits of the peaks are $-31.6\pm 1.6$, $-15.2\pm 1.8$, and $-24.1\pm 1.3$
km~s$^{-1}$ for the three consecutive observations.  We doubt that these 
peaks originate in the Ab component, because the flux ratio is much too small 
for the bright Ab component and because there is no evidence in published optical 
spectra for a narrow-lined spectral component as bright as Ab.  It is unlikely 
that this CCF peak corresponds to a cooler, stripped companion, because the 
associated temperature of the feature is lower than that of the Be star, and 
in all the other cases, we find that the stripped down donor star is hotter than the 
bright mass gainer star.  We also rule out an origin in interstellar lines, 
because we avoided these in forming the CCFs and the peaks appear to have
a variable radial velocity.  Instead, we suspect that this spectral signature
is the result of line absorption from the disk of the Be star.  \citet{gies1998}
found evidence of such absorption features in the Be+sdO system $\phi$~Per 
that appeared when the hot companion was in the foreground in the orbit. 
They surmised that these narrow ``shell'' absorption lines (mainly \ion{Fe}{4}) 
are formed as we view the Be star through heated portions of the outer disk. 
If the weak CCF signal shown in Figure 18 is related to shell absorption lines, 
then the like occcurrence of these in the spectrum of $\phi$~Per suggests that 
HD~214168 may also have a close hot companion.  However, there is no obvious
signal from a hot companion in the CCF residuals (Fig.\ 8). 

\begin{figure*}
\includegraphics[width=\textwidth]{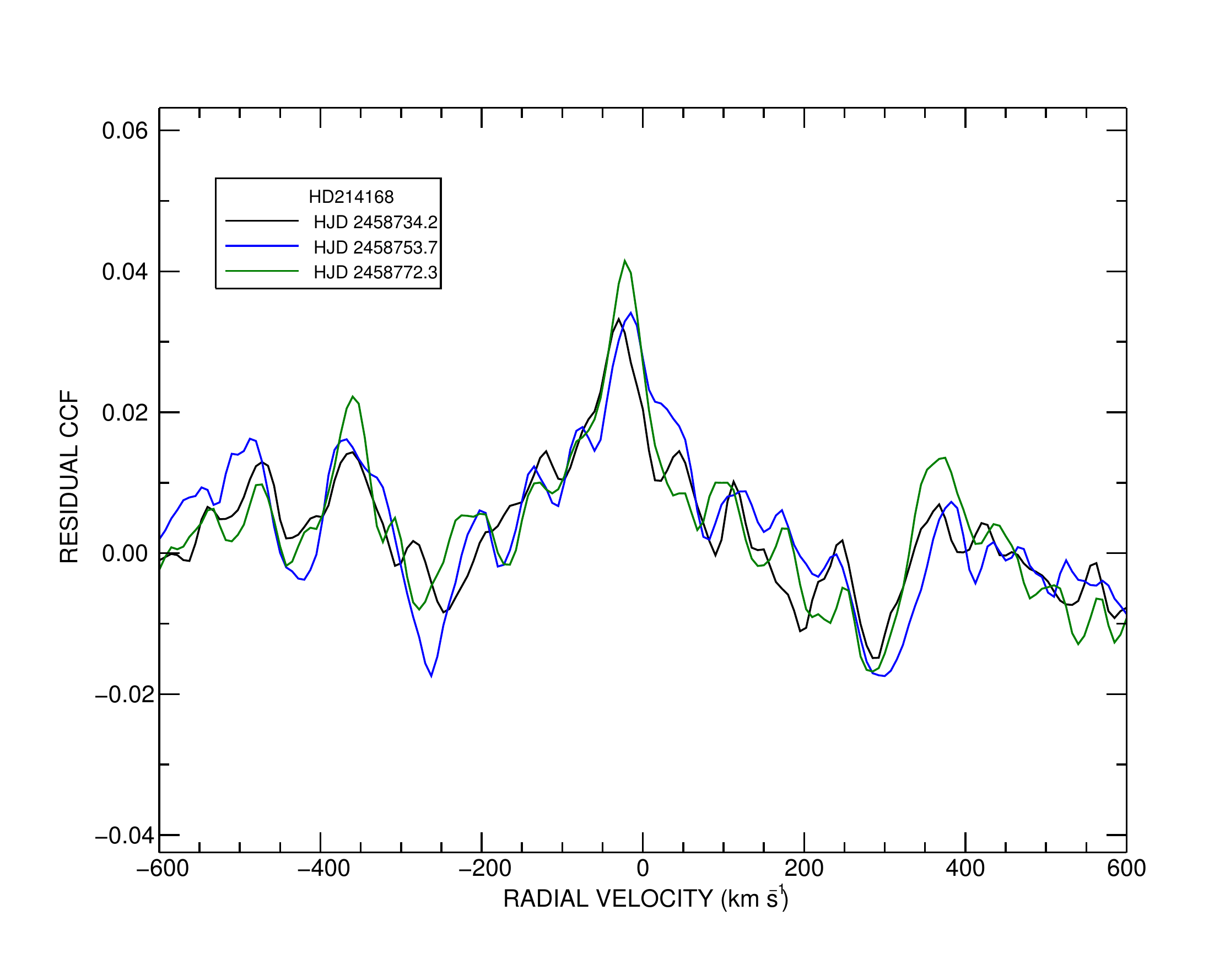}
\caption{CCF residual plot of HD 214168 made through correlation with a cool 
model spectrum with $T_{\rm eff} = 23.3$ kK.}
\label{fig:214168}
\end{figure*}

% Figure 17 here
\placefigure{fig:214168}

We fit the SED of HD~214168 by assuming that the third (faintest) {\it HST}/STIS spectrum 
has a flux representative of the Be star alone, and then we adjusted the 
observed Johnson $UBV$ magnitudes to give the Be star (assumed component Aa)
magnitude by adding $0.56 \pm 0.04$ mag (assuming no significant color 
difference between Aa and Ab).   The fit of the SED together 
with the Gaia DR2 distance yields a stellar radius of $6.0 \pm 1.1 R_\odot$ for Aa.
The radius of Ab from the optical magnitude difference is then  $4.9 \pm 0.9 R_\odot$ 
or somewhat larger if Ab is cooler than Aa.  We caution that these 
values may be slightly overestimated because the FUV flux recorded 
in the third {\it HST}/STIS spectrum may include a small contribution from Ab. 

\citet{wang2018} were able to measure a CCF peak for an sdO companion in 
9 of the 20 {\it IUE} spectra they analysed.  This suggests that there are 
temporal variations that influence the visibility of the sdO star's photospheric flux,
as found in the case of HD~191610.  Thus, the non-detection 
in the CCFs from the {\it HST}/STIS spectra may be due to the making of the observations at a time when the sdO star was obscured from view. 

\bibliography{ms.bib}{}
\bibliographystyle{aasjournal}

%% This command is needed to show the entire author+affiliation list when
%% the collaboration and author truncation commands are used.  It has to
%% go at the end of the manuscript.
%\allauthors

%% Include this line if you are using the \added, \replaced, \deleted
%% commands to see a summary list of all changes at the end of the article.
\listofchanges

\end{document}